%% file: 2P1-LQG-On-The-Edge.tex
\documentclass[english]{article}
\usepackage{mathpazo}
\usepackage[latin9]{inputenc}
\usepackage[letterpaper]{geometry}
\usepackage{babel}
\usepackage{verbatim}
\usepackage{amsmath}
\usepackage{amssymb}
\usepackage{esint}
\usepackage[unicode=true,
 bookmarks=true,bookmarksnumbered=false,bookmarksopen=false,
 breaklinks=false,pdfborder={0 0 1},backref=false,colorlinks=false]
 {hyperref}

\makeatletter
\usepackage{babel}

\usepackage[svgnames]{xcolor}
\usepackage{tikz}
\usetikzlibrary{decorations.markings}
\usetikzlibrary{shapes.geometric}
\usetikzlibrary{calc}
\usetikzlibrary{arrows}

\pgfdeclarelayer{edgelayer}
\pgfdeclarelayer{nodelayer}
\pgfsetlayers{edgelayer,nodelayer,main}

\tikzstyle{none}=[inner sep=0pt]

\tikzstyle{Vertex}=[circle,fill=black,draw=black]
\tikzstyle{Node}=[rectangle,fill=blue,draw=blue]
\tikzstyle{Point}=[circle,fill=red,draw=red,scale=0.33]

\tikzstyle{Edge}=[-,draw=black]
\tikzstyle{Link}=[-,dashed,draw=blue]
\tikzstyle{Segment}=[-,dotted,very thick,draw=red]
\tikzstyle{LinkArrow}=[-,draw=blue,
  postaction={decorate},decoration={markings,
  mark=at position .66 with {\arrow{latex}}}]
\tikzstyle{SegmentArrow}=[-,draw=red,
  postaction={decorate},decoration={markings,
  mark=at position .66 with {\arrow{latex}}}]
\def\centerarc [#1] (#2) (#3:#4:#5); { \draw[#1] ($(#2)+({#5*cos(#3)},{#5*sin(#3)})$) arc (#3:#4:#5); }
\newcommand{\midarrow}{\tikz \draw[-triangle 90] (0,0) -- +(.1,0);}
\newcommand{\midarrowop}{\tikz \draw[-triangle 90] (0,0) -- +(-.1,0);}

\usepackage[margin=1cm]{caption}

\renewcommand{\[}{\begin{equation}}
\renewcommand{\]}{\end{equation}}

\DeclareMathOperator{\e}{e}

\DeclareMathOperator{\ii}{i}

\DeclareMathOperator{\Res}{Res}

\DeclareMathOperator{\tr}{tr}
\DeclareMathOperator{\Tr}{Tr}

\makeatother

\begin{document}

\global\long\def\A{\mathbf{A}}%
\global\long\def\B{\mathbf{B}}%
\global\long\def\C{\mathbf{C}}%
\global\long\def\D{\mathbf{D}}%
\global\long\def\E{\mathbf{E}}%
\global\long\def\F{\mathbf{F}}%
\global\long\def\G{\mathbf{G}}%
\global\long\def\H{\mathbf{H}}%
\global\long\def\I{\mathbf{I}}%
\global\long\def\J{\mathbf{J}}%
\global\long\def\K{\mathbf{K}}%
\global\long\def\LL{\mathbf{L}}%
\global\long\def\M{\mathbf{M}}%
\global\long\def\N{\mathbf{N}}%
\global\long\def\OO{\mathbf{O}}%
\global\long\def\P{\mathbf{P}}%
\global\long\def\Q{\mathbf{Q}}%
\global\long\def\RR{\mathbf{R}}%
\global\long\def\SS{\mathbf{S}}%
\global\long\def\T{\mathbf{T}}%
\global\long\def\U{\mathbf{U}}%
\global\long\def\V{\mathbf{V}}%
\global\long\def\W{\mathbf{W}}%
\global\long\def\X{\mathbf{X}}%
\global\long\def\Y{\mathbf{Y}}%
\global\long\def\Z{\mathbf{Z}}%
\global\long\def\a{\mathbf{a}}%
\global\long\def\b{\mathbf{b}}%
\global\long\def\c{\mathbf{c}}%
\global\long\def\dd{\mathbf{d}}%
\global\long\def\ee{\mathbf{e}}%
\global\long\def\f{\mathbf{f}}%
\global\long\def\g{\mathbf{g}}%
\global\long\def\h{\mathbf{h}}%
\global\long\def\iii{\mathbf{i}}%
\global\long\def\j{\mathbf{j}}%
\global\long\def\k{\mathbf{k}}%
\global\long\def\l{\boldsymbol{l}}%
\global\long\def\el{\boldsymbol{\ell}}%
\global\long\def\m{\mathbf{m}}%
\global\long\def\n{\mathbf{n}}%
\global\long\def\o{\mathbf{o}}%
\global\long\def\p{\mathbf{p}}%
\global\long\def\q{\mathbf{q}}%
\global\long\def\r{\mathbf{r}}%
\global\long\def\s{\mathbf{s}}%
\global\long\def\t{\mathbf{t}}%
\global\long\def\u{\mathbf{u}}%
\global\long\def\v{\mathbf{v}}%
\global\long\def\w{\mathbf{w}}%
\global\long\def\x{\mathbf{x}}%
\global\long\def\y{\mathbf{y}}%
\global\long\def\z{\mathbf{z}}%
\global\long\def\Ga{\boldsymbol{\Gamma}}%
\global\long\def\De{\boldsymbol{\Delta}}%
\global\long\def\Th{\boldsymbol{\Theta}}%
\global\long\def\La{\boldsymbol{\Lambda}}%
\global\long\def\Xii{\boldsymbol{\Xi}}%
\global\long\def\Pii{\boldsymbol{\Pi}}%
\global\long\def\Si{\boldsymbol{\Sigma}}%
\global\long\def\Ph{\boldsymbol{\Phi}}%
\global\long\def\Ps{\boldsymbol{\Psi}}%
\global\long\def\Om{\boldsymbol{\Omega}}%
\global\long\def\al{\boldsymbol{\alpha}}%
\global\long\def\be{\boldsymbol{\beta}}%
\global\long\def\ga{\boldsymbol{\gamma}}%
\global\long\def\de{\boldsymbol{\delta}}%
\global\long\def\vep{\boldsymbol{\varepsilon}}%
\global\long\def\ep{\boldsymbol{\epsilon}}%
\global\long\def\ze{\boldsymbol{\zeta}}%
\global\long\def\et{\boldsymbol{\eta}}%
\global\long\def\th{\boldsymbol{\theta}}%
\global\long\def\io{\boldsymbol{\iota}}%
\global\long\def\ka{\boldsymbol{\kappa}}%
\global\long\def\la{\boldsymbol{\lambda}}%
\global\long\def\muu{\boldsymbol{\mu}}%
\global\long\def\nuu{\boldsymbol{\nu}}%
\global\long\def\xii{\boldsymbol{\xi}}%
\global\long\def\pii{\boldsymbol{\pi}}%
\global\long\def\rhh{\boldsymbol{\rho}}%
\global\long\def\si{\boldsymbol{\sigma}}%
\global\long\def\ta{\boldsymbol{\tau}}%
\global\long\def\ups{\boldsymbol{\upsilon}}%
\global\long\def\ph{\boldsymbol{\phi}}%
\global\long\def\vph{\boldsymbol{\varphi}}%
\global\long\def\ch{\boldsymbol{\chi}}%
\global\long\def\ps{\boldsymbol{\psi}}%
\global\long\def\om{\boldsymbol{\omega}}%
\global\long\def\zr{\mathbf{0}}%
\global\long\def\on{\mathbf{1}}%
\global\long\def\na{\boldsymbol{\nabla}}%
\global\long\def\AAb{\boldsymbol{\mathcal{A}}}%
\global\long\def\BBb{\boldsymbol{\mathcal{B}}}%
\global\long\def\CCb{\boldsymbol{\mathcal{C}}}%
\global\long\def\FFb{\boldsymbol{\mathcal{F}}}%
\global\long\def\LLb{\boldsymbol{\mathcal{L}}}%
\global\long\def\MMb{\boldsymbol{\mathcal{M}}}%
\global\long\def\PPb{\boldsymbol{\mathcal{P}}}%
\global\long\def\QQb{\boldsymbol{\mathcal{Q}}}%
\global\long\def\RRb{\boldsymbol{\mathcal{R}}}%
\global\long\def\eh{\mathbf{\hat{e}}}%
\global\long\def\kh{\mathbf{\hat{k}}}%
\global\long\def\lh{\hat{\boldsymbol{l}}}%
\global\long\def\elh{\hat{\boldsymbol{\ell}}}%
\global\long\def\nh{\mathbf{\hat{n}}}%
\global\long\def\pph{\mathbf{\hat{p}}}%
\global\long\def\rh{\mathbf{\hat{r}}}%
\global\long\def\vh{\mathbf{\hat{v}}}%
\global\long\def\xh{\mathbf{\hat{x}}}%
\global\long\def\yh{\mathbf{\hat{y}}}%
\global\long\def\zh{\mathbf{\hat{z}}}%
\global\long\def\AH{\mathbf{\hat{A}}}%
\global\long\def\BH{\mathbf{\hat{B}}}%
\global\long\def\EH{\mathbf{\hat{E}}}%
\global\long\def\JH{\mathbf{\hat{J}}}%
\global\long\def\LH{\mathbf{\hat{L}}}%
\global\long\def\NH{\mathbf{\hat{N}}}%
\global\long\def\RH{\mathbf{\hat{R}}}%
\global\long\def\SH{\mathbf{\hat{S}}}%
\global\long\def\TH{\mathbf{\hat{T}}}%
\global\long\def\VH{\mathbf{\hat{V}}}%
\global\long\def\XH{\mathbf{\hat{X}}}%
\global\long\def\thh{\hat{\boldsymbol{\theta}}}%
\global\long\def\rhh{\hat{\boldsymbol{\rho}}}%
\global\long\def\phh{\hat{\boldsymbol{\phi}}}%
\global\long\def\omh{\hat{\boldsymbol{\omega}}}%
\global\long\def\hL{\hat{L}}%
\global\long\def\hR{\hat{R}}%
\global\long\def\hh{\hat{h}}%
\global\long\def\hg{\hat{g}}%
\global\long\def\Ab{\bar{A}}%
\global\long\def\Bb{\bar{B}}%
\global\long\def\Cb{\bar{C}}%
\global\long\def\Db{\bar{D}}%
\global\long\def\Eb{\bar{E}}%
\global\long\def\Hb{\bar{H}}%
\global\long\def\Ib{\bar{I}}%
\global\long\def\Jb{\bar{J}}%
\global\long\def\Kb{\bar{K}}%
\global\long\def\Lb{\bar{L}}%
\global\long\def\Nb{\bar{N}}%
\global\long\def\Pb{\bar{P}}%
\global\long\def\Qb{\bar{Q}}%
\global\long\def\Sb{\bar{S}}%
\global\long\def\Tb{\bar{T}}%
\global\long\def\Ub{\bar{U}}%
\global\long\def\Vb{\bar{V}}%
\global\long\def\Xb{\bar{X}}%
\global\long\def\Yb{\bar{Y}}%
\global\long\def\Zb{\bar{Z}}%
\global\long\def\ab{\bar{a}}%
\global\long\def\bb{\bar{b}}%
\global\long\def\cb{\bar{c}}%
\global\long\def\db{\bar{d}}%
\global\long\def\eb{\bar{e}}%
\global\long\def\fb{\bar{f}}%
\global\long\def\gb{\bar{g}}%
\global\long\def\hb{\bar{h}}%
\global\long\def\ib{\bar{i}}%
\global\long\def\jb{\bar{j}}%
\global\long\def\kb{\bar{k}}%
\global\long\def\lb{\bar{l}}%
\global\long\def\elb{\bar{\ell}}%
\global\long\def\mb{\bar{m}}%
\global\long\def\nb{\bar{n}}%
\global\long\def\pb{\bar{p}}%
\global\long\def\qb{\bar{q}}%
\global\long\def\rb{\bar{r}}%
\global\long\def\sb{\bar{s}}%
\global\long\def\tb{\bar{t}}%
\global\long\def\ub{\bar{u}}%
\global\long\def\vbb{\bar{v}}%
\global\long\def\wb{\bar{w}}%
\global\long\def\xb{\bar{x}}%
\global\long\def\yb{\bar{y}}%
\global\long\def\zb{\bar{z}}%
\global\long\def\pab{\bar{\partial}}%
\global\long\def\Gab{\bar{\Gamma}}%
\global\long\def\Pib{\bar{\Pi}}%
\global\long\def\Sib{\bar{\Sigma}}%
\global\long\def\Phb{\bar{\Phi}}%
\global\long\def\Xib{\bar{\Xi}}%
\global\long\def\Psb{\bar{\Psi}}%
\global\long\def\Thb{\bar{\Theta}}%
\global\long\def\alb{\bar{\alpha}}%
\global\long\def\gab{\bar{\gamma}}%
\global\long\def\deb{\bar{\delta}}%
\global\long\def\vepb{\bar{\varepsilon}}%
\global\long\def\epb{\bar{\epsilon}}%
\global\long\def\zeb{\bar{\zeta}}%
\global\long\def\lab{\bar{\lambda}}%
\global\long\def\nub{\bar{\nu}}%
\global\long\def\xib{\bar{\xi}}%
\global\long\def\rhb{\bar{\rho}}%
\global\long\def\sib{\bar{\sigma}}%
\global\long\def\phb{\bar{\phi}}%
\global\long\def\psb{\bar{\psi}}%
\global\long\def\omb{\bar{\omega}}%
\global\long\def\chb{\bar{\chi}}%
\global\long\def\dpp#1{\frac{\partial#1}{\partial p}}%
\global\long\def\dq#1{\frac{\partial#1}{\partial q}}%
\global\long\def\dr#1{\frac{\partial#1}{\partial r}}%
\global\long\def\drr#1{\frac{\partial^{2}#1}{\partial r^{2}}}%
\global\long\def\ds#1{\frac{\partial#1}{\partial s}}%
\global\long\def\dt#1{\frac{\partial#1}{\partial t}}%
\global\long\def\dtt#1{\frac{\partial^{2}#1}{\partial t^{2}}}%
\global\long\def\du#1{\frac{\partial#1}{\partial u}}%
\global\long\def\duu#1{\frac{\partial^{2}#1}{\partial u^{2}}}%
\global\long\def\dv#1{\frac{\partial#1}{\partial v}}%
\global\long\def\dvv#1{\frac{\partial^{2}#1}{\partial v^{2}}}%
\global\long\def\dx#1{\frac{\partial#1}{\partial x}}%
\global\long\def\dxx#1{\frac{\partial^{2}#1}{\partial x^{2}}}%
\global\long\def\dy#1{\frac{\partial#1}{\partial y}}%
\global\long\def\dyy#1{\frac{\partial^{2}#1}{\partial y^{2}}}%
\global\long\def\dz#1{\frac{\partial#1}{\partial z}}%
\global\long\def\dzz#1{\frac{\partial^{2}#1}{\partial z^{2}}}%
\global\long\def\dph#1{\frac{\partial#1}{\partial\phi}}%
\global\long\def\dphph#1{\frac{\partial^{2}#1}{\partial\phi^{2}}}%
\global\long\def\drh#1{\frac{\partial#1}{\partial\rho}}%
\global\long\def\drhrh#1{\frac{\partial^{2}#1}{\partial\rho^{2}}}%
\global\long\def\dth#1{\frac{\partial#1}{\partial\theta}}%
\global\long\def\dthth#1{\frac{\partial^{2}#1}{\partial\theta^{2}}}%
\global\long\def\dzb#1{\frac{\partial#1}{\partial\overline{z}}}%
\global\long\def\dxy#1{\frac{\partial^{2}#1}{\partial x\partial y}}%
\global\long\def\ddr#1{\frac{\mathrm{d}#1}{\mathrm{d}r}}%
\global\long\def\ddrr#1{\frac{\mathrm{d}^{2}#1}{\mathrm{d}r^{2}}}%
\global\long\def\dds#1{\frac{\mathrm{d}#1}{\mathrm{d}s}}%
\global\long\def\ddss#1{\frac{\mathrm{d}^{2}#1}{\mathrm{d}s^{2}}}%
\global\long\def\ddt#1{\frac{\mathrm{d}#1}{\mathrm{d}t}}%
\global\long\def\ddtt#1{\frac{\mathrm{d}^{2}#1}{\mathrm{d}t^{2}}}%
\global\long\def\ddx#1{\frac{\mathrm{d}#1}{\mathrm{d}x}}%
\global\long\def\ddxx#1{\frac{\mathrm{d}^{2}#1}{\mathrm{d}x^{2}}}%
\global\long\def\ddy#1{\frac{\mathrm{d}#1}{\mathrm{d}y}}%
\global\long\def\ddyy#1{\frac{\mathrm{d}^{2}#1}{\mathrm{d}y^{2}}}%
\global\long\def\ddz#1{\frac{\mathrm{d}#1}{\mathrm{d}z}}%
\global\long\def\ddzz#1{\frac{\mathrm{d}^{2}#1}{\mathrm{d}z^{2}}}%
\global\long\def\ddth#1{\frac{\mathrm{d}#1}{\mathrm{d}\theta}}%
\global\long\def\ddthth#1{\frac{\mathrm{d}^{2}#1}{\mathrm{d}\theta^{2}}}%
\global\long\def\ddla#1{\frac{\mathrm{d}#1}{\mathrm{d}\lambda}}%
\global\long\def\ddlala#1{\frac{\mathrm{d}^{2}#1}{\mathrm{d}\lambda^{2}}}%
\global\long\def\ddta#1{\frac{\mathrm{d}#1}{\mathrm{d}\tau}}%
\global\long\def\ddtata#1{\frac{\mathrm{d}^{2}#1}{\mathrm{d}\tau^{2}}}%
\global\long\def\ad{\dot{a}}%
\global\long\def\add{\ddot{a}}%
\global\long\def\bd{\dot{b}}%
\global\long\def\cd{\dot{c}}%
\global\long\def\ddd{\dot{d}}%
\global\long\def\ed{\dot{e}}%
\global\long\def\fd{\dot{f}}%
\global\long\def\gd{\dot{g}}%
\global\long\def\hd{\dot{h}}%
\global\long\def\kd{\dot{k}}%
\global\long\def\pd{\dot{p}}%
\global\long\def\qd{\dot{q}}%
\global\long\def\qdd{\ddot{q}}%
\global\long\def\rd{\dot{r}}%
\global\long\def\rdd{\ddot{r}}%
\global\long\def\sd{\dot{s}}%
\global\long\def\sdd{\ddot{s}}%
\global\long\def\td{\dot{t}}%
\global\long\def\tdd{\ddot{t}}%
\global\long\def\vd{\dot{v}}%
\global\long\def\vdd{\ddot{v}}%
\global\long\def\ud{\dot{u}}%
\global\long\def\udd{\ddot{u}}%
\global\long\def\vd{\dot{v}}%
\global\long\def\vdd{\ddot{v}}%
\global\long\def\wdt{\dot{w}}%
\global\long\def\wdd{\ddot{w}}%
\global\long\def\xd{\dot{x}}%
\global\long\def\xdd{\ddot{x}}%
\global\long\def\yd{\dot{y}}%
\global\long\def\ydd{\ddot{y}}%
\global\long\def\zd{\dot{z}}%
\global\long\def\zdd{\ddot{z}}%
\global\long\def\Ad{\dot{A}}%
\global\long\def\Add{\ddot{A}}%
\global\long\def\Hd{\dot{H}}%
\global\long\def\Id{\dot{I}}%
\global\long\def\Idd{\ddot{I}}%
\global\long\def\Pd{\dot{P}}%
\global\long\def\Qd{\dot{Q}}%
\global\long\def\Qdd{\ddot{Q}}%
\global\long\def\Rd{\dot{R}}%
\global\long\def\Xd{\dot{X}}%
\global\long\def\Xdd{\ddot{X}}%
\global\long\def\Yd{\dot{Y}}%
\global\long\def\Ydd{\ddot{Y}}%
\global\long\def\Zd{\dot{Z}}%
\global\long\def\Zdd{\ddot{Z}}%
\global\long\def\ald{\dot{\alpha}}%
\global\long\def\aldd{\ddot{\alpha}}%
\global\long\def\bed{\dot{\beta}}%
\global\long\def\bedd{\ddot{\beta}}%
\global\long\def\gad{\dot{\gamma}}%
\global\long\def\gadd{\ddot{\gamma}}%
\global\long\def\epd{\dot{\varepsilon}}%
\global\long\def\epdd{\ddot{\varepsilon}}%
\global\long\def\zed{\dot{\zeta}}%
\global\long\def\sid{\dot{\sigma}}%
\global\long\def\sidd{\ddot{\sigma}}%
\global\long\def\thd{\dot{\theta}}%
\global\long\def\thdd{\ddot{\theta}}%
\global\long\def\xid{\dot{\xi}}%
\global\long\def\xidd{\ddot{\xi}}%
\global\long\def\rhod{\dot{\rho}}%
\global\long\def\rhodd{\ddot{\rho}}%
\global\long\def\phd{\dot{\phi}}%
\global\long\def\phdd{\ddot{\phi}}%
\global\long\def\vpd{\dot{\varphi}}%
\global\long\def\vpdd{\ddot{\varphi}}%
\global\long\def\chd{\dot{\chi}}%
\global\long\def\psd{\dot{\psi}}%
\global\long\def\psdd{\ddot{\psi}}%
\global\long\def\Thd{\dot{\Theta}}%
\global\long\def\Thdd{\ddot{\Theta}}%
\global\long\def\Pid{\dot{\Pi}}%
\global\long\def\Pidd{\ddot{\Pi}}%
\global\long\def\ppd{\dot{\mathbf{p}}}%
\global\long\def\ppdd{\ddot{\mathbf{p}}}%
\global\long\def\qqd{\dot{\mathbf{q}}}%
\global\long\def\qqdd{\ddot{\mathbf{q}}}%
\global\long\def\rrd{\dot{\mathbf{r}}}%
\global\long\def\rrdd{\ddot{\mathbf{r}}}%
\global\long\def\xxd{\dot{\mathbf{x}}}%
\global\long\def\xxdd{\ddot{\x}}%
\global\long\def\zzd{\dot{\mathbf{z}}}%
\global\long\def\LLd{\dot{\mathbf{L}}}%
\global\long\def\PPd{\dot{\mathbf{P}}}%
\global\long\def\QQd{\dot{\mathbf{Q}}}%
\global\long\def\RRd{\dot{\mathbf{R}}}%
\global\long\def\RRdd{\ddot{\mathbf{R}}}%
\global\long\def\ZZd{\dot{\mathbf{Z}}}%
\global\long\def\BBB{\mathbb{B}}%
\global\long\def\BBC{\mathbb{C}}%
\global\long\def\BBD{\mathbb{D}}%
\global\long\def\BBF{\mathbb{F}}%
\global\long\def\BBH{\mathbb{H}}%
\global\long\def\BBI{\mathbb{I}}%
\global\long\def\BBR{\mathbb{R}}%
\global\long\def\BBN{\mathbb{N}}%
\global\long\def\BBP{\mathbb{P}}%
\global\long\def\BBS{\mathbb{S}}%
\global\long\def\BBQ{\mathbb{Q}}%
\global\long\def\BBV{\mathbb{V}}%
\global\long\def\BBX{\mathbb{X}}%
\global\long\def\BBY{\mathbb{Y}}%
\global\long\def\BBZ{\mathbb{Z}}%
\global\long\def\AA{\mathcal{A}}%
\global\long\def\BB{\mathcal{B}}%
\global\long\def\CC{\mathcal{C}}%
\global\long\def\DD{\mathcal{D}}%
\global\long\def\EE{\mathcal{E}}%
\global\long\def\FF{\mathcal{F}}%
\global\long\def\GG{\mathcal{G}}%
\global\long\def\HH{\mathcal{H}}%
\global\long\def\II{\mathcal{I}}%
\global\long\def\JJ{\mathcal{J}}%
\global\long\def\KK{\mathcal{K}}%
\global\long\def\LLL{\mathcal{L}}%
\global\long\def\MM{\mathcal{M}}%
\global\long\def\NN{\mathcal{N}}%
\global\long\def\OOO{\mathcal{O}}%
\global\long\def\PP{\mathcal{P}}%
\global\long\def\QQ{\mathcal{Q}}%
\global\long\def\RRR{\mathcal{R}}%
\global\long\def\SSS{\mathcal{S}}%
\global\long\def\TT{\mathcal{T}}%
\global\long\def\UU{\mathcal{U}}%
\global\long\def\VV{\mathcal{V}}%
\global\long\def\WW{\mathcal{W}}%
\global\long\def\XX{\mathcal{X}}%
\global\long\def\YY{\mathcal{Y}}%
\global\long\def\ZZ{\mathcal{Z}}%
\global\long\def\so{\Rightarrow}%
\global\long\def\os{\Leftarrow}%
\global\long\def\to{\rightarrow}%
\global\long\def\ot{\leftarrow}%
\global\long\def\soo{\Longrightarrow}%
\global\long\def\oos{\Longleftarrow}%
\global\long\def\too{\longrightarrow}%
\global\long\def\oot{\longleftarrow}%
\global\long\def\sos{\Leftrightarrow}%
\global\long\def\tot{\leftrightarrow}%
\global\long\def\soos{\Longleftrightarrow}%
\global\long\def\toot{\longleftrightarrow}%
\global\long\def\mt{\mapsto}%
\global\long\def\mtt{\longmapsto}%
\global\long\def\dn{\downarrow}%
\global\long\def\up{\uparrow}%
\global\long\def\updn{\updownarrow}%
\global\long\def\sea{\searrow}%
\global\long\def\nea{\nearrow}%
\global\long\def\nwa{\nwarrow}%
\global\long\def\swa{\swarrow}%
\global\long\def\hk{\hookrightarrow}%
\global\long\def\hkl{\hookleftarrow}%
\global\long\def\nto{\nrightarrow}%
\global\long\def\soosp{\quad\Longrightarrow\quad}%
\global\long\def\soossp{\quad\Longleftrightarrow\quad}%
\global\long\def\su{\subset}%
\global\long\def\us{\supset}%
\global\long\def\se{\subseteq}%
\global\long\def\es{\supseteq}%
\global\long\def\nse{\nsubseteq}%
\global\long\def\nes{\nsupseteq}%
\global\long\def\nin{\notin}%
\global\long\def\sm{\setminus}%
\global\long\def\orr{\vee}%
\global\long\def\and{\wedge}%
\global\long\def\Orr{\bigvee}%
\global\long\def\And{\bigwedge}%
\global\long\def\ex{\exists}%
\global\long\def\fa{\forall}%
\global\long\def\multibrl#1{\left(#1\right.}%
\global\long\def\multibrr#1{\left.#1\right)}%
\global\long\def\multisql#1{\left[#1\right.}%
\global\long\def\multisqr#1{\left.#1\right]}%
\global\long\def\multicul#1{\left\{  #1\right.}%
\global\long\def\multicur#1{\left.#1\right\}  }%
\global\long\def\hf{\frac{1}{2}}%
\global\long\def\trd{\frac{1}{3}}%
\global\long\def\fr{\frac{1}{4}}%
\global\long\def\ff{\frac{1}{5}}%
\global\long\def\sxt{\frac{1}{6}}%
\global\long\def\sv{\frac{1}{7}}%
\global\long\def\ei{\frac{1}{8}}%
\global\long\def\hfp{\frac{\pi}{2}}%
\global\long\def\fp{\frac{\pi}{4}}%
\global\long\def\bl{\bigl|}%
\global\long\def\Bl{\Bigl|}%
\global\long\def\bgl{\biggl|}%
\global\long\def\Bgl{\Biggl|}%
\global\long\def\bk#1#2{\left\langle #1\middle|#2\right\rangle }%
\global\long\def\mc#1#2{\left\{  #1\middle|#2\right\}  }%
\global\long\def\mbr#1#2{\left(#1\middle|#2\right) }%
\global\long\def\bmb#1#2#3{\left(#1\middle|#2\middle|#3\right)}%
\global\long\def\bmk#1#2#3{\left\langle #1\middle|#2\middle|#3\right\rangle }%
\global\long\def\ap{\approx}%
\global\long\def\eqm{\overset{?}{=}}%
\global\long\def\eq{\equiv}%
\global\long\def\neq{\not\equiv}%
\global\long\def\toinf{\xrightarrow[n\to\infty]{}}%
\global\long\def\eqd{\overset{\mathrm{def}}{=}}%
\global\long\def\eqos{\ \hat{=}\ }%
\global\long\def\d{\mathrm{d}}%
\global\long\def\DDD{\mathrm{D}}%
\global\long\def\i{\ii}%
\global\long\def\TTT{t}%
\global\long\def\At{\tilde{A}}%
\global\long\def\Dt{\tilde{D}}%
\global\long\def\Et{\tilde{E}}%
\global\long\def\Ft{\tilde{F}}%
\global\long\def\Ht{\tilde{H}}%
\global\long\def\Jt{\tilde{J}}%
\global\long\def\Lt{\tilde{L}}%
\global\long\def\Nt{\tilde{N}}%
\global\long\def\Vt{\tilde{V}}%
\global\long\def\Xt{\tilde{X}}%
\global\long\def\at{\tilde{a}}%
\global\long\def\bt{\tilde{b}}%
\global\long\def\ct{\tilde{c}}%
\global\long\def\eet{\tilde{e}}%
\global\long\def\ft{\tilde{f}}%
\global\long\def\gt{\tilde{g}}%
\global\long\def\hht{\tilde{h}}%
\global\long\def\jt{\tilde{j}}%
\global\long\def\kt{\tilde{k}}%
\global\long\def\elt{\tilde{\ell}}%
\global\long\def\nt{\tilde{n}}%
\global\long\def\qt{\tilde{q}}%
\global\long\def\st{\tilde{s}}%
\global\long\def\vt{\tilde{v}}%
\global\long\def\xt{\tilde{x}}%
\global\long\def\zt{\tilde{z}}%
\global\long\def\delt{\tilde{\delta}}%
\global\long\def\sit{\tilde{\sigma}}%
\global\long\def\Sit{\tilde{\Sigma}}%
\global\long\def\pst{\tilde{\psi}}%
\global\long\def\pht{\tilde{\phi}}%
\global\long\def\Pht{\tilde{\Phi}}%
\global\long\def\vpt{\tilde{\varphi}}%
\global\long\def\ett{\tilde{\eta}}%
\global\long\def\cht{\tilde{\chi}}%
\global\long\def\zrt{\tilde{0}}%
\global\long\def\ept{\tilde{\epsilon}}%
\global\long\def\Pit{\tilde{\Pi}}%
\global\long\def\omt{\tilde{\omega}}%
\global\long\def\Omt{\tilde{\Omega}}%
\global\long\def\tHH{\widetilde{\mathcal{H}}}%
\global\long\def\mfC{\mathfrak{C}}%
\global\long\def\mfg{\mathfrak{g}}%
\global\long\def\mfh{\mathfrak{h}}%
\global\long\def\mfd{\mathfrak{d}}%
\global\long\def\mfdg{\mathfrak{dg}}%
{} 
\global\long\def\mfig{\mathfrak{ig}}%
\global\long\def\mfih{\mathfrak{ih}}%
\global\long\def\mfdh{\mathfrak{dh}}%

\global\long\def\sl{\!\!\!/}%
\global\long\def\sll{\!\!\!\!/\,}%
\global\long\def\slll{\!\!\!\!\!/\,\,}%
\global\long\def\Wsl{W\slll}%
\global\long\def\Bsl{B\sll}%
\global\long\def\Dsl{D\sll}%
\global\long\def\Asl{A\sl}%
\global\long\def\Zsl{Z\sll}%
\global\long\def\dsl{\partial\sl}%
\global\long\def\ksl{k\sl}%
\global\long\def\lsl{\ell\sl}%
\global\long\def\psl{p\sl}%
\global\long\def\qsl{q\sl}%
\global\long\def\ks{\mathsf{k}}%
\global\long\def\pss{\mathsf{p}}%
\global\long\def\qs{\mathsf{q}}%
\global\long\def\ws{\mathsf{w}}%
\global\long\def\xs{\mathsf{x}}%
\global\long\def\ys{\mathsf{y}}%
\global\long\def\zs{\mathsf{z}}%
\global\long\def\zrs{\mathsf{0}}%
\global\long\def\hv{\vec{h}}%
\global\long\def\pv{\vec{p}}%
\global\long\def\qv{\vec{q}}%
\global\long\def\yv{\vec{y}}%
\global\long\def\Hv{\vec{H}}%
\global\long\def\Mv{\vec{M}}%
\global\long\def\Sv{\vec{S}}%
\global\long\def\Vv{\vec{V}}%
\global\long\def\Xv{\vec{X}}%
\global\long\def\res#1{\underset{#1}{\Res}}%
\global\long\def\pr{\parallel}%
\global\long\def\xx{\times}%
\global\long\def\dg{{^\circ}}%
\global\long\def\sp{,\qquad}%
\global\long\def\qm{\ ?\ }%
\global\long\def\sn#1#2{#1\times10^{#2}}%
\global\long\def\hyp{\thinspace\!_{2}F_{1}}%
\global\long\def\sq{\square}%
\global\long\def\pt{\propto}%
\global\long\def\lrc{\lrcorner\thinspace}%
\global\long\def\pexp{\overrightarrow{\exp}}%
\global\long\def\SUT{\mathrm{SU}\left(2\right)}%
\global\long\def\sut{\mathfrak{su}\left(2\right)}%
\global\long\def\ISUT{\mathrm{ISU}\left(2\right)}%
\global\long\def\isut{\mathfrak{isu}\left(2\right)}%
\global\long\def\DG{\mathrm{DG}}%
\global\long\def\IG{\mathrm{IG}}%
\global\long\def\IH{\mathrm{IH}}%
\global\long\def\DH{\mathrm{DH}}%
\global\long\def\IW{\mathrm{IW}}%
\global\long\def\DW{\mathrm{DW}}%

\title{2+1D Loop Quantum Gravity on the Edge}
\author{\textbf{Laurent Freidel}\thanks{lfreidel@perimeterinstitute.ca} $\ $and
\textbf{Barak Shoshany}\thanks{bshoshany@perimeterinstitute.ca}\\
\emph{ Perimeter Institute for Theoretical Physics,}\\
\emph{31 Caroline Street North, Waterloo, Ontario, Canada, N2L 2Y5}\\
 \\
 \textbf{Florian Girelli}\thanks{fgirelli@uwaterloo.ca}\\
\emph{ Department of Applied Mathematics, University of Waterloo,}\\
\emph{200 University Avenue West, Waterloo, Ontario, Canada, N2L 3G1}}
\maketitle
\begin{abstract}
We develop a new perspective on the discretization of the phase space
structure of gravity in 2+1 dimensions as a piecewise-flat geometry
in 2 spatial dimensions. Starting from a subdivision of the continuum
geometric and phase space structure into elementary cells, we obtain
the loop gravity phase space coupled to a collection of effective
particles carrying mass and spin, which measure the curvature and
torsion of the geometry. We show that the new degrees of freedom associated
to the particle-like elements can be understood as edge modes, which
appear in the decomposition of the continuum theory into subsystems
and do not cancel out in the gluing of cells along codimension 2 defects.
These new particle-like edge modes are gravitationally dressed in
an explicit way. This provides a detailed explanation of the relations
and differences between the loop gravity phase space and the one deduced
from the continuum theory.
\end{abstract}

\tableofcontents{}

\section{Introduction}

One of the key challenges in trying to define a theory of quantum
gravity at the quantum level is to find a regularization that does
not drastically break the fundamental symmetries of the theory. This
is a challenge in any gauge theory, but gravity is especially challenging,
for two reasons. First, one expects that the quantum theory possesses
a fundamental length scale; and second, the gauge group contains diffeomorphism
symmetry, which affects the nature of the space on which the regularization
is applied.

In gauge theories such as QCD, the only known way to satisfy these
requirements\footnote{Other than gauge-fixing before regularization.}
is to put the theory on a lattice, where an effective finite-dimensional
gauge symmetry survives at each scale. One would like to devise such
a scheme in the gravitational context. In this paper, we develop a
step-by-step procedure achieving this in the context of 2+1 gravity.
We initially expected to find the so-called holonomy-flux discretized
phase space, which appears in loop gravity and produces spin networks
after quantization. To our surprise, we discovered that there are
additional degrees of freedom that behave as a collection of particles
coupled to the gravitational degrees of freedom.

In the \textit{loop quantum gravity} (LQG) framework, gravity is quantized
using the canonical approach, with the gravitational degrees of freedom
expressed in terms of the connection and frame field. The quantum
states of geometry are known as \textit{spin networks}. In this framework,
we can show that geometric operators possess a discrete spectrum.
This is, however, only possible after one chooses the quantum states
to have support on a graph. Spin network states can be understood
as describing a quantum version of \textit{discretized} spatial geometry
\cite{Rovelli:2004tv}, and the Hilbert space associated to a graph
can be related, in the classical limit, to a set of discrete piecewise-flat
geometries \cite{Dittrich:2008ar,Freidel:2010aq}. This means that
the LQG quantization scheme consists at the same time of a \emph{quantization}
and a \emph{discretization}; moreover, the quantization of the geometric
spectrum is entangled with the discretization of the fundamental variables.
It has been argued that it is essential to disentangle these two different
features \cite{Freidel:2011ue}, especially when one wants to address
dynamical issues.

In \cite{Freidel:2011ue,Freidel:2013bfa}, it has been argued that
one should understand the discretization as a two step process: a
\emph{subdivision} followed by a \emph{truncation}. In the first step
one subdivides the systems in fundamental cells, and in the second
step one chooses a truncation of degrees of freedom in each cell which
is consistent with the symmetry of the theory. By focusing first on
the classical discretization, before any quantization takes place,
several aspects of the theory can be clarified. Let us mention some
examples: 
\begin{itemize}
\item The discretization scheme allows us to study more concretely how to
recover the continuum geometry out of the discrete geometry which
is the classical picture behind the spin networks \cite{Freidel:2011ue,Freidel:2013bfa}.
In particular, since the discretization is now understood as a truncation
of the continuous degrees of freedom, it is possible to associate
a continuum geometry to the discrete data.
\item It provides a justification for why in the continuum case the momentum
variables are equipped with a zero Poisson bracket, whereas in the
discrete case the momentum variables do not commute with each other
\cite{Freidel:2011ue,Dupuis:2017otn,Dittrich:2014wpa}. These variables
needs to be dressed by the gauge connection and are now understood
as charge generators \cite{Donnelly:2016auv}.
\item It permitted the discovery of duality symmetries that suggests new
discrete variables and a new dual formulation of discrete 2+1 gravity
\cite{Dupuis:2017otn}. While the quantization of spin networks leads
to spin foam models, the quantization of these dual models can be
related to the \textit{Dijkgraaf-Witten model} \cite{clement}. This
also allows one to understand, at the classical level, the presence
of new dual vacua as advocated by Dittrich and Geiller \cite{Dittrich:2014wpa,Dittrich:2014wda}.
\end{itemize}
In this work we revisit these ideas in the context of 2+1 gravity,
and deepen the analysis done in \cite{Freidel:2011ue,Dupuis:2017otn}
by focusing on what happens at the location of the curvature defects.
Since gravity in 2+1 dimensions is equivalent to a Chern-Simons theory
constructed on a Drinfeld double group, we analyze the phase space
truncation of any Chern-Simons theory constructed on a Drinfeld double
group and then specialize the analysis to the case where the double
group is the inhomogeneous ``Poincaré'' group $\DG$ over a ``Lorentz''
group $G$. This corresponds to the case of a zero cosmological constant
when $G$ is a 3D rotation group. The variables of the theory are
a pair of $\mfg$-valued connection $\A$ and frame field $\E$, while
the Chern-Simons connection $\AAb$ is simply the sum $\AAb=\A+\E$.

The aim of the present work is to precisely evaluate such effects
in the 2+1 case. As already emphasized, the procedure of discretization
is now understood as a process of \emph{subdivision} of the underlying
manifold into cells, followed by a \emph{truncation} of the degrees
of freedom in each cell. The truncation we chose, and which is adapted
to 2+1 gravity, is to assume that the geometry is locally \emph{flat}
$\F\left(\A\right)=0$ and \emph{torsionless} $\T=\d_{\A}\E=0$ inside
each cell. This means that the continuum geometric data $(\E,\A)$
is now replaced by a $\DG$ structure, that is, a decomposition of
the 2D manifold $\Sigma$ into cells $c$ with the transition functions
across cells belonging to the group $\DG$ preserving the local flatness
and torsionlessness conditions of the geometry.

We can think of $\DG$ as the local isometry group of our locally-flat
structure, and hence an analogue of the Poincaré group for flat space.
If we allow possible violations of the flatness condition at the vertices
$v$ of the cells, this $\DG$ structure can be understood as a flat
structure on $\Sigma\backslash V$, where $\Sigma$ is the 2D manifold
and $V$ is the set of all vertices. In this context, the locally-flat
geometry on $\Sigma$ is encoded in terms of discrete rotational and
translational holonomies which allow the reconstruction of the flat
connection on the punctured surface.

At the continuum level, the geometric data forms a phase space with
pre-symplectic structure $\Omega=\int_{\Sigma}\delta\E\cdot\delta\A$.
The main advantage of describing the discretization as a truncation
is the fact that one can understand the truncated variables as forming
a reduced phase space. This follows from the fact that the truncation
is implemented in terms of first-class constraints, and is therefore
compatible with the Hamiltonian structure. This philosophy has already
been applied to gravity in \cite{Freidel:2011ue,Dupuis:2017otn}.
However, these works neglected the contributions from curvature and
torsion singularities appearing at the vertices. In the current work
we extend the analysis to include these contributions.

\subsubsection*{A Change of Paradigm}

A key result of this paper is conceptual. In the subdivision process,
some of the bulk degrees of freedom are replaced by \emph{edge mode}
degrees of freedom, which play a key role in the construction of the
full phase space and our understanding of symmetry. The reason this
happens is that we propose to implement the procedure of discretization
as a rewriting of the theory in terms of specific subsystems. Dealing
with subsystems in a gauge theory requires special care with regards
to boundaries, where gauge invariance is naively broken and additional
degrees of freedom must be added in order to restore it.

In other words, what is called a discretization should in fact be
seen as a proper way to extend the phase space by adding extra degrees
of freedom -- a generalization of Goldstone modes needed to restore
the gauge invariance -- that transforms non-trivially under new edge
symmetry transformations. The process of subdivision therefore requires
a canonical extension of the phase space, and converting some momenta
into non-commutative charge generators.

The general philosophy is presented in \cite{Donnelly:2016auv} and
exemplified in the 3+1 gravity context in \cite{Freidel:2015gpa,Freidel:2016bxd,Freidel:2018pvm}.
An intuitive reason behind this fundamental mechanism is also presented
in \cite{Rovelli:2013fga} and the general idea is, in a sense, already
present in \cite{Donnelly:2011hn}. In the 2+1 gravity context, the
edge modes have been studied in great detail in \cite{Geiller:2017xad,Geiller:2017whh}.
This phenomenon even happens when the boundary is taken to be infinity
\cite{Strominger:2017zoo}, where these new degrees of freedom are
the \emph{soft modes}. One point which is important for us is that
these extra degrees of freedom, which possess their own phase space
structure and appear as ``dressings'' of the gravitationally charged
observables, affect the commutation relations of the dressed observables.
In a precise sense, this is what happens with the fluxes in loop gravity:
the ``discretized'' fluxes are dressed by the connection degrees
of freedom, implying a different Poisson structure compared to the
continuum ones. A nice continuum derivation of this fact is also given
in \cite{Cattaneo:2016zsq}.

Once this subdivision and extension of the phase space are done properly,
one has to understand the gluing of subregions as the fusion of edge
modes across the boundaries. If the boundary is trivial, the fusion
merely allows us to extend gauge-invariant observables from one region
to another. When several boundaries meet at a corner, there is now
the possibility to have residual edge degrees of freedom that come
from the fusion product. We witness exactly this phenomenon at the
vertices of our cellular decomposition, where new degrees of freedom,
in addition to the usual loop gravity ones, are present after regluing.

In our analysis we rewrite the gauge degrees of freedom, after truncation,
as a collection of locally-flat ``Poincaré'' connections $\AAb=\A+\E$
in each cell. The choice of collection of cells corresponds to the
\emph{subdivision}, and the imposition of the flatness condition inside
each cell corresponds to the \emph{truncation}. Inside each cell,
we introduce a $\DG$-valued 0-form $\HH_{c}\left(x\right)$ that
parametrizes the flat connection, $\AAb\left(x\right)\equiv\HH_{c}^{-1}\left(x\right)\d\HH_{c}\left(x\right)$.
The quantities $\HH_{c}\left(x\right)$ can be used to reconstruct
the holonomy of the connection, but they contain more information.
This comes from the fact that left and right transformations of the
holonomy have different implementations at the level of the connection:
\begin{eqnarray}
\HH_{c}\left(x\right)\mt\GG_{c}\HH_{c}\left(x\right) & \soosp & \AAb\left(x\right)
\mt\AAb\left(x\right),\\
\HH_{c}\left(x\right)\mt\HH_{c}\left(x\right)\GG\left(x\right) & \soosp & \AAb\left(x\right)\mt\GG^{-1}\left(x\right)\AAb\left(x\right)\GG\left(x\right)+\GG^{-1}\left(x\right)\d\GG\left(x\right).
\end{eqnarray}
In the last case, the standard gauge transformation is expressed as
a right action on $\HH_{c}\left(x\right)$. However, the first transformation,
which acts on the left of $\HH_{c}\left(x\right)$, leaves the connection
invariant. It therefore corresponds to an additional degree of freedom
entering the definition of $\HH_{c}\left(x\right)$ which is not contained
in $\AAb$. One can understand the presence of this additional degree
of freedom as coming from the presence of a boundary, and the left
translation as an edge mode symmetry that has to be implemented when
we reconstruct physical observables.

We can now use the variables $\HH_{c}\left(x\right)$ to reconstruct
our truncated phase space. For each cell, we can express the continuum
symplectic potential in terms of the holonomy variables, and the expression
can be readily seen as only depending on the fields evaluated at the
boundaries of the cell. Summing the symplectic potentials for each
cell simplifies the general expression, and the final symplectic potential
depends only on the fields evaluated across boundaries.

One recovers, in particular, that holonomies from a cell to its neighbor
form a subset of the canonical variables. $\X_{c}^{c'}\in\mfg$ is
the flux and $h_{cc'}\in G$ is the holonomy, while the symplectic
potential reproduces the loop gravity potential $\Theta^{\mathrm{LQG}}=\sum_{(cc')}\Tr\left(\delta h_{cc'}h_{cc'}^{-1}\X_{c}^{c'}\right)$,
which gives the holonomy-flux algebra. We also find, however, that
there are additional contributions coming from the vertices $v$ where
several cells meet. Each vertex carries the phase space structure
$\Theta_{v}$ of a \emph{relativistic particle}, labeled by the edge
modes $\HH_{v}\left(v\right)=\left(h_{v}\left(v\right),\y_{v}\left(v\right)\right)\in G\ltimes\mfg^{*}$
which are non-trivially coupled to the gravity variables through connectors
$h_{vc}$ and $\X_{v}^{c}$.

The mass and spin of the effective relativistic particles are determined
via a generalization of the Gauss constraint and curvature constraint
at the vertices. The appearance of these effective particle degrees
of freedom from pure gravity is quite interesting and unexpected.
The usual loop gravity framework is recovered when the edge mode degrees
of freedom labeled by $\HH_{v}\left(v\right)$ are frozen.

The conceptual shift towards an edge mode interpretation, while not
modifying the mathematical structures at all, provides a different
paradigm to explore some of the key questions of LQG. For example,
the notion of the continuum limit (in a 3+1-dimensional theory) attached
to subregions could be revisited and clarified in light of this new
interpretation, and related to the approach developed in \cite{Delcamp:2018efi,Delcamp:2016yix}.
It also strengthens, in a way, the \emph{spinor approach to LQG} \cite{Girelli:2005ii,Freidel:2009ck,Livine:2011gp},
which allows one to recover the LQG formalism from degrees of freedom,
the spinors, living on the nodes of the graph. These spinors can be
seen as a different parametrization of the edge modes (in a similar
spirit to \cite{Wieland:2018ymr,Wieland:2017ksn}).

Edge modes have recently been studied for the purpose of making proper
entropy calculations in gauge theory or, more generally, defining
local subsystems \cite{Donnelly:2011hn}. Their use could provide
some new guidance on understanding the concept of entropy in LQG.
They are also relevant to the study of specific types of boundary
excitations in condensed matter \cite{Balachandran:1994vi}, which
could generate some interesting new directions to explore in LQG,
just like \cite{Dittrich:2016typ,Delcamp:2018efi}.

\subsubsection*{Comparison to Previous Work}

Discretization (and quantization) of 2+1 gravity was already performed
some time ago in the Chern-Simons formulation \cite{Alekseev:1993rj,Alekseev:1994pa,Alekseev:1994au,Meusburger:2003hc}.
While our results should be equivalent to this formulation, we find
them interesting and relevant for several reasons.

Firstly, we work with the gravitational variables and the associated
geometric quantities, such as torsion and curvature. Our procedure
describes clearly how such objects should be discretized, which is
not obvious in the Chern-Simons picture; see for example \cite{Meusburger:2008bs}
where the link between the combinatorial (quantization) framework
and the LQG one was explored.

Secondly, we are using a different discretization procedure than the
one used in the combinatorial approach. Instead of considering the
reduced graph (flower graph), we use the full graph to generate the
spin network and assume that the equations of motion are satisfied
in the cells and the disks containing the geometric excitations.

Thirdly, one of the most important differences is in the fact that
we \emph{derive} the symplectic structure of the discretized variables
from the continuum one, unlike the combinatorial approach which \emph{postulates}
it (Fock-Rosly formalism). As we will see, there will be some slight
differences in the resulting symplectic potential, which we will comment
on in Sec. \ref{sec:Summary-and-Outlook}.

Finally, our discretization will apply, with the necessary modifications,
to the 3+1 case as well \cite{barak4d}, unlike the combinatorial
approach.

\subsubsection*{Outline}

The paper is organized as follows. In Sec. \ref{sec:Piecewise-Flat-Manifolds-in},
we recall the necessary ingredients to define piecewise-homogeneous
geometries. We construct the corresponding discrete Chern-Simons connection
and establish the first main result: the expression of the Chern-Simons
symplectic structure around a curvature defect. In Sec. \ref{sec:Gravity:-Specializing-to},
we focus explicitly on a geometric structure suitable for 2+1 gravity
with zero cosmological constant and characterize the connection in
terms of the holonomies, which are the relevant variables to characterize
the discrete geometry. In Sec. \ref{sec:Discretizing-the-Symplectic},
we explicitly determine the phase space structure for the piecewise-flat
geometry by discretizing the symplectic potential. We obtain a structure
corresponding to classical spin networks coupled to torsion or curvature
excitations which behave like relativistic particles. In Sec. \ref{sec:The-Gauss-and},
we discuss constraints and symmetries.

\section{\label{sec:Piecewise-Flat-Manifolds-in}Piecewise-$\protect\DG$-Flat
Manifolds in Two Dimensions}

In this section we define the concept of a piecewise-$\DG$-flat manifold
. We describe the cellular decomposition, which is the discrete structure
we impose on our manifold. We also define the (continuous) connection
relevant to the piecewise-flat geometry, and show how it can be expressed
in terms of (discrete) holonomies. This allows us to construct the
relevant phase space structure in the next section.

\subsection{The Cellular Decomposition}

Consider a 2-dimensional manifold $\Sigma$ without boundary. We introduce
a \emph{cellular decomposition} $\Delta$ of $\Sigma$ made of 0-cells
(\emph{vertices}) denoted $v$, 1-cells (\emph{edges}) denoted $e$
and 2-cells (\emph{cells}) denoted $c$. The \emph{1-skeleton} $\Gamma\su\Delta$
is the set of all vertices and edges of $\Delta$. The dual \emph{spin
network graph} $\Gamma^{*}$ is composed of \emph{nodes} $c^{*}$
connected by \emph{links} $e^{*}$, such that each node $c^{*}\in\Gamma^{*}$
is dual to a 2-cell $c\in\Gamma$ and each link $e^{*}\in\Gamma^{*}$
is dual to an edge (1-cell) $e\in\Gamma$.

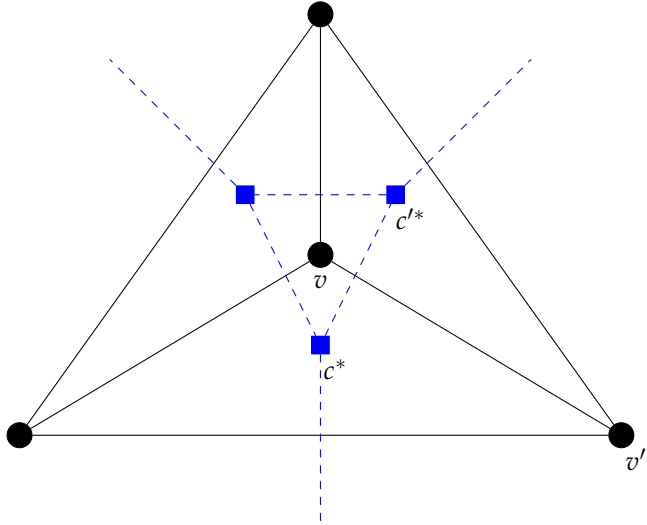
\begin{figure}[!h]
\begin{centering}
\input{Figure-1-Triangle.tex} 
\par\end{centering}
\caption{\label{fig:Triangle}A simple piece of the cellular decomposition
$\Delta$, in black, and its dual spin network $\Gamma^{*}$, in blue.
The vertices $v$ of the 1-skeleton $\Gamma\protect\su\Delta$ are
shown as black circles, while the nodes $c^{*}$ of $\Gamma^{*}$
are shown as blue squares. The edges $e\in\Gamma$ are shown as black
solid lines, while the links $e^{*}\in\Gamma^{*}$ are shown as blue
dashed lines. In particular, two nodes $c^{*}$ and $c^{\prime*}$,
connected by a link $e^{*}=\left(cc'\right)^{*}$, are labeled, as
well as two vertices $v$ and $v'$, connected by an edge $e=\left(vv'\right)=\left(cc'\right)=c\cap c'$,
which is dual to the link $e^{*}$.}
\end{figure}

The edges $e\in\Gamma$ are oriented, and we use the notation $e=\left(vv'\right)$
to mean that the edge $e$ begins at the vertex $v$ and ends at the
vertex $v'$. The inverse edge of $e$, denoted by $e^{-1}$, is the
same edge with reverse orientation: $e^{-1}=\left(v'v\right)$. The
links $e^{*}\in\Gamma^{*}$ are also oriented, and we use the notation,
$e^{*}=\left(cc'\right)^{*}$ to denote that the link $e^{*}$ connects
the nodes $c^{*}$ and $c^{\prime*}$. If the link $e^{*}=\left(cc'\right)^{*}$
is dual to the edge $e=\left(vv'\right)$, as in Fig. \ref{fig:Triangle},
one can also write $e=\left(cc'\right)\equiv c\cap c'$, which means
that the edge $e$ is the (oriented) intersection of the cells $c$
and $c'$. The orientation is such that $e$ is a counterclockwise
rotation of the dual edge $(cc')^{*}$. In Fig. \ref{fig:Triangle}
we show a simple triangulation; however, the cells can be general
polygons.

Let $\DG$ be a Lie group and $\mfdg$ be its Lie algebra. In this
section, $\DG$ represents the Chern-Simons group; later we will specialize
to the case where it is a Drinfeld double. We use a calligraphic font,
e.g. $\HH$, to denote $\DG$-valued differential forms and a bold
calligraphic font, e.g. $\AAb$, to denote $\mfdg$-valued differential
forms. We define a $\mfdg$-valued connection 1-form $\AAb$ on $\Sigma$
and its $\mfdg$-valued curvature 2-form\footnote{The \emph{graded commutator} of two Lie-algebra-valued differential
forms $\AAb,\boldsymbol{\mathcal{B}}$ is given by 
\begin{equation}
\left[\AAb,\boldsymbol{\mathcal{B}}\right]\equiv\AAb\wedge\boldsymbol{\mathcal{B}}-\left(-1\right)^{\deg\AAb\deg\boldsymbol{\mathcal{B}}}\boldsymbol{\mathcal{B}}\wedge\AAb,\label{eq:graded-commutator-index-free}
\end{equation}
where $\deg$ is the degree of the form. In this case, $\left[\AAb,\AAb\right]=2\AAb\wedge\AAb$
since $\deg\AAb=1$.} $\FFb$: 
\begin{equation}
\FFb\left(\AAb\right)\equiv\d\AAb+\hf\left[\AAb,\AAb\right].\label{eq:curvature-def}
\end{equation}
In order to define a piecewise-$\DG$-flat geometry on $\Sigma$,
we assume that $\AAb$ is flat ($\FFb=0$) inside each cell $c$,
and the curvature is restricted to the vertices $v$. We make this
notion precise in the following sections.

\subsection{The $\protect\DG$ Connection Inside the Cells}

As mentioned above, we make the assumption that the curvature inside
each cell $c$ vanishes. Since $c$ is simply connected, this means
that the connection 1-form $\AAb$ can be expressed at any point $x\in c$
as a (left-invariant) \emph{Maurer-Cartan form}: 
\begin{equation}
\AAb\bl_{c}=\HH_{c}^{-1}\d\HH_{c},\label{eq:A-from-H}
\end{equation}
where $\HH_{c}$ is a $\DG$-valued 0-form and the notation $\bl_{c}$
means the relation is valid in the interior\footnote{That is, inside any open set that does not intersect the boundary
of $c$.} of the cell $c$. It is easy to see that, indeed, $\FFb\bl_{c}=0$
for this connection.

Let $\GG$ be a $\DG$-valued 0-form. \emph{Right translations} 
\begin{equation}
\HH_{c}\left(x\right)\mt\HH_{c}\left(x\right)\GG\left(x\right),\label{eq:right-tr-c}
\end{equation}
are \textit{gauge transformations} that affect the connection in the
usual way: 
\begin{equation}
\AAb\mt\GG^{-1}\AAb\GG+\GG^{-1}\d\GG.\label{eq:gauge-tr-A}
\end{equation}
We can also consider \emph{left translations} 
\begin{equation}
\HH_{c}\left(x\right)\mt\GG_{c}\HH_{c}\left(x\right),\label{eq:left-tr-c}
\end{equation}
acting on $\HH_{c}$ with a constant group element $\GG_{c}\in\DG$.
These transformations leave the connection invariant, $\AAb\mt\AAb$.
On the one hand, they label the redundancy of our parametrization
of $\AAb$ in terms of $\HH_{c}$. On the other hand, these transformations
can be understood as \emph{symmetries} of our parametrization in terms
of group elements that stem from the existence of new degrees of freedom
in $\HH_{c}$ beyond the ones in the connection $\AAb$.

This situation is similar to the situation that arises any time one
considers a gauge theory in a region with boundaries \cite{Donnelly:2016auv}.
As shown in \cite{Donnelly:2016auv}, when we subdivide a region of
space we need to add new degrees of freedom at the boundaries of the
subdivision in order to restore gauge invariance. These degrees of
freedom are the \emph{edge modes}, which carry a non-trivial representation
of the boundary symmetry group that descends from the bulk gauge transformations.\footnote{As shown in \cite{Freidel:2018fsk}, this group also contains the
duality group.}

Now, we can invert \eqref{eq:A-from-H} and write $\HH_{c}$ using
a \emph{path-ordered exponential} as follows: 
\begin{equation}
\HH_{c}\left(x\right)=\HH_{c}\left(c^{*}\right)\,\pexp\int_{c^{*}}^{x}\AAb,\label{eq:H-from-A}
\end{equation}
where $\HH_{c}\left(c^{*}\right)$, the value of $\HH_{c}$ at the
node $c^{*}$, is the extra information contained in the edge mode
field $\HH_{c}$ that cannot be obtained from the connection $\AAb$.
Left translations can thus be understood as simply translating the
value of $\HH_{c}\left(c^{*}\right)$ without affecting the value
of $\AAb$.

\subsection{\label{subsec:Connection-Vertices}The $\protect\DG$ Connection
Inside the Disks}

\subsubsection{\label{punc}The Punctured Disk $v^{*}$}

The next step in defining our piecewise-$\DG$-flat geometry is to
parametrize the connection around the vertices. In our discrete geometrical
setting, the set of vertices is the locus where the curvature is concentrated.
It is therefore important to understand the local geometry of the
connection in a neighborhood of the vertices $v\in\Gamma$.

An open set containing $v$ forms the interior of a disk, denoted
$D_{v}$. In order to describe the connection around $v$ in a regular
manner, it will be necessary to excise an infinitesimal neighborhood
of $v$ and consider $A_{v}\equiv D_{v}\backslash\left\{ v\right\} $,
which has the topology of an annulus. It will also be necessary to
introduce a \textit{cut} denoted $C_{v}$ that runs from $v$ to the
boundary of $D_{v}$. This \emph{cut and punctured disk }will be denoted
from now on as $v^{*}\equiv A_{v}\backslash C_{v}$. It will sometimes
be referred to simply as the \emph{punctured disk }for simplicity.

In order to understand the geometry at play on $v^{*}$, it is convenient
to think of the punctured disk as the interior of a cut annulus where
the boundary around $v$ is shrunk to a point. In order to do so,
let us introduce Cartesian coordinates that parametrize the cut annulus.
It is isomorphic to a rectangle with coordinates $\left(r_{v},\phb_{v}\right)$
such that $r_{v}\in\left(0,R\right)$ and $\phb_{v}\in\left[\alb_{v}-\pi,\alb_{v}+\pi\right]$
where $R,\alb_{v}\in\BBR$, and where the line at $\phb=\alb_{v}-\pi$
is identified with the line at $\phb=\alb_{v}+\pi$. We also identify
the entire line at $r_{v}=0$ with the vertex $v$, in the sense that
any function $f\left(r_{v},\phi_{v}\right)$ evaluated at $r_{v}=0$
reduces to a constant value $f\left(v\right)$ regardless of the value
of $\phi_{v}$. Then $v^{*}$ indeed describes a punctured disk of
radius $R$,\footnote{We assume that $R$ is chosen small enough so that the intersection
of any two disks $v^{*}\cap v^{\prime*}$ is empty.} with the puncture located at $v$.

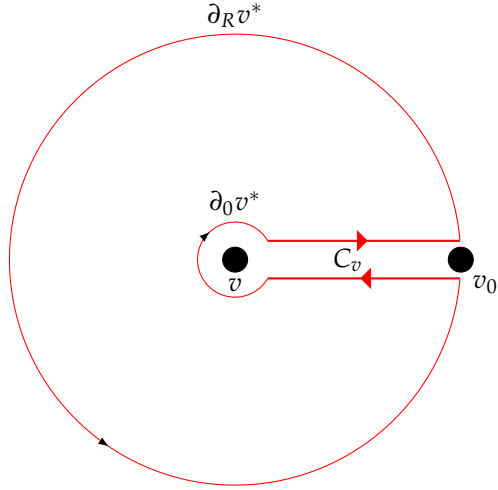
\begin{figure}[!h]
\begin{centering}
\input{Figure-2-Disk.tex} 
\par\end{centering}
\caption{\label{fig:Disk}The punctured disk $v^{*}$. The figure shows the
vertex $v$, cut $C_{v}$, inner boundary $\partial_{0}v^{*}$, outer
boundary $\partial_{R}v^{*}$, and reference point $v_{0}$.}
\end{figure}

The boundary $\partial v^{*}$ consists of two curves of length $2\pi$,
one at $r_{v}=0$ and another at $r_{v}=R$. We will call the one
at $r_{v}=0$ the \emph{inner boundary }$\partial_{0}v^{*}$ and the
one at $r_{v}=R$ the \emph{outer boundary} $\partial_{R}v^{*}$.
In other words 
\begin{equation}
\partial v^{*}=\partial_{0}v^{*}\cup\partial_{R}v^{*}\sp\partial_{0}v^{*}\equiv\mc{\left(r_{v},\phb_{v}\right)}{r_{v}=0}\sp\partial_{R}v^{*}\equiv\mc{\left(r_{v},\phb_{v}\right)}{r_{v}=R}.\label{eq:inner-outer-boundary}
\end{equation}
We will also define a point $v_{0}$ at $r_{v}=R$ and $\phb_{v}=\alb_{v}-\pi$;
then the line at $\phb_{v}=\alb_{v}-\pi$, which extends from $v$
to $v_{0}$, is the \emph{cut} $C_{v}$ where we identified the two
edges of the rectangle. Note that both $v$ and $v_{0}$ are on $\partial v^{*}$
and not inside $v^{*}$. The punctured disk is shown in Fig. \ref{fig:Disk}.

For brevity of notation, we define a reduced angle function $\phi_{v}$
such that 
\[
\phi_{v}\equiv\frac{\phb_{v}}{2\pi},\quad\alpha_{v}\equiv\frac{\alb_{v}}{2\pi}\qquad\phi_{v}\in\left[\alpha_{v}-\tfrac{1}{2},\alpha_{v}+\tfrac{1}{2}\right),
\]
which will be used from now on.

\subsubsection{The Distributional Curvature}

We have assumed that the curvature is concentrated at the vertex.
Loosely, this means that the curvature on the full (non-punctured)
disk $D_{v}$ is distributional: 
\begin{equation}
\FFb\bl_{D_{v}}=\PPb_{v}\thinspace\delta\left(v\right),\label{eq:naive-curv}
\end{equation}
where $\PPb_{v}$ is some constant element of the Lie algebra $\mfdg$
and $\delta(v)$ is the 2-form Dirac distribution concentrated at
$v$, defined such that, for any 0-form $f$, 
\[
\int_{D_{v}}f\delta\left(v\right)\equiv f\left(v\right).
\]
However, such a formulation is too singular for our purpose. Moreover,
it contains some ambiguities. In particular, under the gauge transformation
\eqref{eq:gauge-tr-A} the curvature at $v$ is conjugated: $\PPb_{v}\mt\GG^{-1}\PPb_{v}\GG$.

It is possible to partially fix this ambiguity by choosing a \emph{Cartan
subgroup} $\DH\subset\DG$, that is, an Abelian subgroup of $\DG$
which can serve as a reference for conjugacy classes. Then we demand
that $\PPb_{v}$ is conjugate to an element $\MMb_{v}\in\mfdh$ in
the corresponding Lie subalgebra. The gauge symmetry is still acting
on $\DH$ by Weyl transformations $\WW\in\DW$, where $\DW$ is the
subgroup of residual transformations $\HH\mt\WW^{-1}\HH\WW$ which
map $\DH$ onto itself. The quotient $\DH/\DW$ then labels the sets
on conjugacy classes. This means that to every vertex $v\in\Gamma$
we attach a conjugacy class labeled by $\MMb_{v}\in\mfdh$.

\subsubsection{Properly Defining the Connection and Curvature}

The proper mathematical formulation of the naive condition \eqref{eq:naive-curv}
is to demand that we have instead a flat connection on the punctured
disk $v^{*}$. That is, 
\[
\FFb\bl_{v^{*}}=0.
\]
Now, $v^{*}$ is not simply connected; it possesses a non-trivial
homotopy group $\pi_{1}(v^{*})=\mathbb{Z}$ labeling the winding modes.
This means that the connection $\AAb\bl_{v^{*}}$ possesses a non-trivial
holonomy, 
\[
\mathrm{Hol}_{v}=\pexp\oint_{S_{v}}\AAb,
\]
where $S_{v}$ is any circle in $v^{*}$ encircling the vertex $v$
once and not encircling any other vertices, and which starts at the
cut $C_{v}$. We demand that this holonomy be in the conjugacy class
labeled by $\MMb_{v}$. We are therefore looking for a connection
on $v^{*}$ which satisfies 
\begin{equation}
\FFb\bl_{v^{*}}=0\sp\left[\mathrm{Hol}_{v}\right]=\left[\exp\MMb_{v}\right],\label{eq:Feq0,hol}
\end{equation}
where the brackets $\left[\cdot\right]$ denote the equivalence class
under conjugation; that is, $\mathrm{Hol}_{v}$ is related to $\exp\MMb_{v}$
via conjugation with some element $\WW\in\DW$.

Such a connection can be conveniently written in terms of a $\DG$-valued
0-form $\HH_{v}$ and an element $\MMb_{v}\in\mfdh$ in the Cartan
subgroup as 
\begin{equation}
\AAb\bl_{v^{*}}\equiv\left(\e^{\MMb_{v}\phi_{v}}\HH_{v}\right)^{-1}\d\left(\e^{\MMb_{v}\phi_{v}}\HH_{v}\right)=\HH_{v}^{-1}\MMb_{v}\HH_{v}\thinspace\d\phi_{v}+\HH_{v}^{-1}\d\HH_{v}.\label{eq:connection-v-calM}
\end{equation}
It is important to note that $\HH_{v}$ is defined on the full disk
$D_{v}$. In particular, it is periodic when going around $v$ and
its value $\HH_{v}\left(v\right)$ at $v$ is well defined, while
$\phi_{v}$ is defined only on the cut disk and $\d\phi_{v}$ is defined
on the punctured disk.

As for the cells, we again see that gauge transformations are given
by right translations 
\begin{equation}
\HH_{v}\left(x\right)\mt\HH_{v}\left(x\right)\GG\left(x\right),\label{eq:right-tr-v}
\end{equation}
while left translations by a constant element $\GG_{v}$ in the Cartan
subgroup $\DH$ (which thus commutes with $\MMb_{v}$), 
\begin{equation}
\HH_{v}\left(x\right)\mt\GG_{v}\HH_{v}\left(x\right),\label{eq:left-tr-v}
\end{equation}
leave the connection invariant.

\subsubsection{Calculating the Curvature}

The curvature can now be obtained in a well-defined way as follows.
First, we note that $\AAb\bl_{v^{*}}$ is, in fact, the gauge transformation
of a \textit{Lagrangian connection}\textit{\emph{ $\LLb_{v}$ by a
$\DG$-valued 0-form $\HH_{v}$: 
\[
\AAb\bl_{v^{*}}\equiv\HH_{v}^{-1}\LLb\HH_{v}+\HH_{v}^{-1}\d\HH_{v}\sp\LLb_{v}\equiv\MMb_{v}\thinspace\d\phi_{v}\sp\left[\LLb_{v},\LLb_{v}\right]=0.
\]
Now, the curvature of $\LLb_{v}$ is given by 
\[
\FFb\left(\LLb_{v}\right)=\d\LLb_{v}=\MMb_{v}\thinspace\d^{2}\phi_{v}.
\]
Of course, since the exterior derivative satisfies $\d^{2}=0$ on
$v^{*}$, the curvature vanishes on $v^{*}$, as required by }}\eqref{eq:Feq0,hol}.
However, since $\phi_{v}$ is not well defined at the origin $v\in D_{v}$,
the term $\d^{2}\phi_{v}$ might not vanish at $v$ itself. Let us
thus perform the following integral over the full disk $D_{v}$: 
\[
\int_{D_{v}}\FFb\left(\LLb_{v}\right)=\int_{\partial D_{v}}\LLb_{v}=\MMb_{v}\int_{\partial D_{v}}\d\phi_{v}=\MMb_{v},
\]
since the integral over the circle is just 1. Since $\FFb\left(\LLb_{v}\right)=0$
everywhere on $D_{v}$ except at the origin, and yet its integral
over $D_{v}$ is equal to the finite quantity $\MMb_{v}$, we are
well within our rights to declare that the curvature takes the form
\[
\FFb\left(\LLb_{v}\right)=\MMb_{v}\thinspace\delta\left(v\right).
\]
Next, we gauge-transform $\LLb_{v}\mt\AAb\bl_{v^{*}}$, obtaining
the expression \eqref{eq:connection-v-calM}. Then the curvature transforms
in the usual way: 
\[
\FFb\left(\LLb_{v}\right)\mt\FFb\left(\AAb\right)\bl_{D_{v}}=\HH_{v}^{-1}\FFb\left(\LLb_{v}\right)\HH_{v}\equiv\PPb_{v}\thinspace\delta\left(v\right),
\]
where we have defined 
\begin{equation}
\PPb_{v}\equiv\HH_{v}^{-1}\left(v\right)\MMb_{v}\HH_{v}\left(v\right).\label{eq:bigm}
\end{equation}
Thus, Eq. \eqref{eq:naive-curv} is justified; the curvature may be
thought of as taking the form $\FFb\bl_{D_{v}}=\PPb_{v}\thinspace\delta\left(v\right)$,
with the element $\HH_{v}(v)$ parametrizing the representative of
the conjugacy class.

\subsubsection{Holonomies}

Furthermore, for some subset $K_{v}\se D_{v}$ we have 
\[
\int_{K_{v}}\FFb\left(\LLb_{v}\right)=\int_{\partial K_{v}}\LLb_{v}=\MMb_{v},
\]
and by exponentiating and taking the gauge transformation $\LLb_{v}\mt\AAb\bl_{v^{*}}$
we see that the holonomy of the connection along the loop $\partial K_{v}$
starting at some point $x\in\partial K_{v}$ and winding once around
$v$ is given by 
\begin{equation}
\pexp\oint_{\partial K_{v}}\AAb=\HH_{v}^{-1}\left(x\right)\e^{\MMb_{v}}\HH_{v}\left(x\right),\label{eq:Holv}
\end{equation}
which is indeed conjugate to $\e^{\MMb_{v}}$. The advantage of the
parametrization in terms of $\HH_{v}$ is that even if the notion
of a loop starting at $v$ and encircling $v$ once is ill defined,
the right-hand side of \eqref{eq:Holv} is still well defined when
$x=v$.

The pair $(\MMb_{v},\HH_{v})$ determines the holonomy, but the reverse
is not true. The Cartan subgroup $\DH$ acts on the left of $\HH_{v}$
as a symmetry group $\HH_{v}\left(x\right)\mt\GG_{v}\HH_{v}\left(x\right)$,
with $\GG_{v}\in\DH$ constant, which leaves the connection and the
holonomy invariant. However, there is also a left action of the Weyl
group $\DW$: 
\[
\left(\MMb_{v},\HH_{v}\right)\mt\left(\WW^{-1}\MMb_{v}\WW,\WW^{-1}\HH_{v}\right)\sp\WW\in\DW,
\]
which does not leave $\MMb_{v}$ invariant, but fixes its conjugacy
class.

\subsection{Continuity Conditions Between Cells}

Let us consider the link $e^{*}=\left(cc'\right)^{*}$ connecting
two adjacent nodes $c^{*}$ and $c^{\prime*}$. This link is dual
to the edge $e=\left(cc'\right)=c\cap c'$, which is the boundary
between the two adjacent cells $c$ and $c'$. The connection is defined
in the union $c\cup c'$, while in each cell its restriction is encoded
in $\AAb\bl_{c}$ and $\AAb\bl_{c'}$ as defined above, in terms of
$\HH_{c}$ and $\HH_{c'}$ respectively.

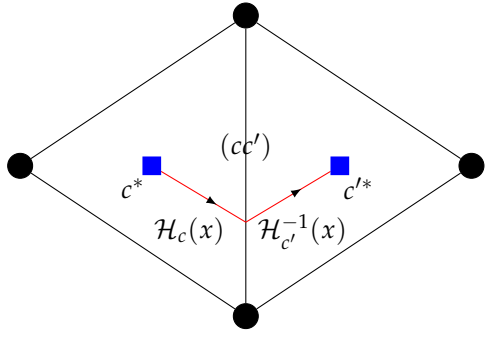
\begin{figure}[!h]
\begin{centering}
\input{Figure-3-ContinuityNN.tex} 
\par\end{centering}
\caption{\label{fig:ContinuityNN}To get from the node $c^{*}$ to the adjacent
node $c^{\prime*}$, we use the group element $\protect\HH_{cc'}$.
First, we choose a point $x$ somewhere on the edge $\left(cc'\right)=c\cap c'$.
Then, we take $\protect\HH_{c}\left(x\right)$ from $c^{*}$ to $x$,
following the first red arrow. Finally, we take $\protect\HH_{c'}^{-1}\left(x\right)$
from $x$ to $c^{\prime*}$, following the second red arrow. Thus
$\protect\HH_{cc'}=\protect\HH_{c}\left(x\right)\protect\HH_{c'}^{-1}\left(x\right)$.
Note that any $x\in c\cap c'$ will do, since the connection is flat
and thus all paths are equivalent.}
\end{figure}

The continuity equation on the edge $\left(cc'\right)$ between the
two adjacent cells\footnote{Strictly speaking, one should consider open neighborhoods $U_{c},U_{c'}$
of $c$ and $c'$ and consider the overlap condition on the open set
$U_{c}\cap U_{c'}$. We will not dwell too much on this subtlety here,
since this is not the main point of our paper, but keep in mind that
if necessary one might have to resort to open cell overlaps instead
of edges.} reads 
\[
\AAb\bl_{c}=\HH_{c}^{-1}\d\HH_{c}=\HH_{c'}^{-1}\d\HH_{c'}=\AAb\bl_{c'}\sp\textrm{on }\left(cc'\right)=c\cap c'.
\]
Since the connections match, this means that the group elements $\HH_{c}$
and $\HH_{c'}$ differ only by the action of a left symmetry element.
This implies that there exists a group element $\HH_{cc'}\in\DG$
which is independent of $x$ and provides the change of variables
between the two parametrizations $\HH_{c}\left(x\right)$ and $\HH_{c'}\left(x\right)$
on the overlap: 
\[
\HH_{c'}\left(x\right)=\HH_{c'c}\HH_{c}\left(x\right)\sp x\in\left(cc'\right)=c\cap c'.
\]
Note that $\HH_{c'c}=\HH_{cc'}^{-1}$. Furthermore, $\HH_{cc'}$ can
be decomposed as 
\begin{equation}
\HH_{cc'}=\HH_{c}\left(x\right)\HH_{c'}^{-1}\left(x\right),\label{eq:Hcc}
\end{equation}
as illustrated in Fig. \ref{fig:ContinuityNN}. The quantity $\HH_{cc'}$
is invariant under the right gauge transformation \eqref{eq:right-tr-c},
since it is independent of $c$. However, it is not invariant under
the left symmetry \eqref{eq:left-tr-c} performed at $c$ and $c'$,
under which we obtain
\begin{equation}
\HH_{c'c}\mt\GG_{c'}\HH_{c'c}\GG_{c}^{-1}.\label{eq:left-tr-Hcpc}
\end{equation}
Since this symmetry leaves the connection invariant, this means that
$\HH_{cc'}$ is \emph{not} the holonomy from $c^{*}$ to $c^{\prime*}$,
along the link $(cc')^{*}$, as it is usually assumed. Instead, from
\eqref{eq:H-from-A} and \eqref{eq:Hcc} we have that 
\begin{equation}
\HH_{cc'}=\HH_{c}\left(c^{*}\right)\left(\pexp\int_{c^{*}}^{c^{\prime*}}\AAb\right)\HH_{c'}^{-1}\left(c'^{*}\right)\label{eq:H_ccp-pathexp}
\end{equation}
is a dressed gauge-invariant observable. It will be referred to as
a \emph{discrete holonomy}, while it is understood that it is a gauge-invariant
version of the holonomy.

The map $\AAb\mt\left\{ \HH_{cc'}\right\} $ can be formalized as
follows. Let $V_{\Gamma},E_{\Gamma}$ be the sets of vertices and
edges, respectively, of the 1-skeleton $\Gamma$. Then we can either
define the space $\PP\left(\Sigma,\Gamma,\DG\right)/\DG$ of $\DG$-flat
connections on $\Sigma\backslash V_{\Gamma}$ (that is, on the 2-dimensional
manifold $\Sigma$ with a puncture at each vertex) modulo gauge transformation,
or we can define the space 
\[
\DD\left(\Sigma,\Gamma,\DG\right)=\DG^{E_{\Gamma}}/\DG^{V_{\Gamma}}=\left\{ \HH_{cc'}\in\DG\right\} /\left\{ \GG_{c}\in\DG\right\} 
\]
of discrete holonomies at each edge $e=c\cap c'$ modulo global symmetries.
The main claim we want to expand upon is that the map $\AAb/\GG\mt\{\HH_{cc'}\}/\{\GG_{c}\}$
provides an isomorphism between these two structures.

This means that we can think of the space of discrete holonomies alone,
before the quotient by the symmetry, as a definition of the space
of flat $\DG$-connections modulo gauge transformations that vanish
at the vertices of $\Gamma$. The latter space is only formally defined,
while the space $\{\HH_{cc'}\}$ of discrete connections is well defined.

\subsection{\label{subsec:Continuity-Conditions-Disks}Continuity Conditions
Between Disks and Cells}

\begin{figure}[!h]
\begin{centering}
\input{Figure-4-ContinuityNV.tex} 
\par\end{centering}
\caption{\label{fig:ContinuityNV}To get from the vertex $v$ to the node $c^{*}$,
we use the group element $\protect\HH_{vc}$. First, we choose a point
$x$ somewhere on the arc $\left(vc\right)=\partial v^{*}\cap c$.
Then, we use $\e^{\protect\MMb_{v}\phi_{v}\left(x\right)}$ to rotate
from the cut $C_{v}$ to the angle corresponding to $x$ (rotation
not illustrated). Next, we take $\protect\HH_{v}\left(x\right)$ from
$v$ to $x$, following the first red arrow. Finally, we take $\protect\HH_{c}^{-1}\left(x\right)$
from $x$ to $c^{*}$, following the second red arrow. Thus $\protect\HH_{vc}=\e^{\protect\MMb_{v}\phi_{v}\left(x\right)}\protect\HH_{v}\left(x\right)\protect\HH_{c}^{-1}\left(x\right)$.}
\end{figure}
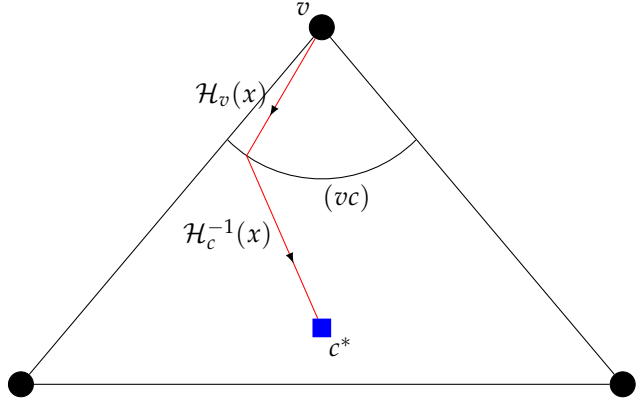

A similar discussion applies when one looks at the overlap $v^{*}\cap c$
between a punctured disk $v^{*}$ and a cell $c$. The boundary of
this region consists of two \emph{truncated edges} of length $R$
(the coordinate radius of the disk) touching $v$, plus an arc connecting
the two edges, which lies on the boundary of the disk $v^{*}$. In
the following we denote this arc\footnote{The arc $\left(vc\right)$ is dual to the line segment $\left(vc\right)^{*}$
connecting the vertex $v$ with the node $c^{*}$, just as the edge
$e$ is dual to the link $e^{*}$.} by $\left(vc\right)$. It is clear that the union of all such arcs
around a vertex $v$ reconstructs the outer boundary $\partial_{R}v^{*}$
of the disk, as defined in \eqref{eq:inner-outer-boundary}: 
\[
\left(vc\right)\equiv\partial_{R}v^{*}\cap c,\qquad\partial_{R}v^{*}=\bigcup_{c\ni v}\left(vc\right),
\]
where $c\ni v$ means ``all cells $c$ which have the vertex $v$
on their boundary''. It is also useful to introduce the \emph{truncated
cells}: 
\[
\ct\equiv c\backslash\bigcup_{v\in c}D_{v},
\]
where $v\in c$ means ``all vertices $v$ on the boundary of the
cell $c$''. In other words, $\ct$ is the complement of the union
of disks $D_{v}$ intersecting $c$. The union of all truncated cells
reconstructs the manifold $\Sigma$ minus the disks: 
\[
\bigcup_{c}\ct=\Sigma\backslash\bigcup_{v}D_{v}.
\]
In the intersection $v^{*}\cap c$ we have two different descriptions
of the connection $\AAb$. On $c$ it is described by the $\DG$-valued
0-form $\HH_{c}$, and on $v^{*}$ it is described by a $\DG$-valued
0-form $\HH_{v}$. The fact that we have a single-valued connection
is expressed in the continuity conditions 
\begin{equation}
\AAb\bl_{v^{*}}=\HH_{v}^{-1}\MMb_{v}\HH_{v}\thinspace\d\phi_{v}+\HH_{v}^{-1}\d\HH_{v}=\HH_{c}^{-1}\d\HH_{c}=\AAb\bl_{c}\sp\textrm{on }\left(vc\right)=\partial v^{*}\cap c.\label{continuity-vc}
\end{equation}
The relation between the two connections can be integrated. It means
that the elements $\HH_{v}(x)$ and $\HH_{c}(x)$ differ by the action
of the left symmetry group. In practice, this means that the integrated
continuity relation involves a (discrete) holonomy $\HH_{cv}$: 
\begin{equation}
\HH_{c}\left(x\right)=\HH_{cv}\e^{\MMb_{v}\phi_{v}\left(x\right)}\HH_{v}\left(x\right)\sp x\in\left(vc\right)\label{eq:continuity-H_cv}
\end{equation}
where $\phi_{v}\left(x\right)$ is the angle corresponding to $x$
with respect to the cut $C_{v}$. Isolating $\HH_{vc}\equiv\HH_{cv}^{-1}$,
we find 
\[
\HH_{vc}=\e^{\MMb_{v}\phi_{v}\left(x\right)}\HH_{v}\left(x\right)\HH_{c}^{-1}\left(x\right),
\]
which is illustrated in Fig. \ref{fig:ContinuityNV}.

The quantity $\HH_{vc}$ is invariant under right gauge transformations
\eqref{eq:right-tr-c} and \eqref{eq:right-tr-v}:
\begin{equation}
\HH_{c}\left(x\right)\mt\GG_{c}\HH_{c}\left(x\right)\sp\HH_{v}\left(x\right)\mt\GG_{v}\HH_{v}\left(x\right).\label{eq:left-c-v}
\end{equation}
However, under left symmetry transformations \eqref{eq:left-tr-c}
and \eqref{eq:left-tr-v}, the connection is left invariant, and we
get 
\begin{equation}
\HH_{vc}\mt\GG_{v}\HH_{vc}\GG_{c}^{-1}.\label{eq:left-tr-Hvc}
\end{equation}

Note also that the translation in $\phi_{v}\left(x\right)$ can be
absorbed into the definition of $\HH_{v}$, so that the transformation
\begin{equation}
\phi_{v}\left(x\right)\mt\phi_{v}\left(x\right)+\beta_{v}\sp\HH_{v}\left(x\right)\mt\e^{-\MMb_{v}\beta_{v}}\HH_{v}\left(x\right),\label{transphi}
\end{equation}
is also a symmetry under which \eqref{eq:continuity-H_cv} is invariant.
The connection $\AAb\bl_{v^{*}}$ is also invariant under this symmetry.

\subsection{Summary}

In conclusion, the connection $\AAb$ is defined on every point of
the manifold $\Sigma$ as follows. Inside each cell $c$, we have
a flat connection $\AAb\bl_{c}$. Since the cell $c$ is simply connected,
this connection can be written in terms of a $\DG$-valued 0-form
$\HH_{c}$: 
\[
\AAb\bl_{c}=\HH_{c}^{-1}\d\HH_{c}.
\]
Inside each punctured disk $v^{*}$, we have a flat connection $\AAb\bl_{v^{*}}$
parametrized by a $\DG$-valued 0-form $\HH_{v}$ and a constant element
$\MMb_{v}$ of the Cartan subalgebra $\mfdh$: 
\[
\AAb\bl_{v^{*}}=\HH_{v}^{-1}\MMb_{v}\HH_{v}\thinspace\d\phi_{v}+\HH_{v}^{-1}\d\HH_{v}.
\]
The continuity of the connection $\AAb$ along the boundaries between
cells and other cells or disks is expressed in the relations 
\[
\HH_{cc'}=\HH_{c}\left(x\right)\HH_{c'}^{-1}\left(x\right)\sp x\in\left(cc'\right)\equiv c\cap c',
\]
\[
\HH_{vc}=\e^{\MMb_{v}\phi_{v}\left(x\right)}\HH_{v}\left(x\right)\HH_{c}^{-1}\left(x\right)\sp x\in\left(vc\right)\equiv\partial v^{*}\cap c.
\]

\subsection{The Chern-Simons Symplectic Potential}

The first goal of this paper is the construction of the symplectic
potential for the Chern-Simons connection in terms of the discrete
data $\left\{ \HH_{cc'}\right\} _{c,c'\in\Delta}$ or the discrete
data $\left\{ \HH_{cv},\MMb_{v}\right\} _{(c^{*},v)\in\Gamma^{*}\times\Gamma}$.
In the continuum, the Chern-Simons symplectic structure is given by\footnote{The dot product is defined for any two Lie algebra elements $\AAb,\BBb$
with components $\AA^{i}\equiv\Tr\left(\AAb\ta^{i}\right)$ and $\BB^{i}\equiv\Tr\left(\BBb\ta^{i}\right)$,
where $\ta^{i}$ are the generators of the Lie algebra, as 
\[
\AAb\cdot\BBb\equiv\Tr\left(\AAb\wedge\BBb\right)=\AA^{i}\wedge\BB_{i}.
\]
} 
\[
\Omega_{\Sigma}\left(\AAb\right)=\int_{\Sigma}\omega\left(\AAb\right),\qquad\omega\left(\AAb\right)=\delta\AAb\cdot\delta\AAb.
\]
We are interested in the computation of the Chern-Simons symplectic
potential for a disk $D$, which has the symplectic form $\Omega_{D}\left(\AAb\right)\equiv\int_{D}\omega\left(\AAb\right)$.
We refer the reader to \cite{Alekseev:1993rj,Meusburger:2005mg} for
earlier references exploring the same question.

Omitting the subscript $v$ for brevity, we consider the case where
the connection $\AAb$ inside $D$ can be written as the gauge transformation
of a \textit{Lagrangian connection}\textit{\emph{ $\LLb$ by a $\DG$-valued
0-form $\HH$: 
\[
\AAb\equiv\HH^{-1}\LLb\HH+\HH^{-1}\d\HH,
\]
where we assume that $\LLb$ belongs to a Lagrangian subspace: This
means that $\delta\LLb\cdot\delta\LLb=0$, so that $\omega\left(\LLb\right)=0$,
and where we have omitted the subscripts $v$ for brevity. One first
evaluates the variation 
\[
\delta\AAb=\HH^{-1}\left(\delta\LLb+\d_{\LLb}\De\HH\right)\HH,
\]
where $\d_{\LLb}$ denotes the }}\textit{covariant differential}\textit{\emph{
$\d_{\LLb}\equiv\d+[\LLb,\ ]$, and we have used the shorthand notation}}
\[
\De\HH\equiv\delta\HH\HH^{-1}
\]
for the right-invariant Maurer-Cartan variational form, described
in more detail in Appendix \ref{sec:The-Maurer-Cartan-Form}. \textit{\emph{Under
this assumption, we can evaluate the Chern-Simons symplectic form:}}
\[
\omega\left(\AAb\right)=2\delta\left(\FFb\left(\LLb\right)\cdot\De\HH\right)+\d\left(\De\HH\cdot\d\De\HH\right)-2\d\delta\left(\LLb\cdot\De\HH\right),
\]
\textit{\emph{where $\FFb\left(\LLb\right)\equiv\d\LLb+\hf\left[\LLb,\LLb\right]$
is the curvature of $\LLb$. The derivation of this important formula
is spelled out in Appendix }}\ref{sec:Evaluation-of-CS}\textit{\emph{.
Furthermore, in the particular case considered above, the }}Lagrangian
connection satisfies 
\[
\LLb\equiv\MMb\thinspace\d\phi\sp\FFb\left(\LLb\right)=\MMb\thinspace\delta\left(v\right)\sp\MMb\in\mfdh.
\]
In this case, the symplectic form associated with a disk $D$ centered
at $v$ may be further simplified to

\[
{\Omega_{D}\left(\AAb\right)=\oint_{\partial D}\De\tHH\cdot\d\De\tHH-2\oint_{\partial D}\delta\left(\HH^{-1}\left(v\right)\MMb\HH\left(v\right)\cdot\De\tHH\right)\d\phi},
\]
where we have defined $\tHH\left(x\right)\equiv\HH^{-1}\left(v\right)\HH\left(x\right)$,
such that $\tHH\left(v\right)=1$, and used the ``Leibniz rule''
\ref{eq:Delta-Leibniz-rule} for the Maurer-Cartan form, 
\[
\De\tHH(x)=\HH^{-1}\left(v\right)\left(\De\HH(x)-\De\HH(v)\right)\HH\left(v\right).
\]
This constitutes the first main technical result of this paper.

The goal of this paper is to study in depth this formula and the consequences
it has when one starts to glue together different regions associated
to a collection of topological disks. In general this is a formidable
task, and we will pursue it under simplifying assumptions. Mainly,
we will need to choose a group $\DG$ such that the first term $\oint_{\partial D}\De\tHH\cdot\d\De\tHH$
can be written as an exact variational form, and thus the symplectic
form can be written as the variation of a symplectic potential, $\Omega_{D}=\delta\Theta$.

\section{\label{sec:Gravity:-Specializing-to}Gravity: Specializing to $G\ltimes\protect\mfg^{*}$}

\subsection{Introduction}

In order to move forward and explore the discretization of the Chern-Simons
symplectic form, we will focus on theories for which the Chern-Simons
group $\DG$ is specifically chosen to be a Drinfeld double which
is locally of the form $\DG\equiv G\times G^{*}$, with $G^{*}$ a
group dual to $G$. At the Lie algebra level, this means that the
Lie algebra of the double, $\mfdg$, possesses the structure of a
Manin triple: $\mfdg=\mfg\oplus\mfg^{*}$ is equipped with a non-degenerate
pairing $\langle\thinspace,\thinspace\rangle$ which is ad-invariant,
i.e. $\langle[\A,\B],\C\rangle=\langle\A,[\B,\C]\rangle$, and which
is such that both $\mfg$ and $\mfg^{*}$ are isotropic; i.e. the
pairing restricted to $\mfg$ or $\mfg^{*}$ is null.

It turns out that all theories of Euclidean gravity in 2+1 dimensions
correspond to the Chern-Simons theory of a double where the factor
$G$ is simply the group $\SUT$, while the factor $G^{*}$ is another
$\SUT$, the 2D Borel group $\mathrm{AN_{2}}$, or simply the Abelian
group $\mathbb{R}^{3}$, depending on the sign of the cosmological
constant.

In this work we will focus, for simplicity, on the case of a zero
cosmological constant. This means that we will restrict ourselves
to the study of a double of trivial topology, where $G$ is a simple
group and $G^{*}=\mfg^{*}$ is an Abelian group. In this case, the
double is simply the semi-direct product 
\[
\DG\equiv G\ltimes\mfg^{*}.
\]
Note that $\DG$ is also isomorphic to the cotangent bundle $T^{*}G$,
which shows that the natural pairing on $\DG$ descends from the duality
pairing of vector and covector fields $TG\times T^{*}G$, since the
Lie algebra of $G$ can be viewed as the set of right-invariant vector
fields.

When $G=\SUT$ or $G=\mathrm{SU}(1,1)$, this group reduces to the
3D \emph{Euclidean group} $\ISUT$ or 2+1D \emph{Poincaré group} $\mathrm{ISU}(1,1)$,
respectively, which is the group of isometries of 2+1D flat gravity.\footnote{See \cite{Meusburger:2003ta,Meusburger:2008dc,Ballesteros:2013dca}
for more details on the correspondence between doubles and 2+1D gravity} Since we work at the classical level, none of our derivations depends
on the fact that the $G=\SUT$, so we will keep $G$ general; we just
need $G$ to be equipped with a non-trivial trace, denoted $\Tr$
and incorporated into the dot product: 
\[
\A\cdot\B\equiv\Tr\left(\A\wedge\B\right)\equiv A^{i}\wedge B_{i}\sp A^{i}\equiv\Tr\left(\A\ta^{i}\right)\sp B^{i}\equiv\Tr\left(\B\ta^{i}\right),
\]
where $\ta^{i}$ are the generators of the Lie algebra. Keeping in
mind the applications to 2+1D gravity, we will call the group $\DG$
the ``Euclidean'' group for reference.

\subsection{The ``Euclidean'' Group $\protect\DG$}

The transformations between cells will be given by $\DG$ group elements,
and the connection 1-form $\AAb$ will be valued in the Lie algebra
$\mfdg$. This algebra is generated by the rotation generators $\J_{i}$
and the translation generators $\P_{i}$, where $i=1,\ldots,\dim\mfg$.
The generators have the Lie brackets and Killing form 
\begin{equation}
\left[\P_{i},\P_{j}\right]=0\sp\left[\J_{i},\J_{j}\right]=C_{ij}{}^{k}\J_{k}\sp\left[\J_{i},\P_{j}\right]=C_{ij}{}^{k}\P_{k}\sp\langle\J_{i},\P_{j}\rangle=\delta_{ij}.\label{eq:IG-generators}
\end{equation}
Here $C_{ij}{}^{k}$ denotes the structure constants\footnote{For $\mfg=\sut$ we have $C_{ijk}=\epsilon_{ijk}$, the Levi-Civita
tensor, and $\delta_{ij}$ the Euclidean metric. In this case we can
define $\ta_{i}\equiv-\i\si_{i}/2$ to be the generators of $\sut$,
where $\si_{i}$ are the Pauli matrices. The generators satisfy the
algebra $\left[\ta_{i},\ta_{j}\right]=\epsilon_{ij}^{k}\ta_{k}$.
The normalized trace is $\Tr\equiv-2\tr$ where $\tr$ is the usual
matrix trace, and it satisfies $\Tr\left(\ta_{i}\ta_{j}\right)=\delta_{ij}$.} of $\mfg$. We see that both the rotation algebra $\mfg$ generated
by $\J_{i}$ and the translation algebra $\mfg^{*}$ generated by
$\P_{i}$ are subalgebras of $\mfdg$. However, $\mfg$ is non-Abelian,
while $\mfg^{*}$ is Abelian and a normal subalgebra. The pairing
$\langle\,,\,\rangle$ identifies the translation subalgebra with
the dual of the Lie algebra (which is why we denote the translation
algebra by $\mfg^{*}$). The metric $\delta_{ij}$ involved in the
definition of the pairing is a Killing metric; the tensor $C_{ijk}\equiv C_{ij}{}^{l}\delta_{lk}$
is fully anti-symmetric.

By exponentiating this algebra, we get 
\[
\DG\cong G\ltimes\mfg^{*},
\]
where $\mfg^{*}$ is an Abelian normal subgroup. This means that every
element $\HH\in\DG$ may be uniquely decomposed, using the so-called
\textit{Cartan decomposition}, into a pair 
\[
\HH\equiv\e^{\y}h\sp\left(h,\y\right)\in G\xx\mfg^{*}.
\]
To avoid confusion, throughout the paper we will be using a calligraphic
font (e.g. $\HH$) for $\DG$ elements, bold calligraphic font (e.g.
$\MMb$) for $\mfdg$ elements, Roman font (e.g. $h$) for $G$ elements
and bold Roman font (e.g. $\y$) for $\mfg^{*}$ elements.

The product rule is such that 
\[
\e^{\y}\e^{\y'}=\e^{\y+\y'}\sp h\e^{\y}=\e^{h\y h^{-1}}h.
\]
This means that products and inverse elements of $\DG$ elements are
given in terms of $G$ and $\mfg^{*}$ elements by: 
\begin{equation}
\HH\HH'=\e^{\y}h\e^{\y'}h'=\e^{\y+h\y'h^{-1}}hh',\label{eq:IG-product}
\end{equation}
\begin{equation}
\HH^{-1}=h^{-1}\e^{-\y}=\e^{-h^{-1}\y h}h^{-1},\label{eq:IG-inverse}
\end{equation}
and by combining them together we get 
\begin{equation}
\HH^{-1}\HH'=h^{-1}\e^{-\y}\e^{\y'}h'=\e^{h^{-1}\left(\y'-\y\right)h}h^{-1}h'.\label{eq:IG-combined}
\end{equation}

\subsection{\label{subsec:con-ff-inside-cd}The $G$ Connection and Frame Field
Inside the Cells}

Our $\mfdg$-valued connection 1-form $\AAb$ can be decomposed in
terms of the generators of the algebra as follows: 
\[
\AAb\equiv A^{i}\J_{i}+E^{i}\P_{i}.
\]
Bearing in mind the Chern-Simons formulation of gravity, it will be
convenient to interpret $\A\equiv A^{i}\J_{i}$ as a $\mfg$-valued
connection 1-form and $\E\equiv E^{i}\P_{i}$ as a $\mfg^{*}$-valued
\emph{(co)frame field}. Accordingly, the curvature 2-form $\FFb$
of the $\DG$ connection $\AAb$ can also be decomposed into the sum
$\FFb=\F+\T$ of a ``rotational'' curvature $\F$ (referred to as
the $G$-curvature) and ``translational'' curvature (referred to
as the $G$-torsion). The $G$ curvature $\F$ and torsion $\T$ are
2-forms defined as 
\[
\F\left(\A\right)\equiv\d\A+\hf\left[\A,\A\right],\qquad\T\left(\A,\E\right)\equiv\d_{\A}\E\equiv\d\E+\left[\A,\E\right].
\]
If $\AAb$ is flat, $\FFb\left(\AAb\right)=0$, we automatically get
a vanishing $G$ curvature and torsion: 
\[
\F\left(\A\right)=\T\left(\A,\E\right)=0.
\]
In other words, a flat $\DG$ geometry corresponds to a flat and torsionless
$G$ geometry.

Now, recall from \eqref{eq:A-from-H} that inside the cell $c$ we
have 
\[
\AAb\bl_{c}=\HH_{c}^{-1}\d\HH_{c},
\]
where $\HH_{c}$ is a $\DG$-valued 0-form. The group element $\HH_{c}^{-1}\left(c\right)\HH_{c}(x)$
defines the $\DG$ holonomy of the connection $\AAb$ from $c^{*}$
to $x$. We can decompose $\HH_{c}$ in terms of a rotation $h_{c}$
and a translation $\y_{c}$ using the Cartan decomposition: 
\[
\HH_{c}\equiv\e^{\y_{c}}h_{c}\sp h_{c}\in G\sp\y_{c}\in\mfg.
\]
Plugging it into $\AAb$, we get 
\[
\AAb\bl_{c}=h_{c}^{-1}\d\y_{c}h_{c}+h_{c}^{-1}\d h_{c}.
\]
It is easy to see that the first term is a pure rotation, that is,
proportional to $\J_{i}$, while the second term is a pure translation,
that is, proportional to $\P_{i}$. Recalling that $\AAb=A^{i}\J_{i}+E^{i}\P_{i}$,
we deduce that the corresponding $\mfg$-valued connection and frame
field are 
\begin{equation}
\A\bl_{c}=h_{c}^{-1}\d h_{c}\sp\E\bl_{c}=h_{c}^{-1}\d\y_{c}h_{c}.\label{eq:AcEc}
\end{equation}
As before, we have two types of transformations. Gauge transformations
(right translations), labeled by $\DG$-valued 0-forms $(g,\x)$ and
given by 
\[
\A\mt g^{-1}\A g+g^{-1}\d g,\qquad\E\mt g^{-1}\left(\E+\d_{\A}\x\right)g,
\]
act on $(h_{c},\y_{c})$ as follows from \eqref{eq:right-tr-c}: 
\begin{equation}
h_{c}(x)\mt h_{c}(x)g(x),\qquad\y_{c}(x)\mt\y_{c}(x)+(h_{c}\x h_{c}^{-1})(x).\label{eq:gauge-c}
\end{equation}
Symmetry transformations (left translations), labeled by constant
$\DG$ elements $\left(g_{c},\z_{c}\right)$ assigned to the cell
$c$, leave the connection and frame invariant, $(\A,\E)\mt(\A,\E)$,
and act on $(h_{c},\y_{c})$ as follows from \eqref{eq:left-tr-c}:
\begin{equation}
h_{c}(x)\mt g_{c}h_{c}(x),\qquad\y_{c}(x)\mt\z_{c}+g_{c}\y_{c}(x)g_{c}^{-1}.\label{eq:symmetry-c}
\end{equation}

\subsection{The $G$ Connection and Frame Field Inside the Disks}

From \eqref{eq:connection-v-calM}, the connection $\AAb$ inside
the punctured disk $v^{*}$ is labeled by a $\DG$-valued 0-form $\HH_{v}$
and an element $\MMb_{v}$ of the Cartan subalgebra $\mfdh$. We may
decompose $\MMb_{v}$ as follows: 
\begin{equation}
\MMb_{v}=\M_{v}+\SS_{v}\sp\M_{v}\in\mfh\sp\SS_{v}\in\mfh^{*},\label{eq:M-M-S}
\end{equation}
where $\mfh$ is the Cartan subalgebra of $\mfg$. The connection
$\AAb$ inside the punctured disk $v^{*}$ is then given by 
\[
\AAb\bl_{v^{*}}=\HH_{v}^{-1}\left(\M_{v}+\SS_{v}\right)\HH_{v}\thinspace\d\phi_{v}+\HH_{v}^{-1}\d\HH_{v}.
\]
Using the Cartan decomposition 
\[
\HH_{v}\equiv\e^{\y_{v}}h_{v},
\]
we can unpack the $\mfdg$-valued connection into the corresponding
$\mfg$-valued connection and frame field in $v^{*}$: 
\begin{equation}
\A\bl_{v^{*}}=h_{v}^{-1}\M_{v}h_{v}\thinspace\d\phi_{v}+h_{v}^{-1}\d h_{v}\sp\E\bl_{v^{*}}=h_{v}^{-1}\left(\left(\SS_{v}+\left[\M_{v},\y_{v}\right]\right)\d\phi_{v}+\d\y_{v}\right)h_{v}.\label{eq:AvEv}
\end{equation}
We may similarly decompose the momentum $\PPb_{v}$ defined in \eqref{eq:bigm}:
\[
\PPb_{v}\equiv\HH_{v}^{-1}\left(v\right)\MMb_{v}\HH_{v}\left(v\right)=\p_{v}+\j_{v},
\]
where $\p_{v},\j_{v}\in\mfg$ represent the \emph{momentum }and \emph{angular
momentum }respectively: 
\begin{equation}
\p_{v}\equiv h_{v}^{-1}\M_{v}h_{v},\qquad\j_{v}\equiv h_{v}^{-1}\left(\SS_{v}+\left[\M_{v},\y_{v}\right]\right)h_{v}.\label{eq:momenta}
\end{equation}
Then, by decomposing \eqref{eq:naive-curv}, we may obtain the (naive)
distributional $\mfg$-valued curvature and torsion 2-forms: 
\begin{equation}
\F\bl_{v^{*}}=\p_{v}\thinspace\delta\left(v\right),\qquad\T\bl_{v^{*}}=\j_{v}\,\delta\left(v\right).\label{eq:F-T-delta}
\end{equation}
Finally, we see that gauge transformations (right translations) \eqref{eq:right-tr-v}
labeled by $\DG$-valued 0-forms $(g,\x)$ act on $(h_{v},\y_{v})$
the same way they act on $(h_{c},\y_{c})$: 
\begin{equation}
h_{v}(x)\mt h_{v}(x)g(x),\qquad\y_{v}(x)\mt\y_{v}(x)+(h_{v}\x h_{v}^{-1})(x).\label{eq:gauge-v}
\end{equation}
The symmetry transformations (left translations) \eqref{eq:left-tr-v}
labeled by constant elements $(g_{v},\z_{v})$ of the Cartan subgroup
$\DH$, leave $\M_{v}$ and $\SS_{v}$ invariant, while they transform
$(h_{v},\y_{v})$ by 
\begin{equation}
h_{v}(x)\mt g_{v}h_{v}(x),\qquad\y_{v}(x)\mt\z_{v}+g_{v}\y_{v}(x)g_{v}^{-1}.\label{eq:symmetry-v}
\end{equation}
We discuss these two types of transformations in the context of the
relativistic particle at $v$ in Appendix \ref{sec:The-Relativistic-Particle}.

As we have seen, the boundary conditions between cells and disks can
be expressed as continuity equations either across the edge $\left(cc'\right)\equiv c\cap c'$
bounding two cells or across the arc $\left(vc\right)\equiv\partial v^{*}\cap c$
bounding the interface of a disk and a cell. Although both continuity
conditions are similar, the one across the arcs is more involved.

\subsection{Continuity Conditions Between Cells}

For $x\in\left(cc'\right)=c\cap c'$ we have from \eqref{eq:Hcc}
\begin{equation}
\HH_{c'}\left(x\right)=\HH_{c'c}\HH_{c}\left(x\right)\sp x\in\left(cc'\right).\label{eq:continuity-cc}
\end{equation}
The holonomies $\HH_{cc'}$ and $\HH_{c}$ can be decomposed into
a rotational and translational part using the Cartan decomposition,
as usual\footnote{The index placement in $\HH_{cc'}$ reflects that this is a transformation
mapping objects at $c$ to objects at $c'$, while $\y_{c}^{c'}$
denotes a transformation based at $c$.}: 
\[
\HH_{cc'}=\e^{\y_{c}^{c'}}h_{cc'}\sp\HH_{c}=\e^{\y_{c}}h_{c}.
\]
Note that under the exchange of indices we have, from \eqref{eq:IG-inverse},
\[
\HH_{cc'}=\HH_{c'c}^{-1}\sp h_{cc'}=h_{c'c}^{-1},\qquad\y_{c}^{c'}=-h_{cc'}\y_{c'}^{c}h_{c'c}.
\]
Using these quantities and the rules \eqref{eq:IG-product} and \eqref{eq:IG-combined},
\eqref{eq:continuity-cc} can be split into a rotational and translational
part: 
\begin{equation}
h_{c'}\left(x\right)=h_{c'c}h_{c}\left(x\right)\sp\y_{c'}\left(x\right)=h_{c'c}\left(\y_{c}\left(x\right)-\y_{c}^{c'}\right)h_{cc'},\label{eq:continuity-cc-split}
\end{equation}
for $x\in\left(cc'\right)$. These relations show that the rotational
and translational holonomies $(h_{cc'},\y_{c}^{c'})$ are \emph{invariant}
under the gauge transformation \eqref{eq:gauge-c} (since it is independent
of $c$). On the other hand, the discrete ``holonomies'' $(h_{cc'},\y_{c}^{c'})$
transform non-trivially under the global symmetries \eqref{eq:symmetry-c}:
\begin{equation}
h_{c'c}\mt\tilde{h}_{c'c}\equiv g_{c'}h_{c'c}g_{c}^{-1}\sp\y_{c}^{c'}\mt\tilde{\y}_{c}^{c'}\equiv g_{c}\y_{c}^{c'}g_{c}^{-1}+\z_{c}-\tilde{h}_{cc'}\z_{c'}\tilde{h}_{c'c}.\label{eq:sym-cpc}
\end{equation}
This may also be obtained from the Cartan decomposition of \eqref{eq:left-tr-Hcpc},
$\tilde{\HH}_{c'c}\equiv\GG_{c'}\HH_{c'c}\GG_{c}^{-1}$.

\subsection{Continuity Conditions Between Disks and Cells}

For $x\in\left(vc\right)=\partial v^{*}\cap c$ one has the continuity
condition \eqref{eq:continuity-H_cv}: 
\begin{equation}
\HH_{c}\left(x\right)=\HH_{cv}\e^{\MMb_{v}\phi_{v}\left(x\right)}\HH_{v}\left(x\right)\sp x\in\left(vc\right).\label{eq:continuity-cv}
\end{equation}
The holonomies $\HH_{cv}$ and $\HH_{v}$ can be decomposed into a
rotational and translational part as we did for $\HH_{cc'}$ and $\HH_{c}$
above, and given \eqref{eq:M-M-S}, we can write 
\[
\e^{\MMb_{v}\phi_{v}\left(x\right)}=\e^{\SS_{v}\phi_{v}\left(x\right)}\e^{\M_{v}\phi_{v}\left(x\right)},
\]
where $\SS_{v}\phi_{v}\left(x\right)$ is the translational part and
$\e^{\M_{v}\phi_{v}\left(x\right)}$ the rotational part.\footnote{In comparison, when we decomposed $\HH\equiv\e^{\y}h$, the translational
part was $\y$ and the rotational part was $h$.} Using these quantities, the continuity relations \eqref{eq:continuity-cv}
can be split into a rotational and translational part: 
\begin{equation}
h_{c}\left(x\right)=h_{cv}\e^{\M_{v}\phi_{v}\left(x\right)}h_{v}\left(x\right)\sp x\in\left(vc\right),\label{eq:continuity-h_cv}
\end{equation}
\begin{equation}
\y_{c}\left(x\right)=h_{cv}\left(\e^{\M_{v}\phi_{v}\left(x\right)}\left(\y_{v}\left(x\right)+\SS_{v}\phi_{v}\left(x\right)\right)\e^{-\M_{v}\phi_{v}\left(x\right)}-\y_{v}^{c}\right)h_{vc}\sp x\in\left(vc\right).\label{eq:continuity-y_cv}
\end{equation}
The quantities $h_{vc}$ and $\y_{v}^{c}$ are invariant under the
gauge transformation (right translation) \eqref{eq:gauge-c} and \eqref{eq:gauge-v}.
However, under the symmetry transformation (left translation) \eqref{eq:symmetry-c}
and \eqref{eq:symmetry-v}, with $g_{c}\in G$ and $g_{v}$ in the
Cartan subgroup $H$, 
\[
h_{c}\mt g_{c}h_{c},\qquad\y_{c}\mt\z_{c}+g_{c}\y_{c}g_{c}^{-1},
\]
\begin{equation}
h_{v}\mt g_{v}h_{v},\qquad\y_{v}\mt\z_{v}+g_{v}\y_{v}g_{v}^{-1},\label{eq:hv-gv-left-tr}
\end{equation}
we have 
\begin{equation}
h_{vc}\mt\hht_{vc}\equiv g_{v}h_{vc}g_{c}^{-1}\sp\y_{v}^{c}\mt\tilde{\y}_{v}^{c}\equiv g_{v}\y_{v}^{c}g_{v}^{-1}+\z_{v}-\hht_{vc}\z_{c}\hht_{cv},\label{eq:sym-vc}
\end{equation}
where we have used the fact that $\e^{\M_{v}}$, $\SS_{v}$, $g_{v}$
and $\z_{v}$ all commute with each other. This follows from \eqref{eq:left-tr-Hvc},
$\HH_{vc}\mt\GG_{v}\HH_{vc}\GG_{c}^{-1}$, using the Cartan decomposition.

Note also that the continuity relation on $\left(vc\right)$ and the
connection $\AAb$ are invariant under the symmetry transformation
\eqref{transphi}: 
\[
\phi_{v}\left(x\right)\mt\phi_{v}\left(x\right)+\beta_{v}\sp\HH_{v}\left(x\right)\mt\e^{-\MMb_{v}\beta_{v}}\HH_{v}\left(x\right).
\]
This transformation of $\HH_{v}\left(x\right)$ decomposes via \eqref{eq:IG-combined}
as follows: 
\begin{equation}
\phi_{v}\left(x\right)\mt\phi_{v}\left(x\right)+\beta_{v}\sp\y_{v}\left(x\right)\mt\e^{-\M_{v}\beta_{v}}\left(\y_{v}\left(x\right)-\SS_{v}\beta_{v}\right)\e^{\M_{v}\beta_{v}}\sp h_{v}\left(x\right)\mt\e^{-\M_{v}\beta_{v}}h_{v}\left(x\right).\label{eq:beta-symmetry-decomp}
\end{equation}
Of course, by construction, the relations \eqref{eq:continuity-h_cv}
and \eqref{eq:continuity-y_cv} are invariant under this transformation.
This transformation turns out to be a special case of a more general
class of transformations, as shown in Appendix \ref{sec:The-Relativistic-Particle}.

\section{\label{sec:Discretizing-the-Symplectic}Discretizing the Symplectic
Potential for 2+1 Gravity}

\subsection{The Symplectic Potential}

\subsubsection{The $BF$ Action}

Now that we have expressed the connection and frame field in terms
of discretized variables, we would like to construct the phase space
structure. For this we take inspiration from the 2+1 gravity action,
as given by $BF$ theory. The $BF$ action is\footnote{We view both $\E$ and $\A$ as elements of $\mfg$, and we define
the normalized trace $\Tr$, which satisfies 
\begin{equation}
\Tr\left(\ta_{i}\ta_{j}\right)=\delta_{ij},\label{eq:trace-tau-i-j}
\end{equation}
where $\ta_{i}$ are the generators of $\mfg$. Then the dot product
is defined as before, $\A\cdot\B\equiv\Tr\left(\A\wedge\B\right)=A^{i}\wedge B_{i}$.} 
\[
S=\int_{M}\E\cdot\F\left(\A\right),
\]
where $M$ is a 2+1-dimensional spacetime manifold, and the symplectic
potential is 
\[
\Theta=-\int_{\Sigma}\E\cdot\delta\A,
\]
where $\Sigma$ is a spatial slice. To get the discretized version
of the symplectic potential, we are going to express the different
components in terms of their associated holonomies, in the truncated
cells $\ct\equiv c\backslash\bigcup_{v\in c}v^{*}$ and punctured
disks $v^{*}$ . In other words, we define 
\[
\Theta_{c}\equiv-\int_{\ct}\E\cdot\delta\A,\qquad\Theta_{v^{*}}\equiv-\int_{v^{*}}\E\cdot\delta\A,
\]
and since $\Sigma\backslash V_{\Gamma}=\cup_{c}\tilde{c}\cup_{v}v^{*}$
the total symplectic potential is simply the sum over all cells $c$
and vertices $v$: 
\[
\Theta=\sum_{c}\Theta_{c}+\sum_{v}\Theta_{v^{*}}.
\]
We will first evaluate $\Theta_{c}$ and $\Theta_{v^{*}}$ independently
using the flatness condition, and then take advantage of the simplification
that occurs thanks to the fact that the transition maps between cells
and adjacent cells or disks are $\DG$ transformations.

For a quicker derivation using non-periodic variables, please see
Appendix \ref{sec:A-Quicker-Derivation}.

\subsubsection{Evaluation of $\Theta_{c}$}

For $\Theta_{c}$, we have from \eqref{eq:AcEc} that $\A\bl_{c}=h_{c}^{-1}\d h_{c}$
and $\E\bl_{c}=h_{c}^{-1}\d\y_{c}h_{c}$, and we find 
\[
\delta\A\bl_{c}=h_{c}^{-1}\left(\d\De h_{c}\right)h_{c}.
\]
Thus\footnote{We call this choice the \emph{LQG polarization}. One could alternatively
write this expression as a total differential using $\d\y_{c}\cdot\d\De h_{c}=\d\left(\y_{c}\cdot\d\De h_{c}\right)$,
that is, with $\d$ on $\De h_{c}$ instead of $\y_{c}$. This leads
to the ``dual polarization'', which we will explore in \cite{barak3d}.
, } 
\[
-\E\cdot\delta\A=-\d\y_{c}\cdot\d\De h_{c}=\d\left(\d\y_{c}\cdot\De h_{c}\right),
\]
and we may integrate to get $\Theta_{c}$ as an integral over the
boundary of the truncated cell 
\[
\Theta_{c}=\int_{\partial\ct}\d\y_{c}\cdot\De h_{c}.
\]
In order to integrate this further, we will need to use the continuity
conditions, which we will do in Sec. \ref{subsec:Rearranging-sums}.

\subsubsection{Evaluation of $\Theta_{v^{*}}$}

For $\Theta_{v^{*}}$, we have from \eqref{eq:AvEv} that $\A\bl_{v^{*}}=h_{v}^{-1}\d h_{v}+h_{v}^{-1}\M_{v}h_{v}\thinspace\d\phi_{v}$,
and similarly for the frame field $\E\bl_{v^{*}}=h_{v}^{-1}\left(\d\y_{v}+\left(\SS_{v}+\left[\M_{v},\y_{v}\right]\right)\d\phi_{v}\right)h_{v}$.
Therefore one finds 
\[
\delta\A\bl_{v^{*}}=h_{v}^{-1}\left(\d\De h_{v}+\left(\delta\M_{v}+\left[\M_{v},\De h_{v}\right]\right)\d\phi_{v}\right)h_{v}.
\]
Thus, after some simplification (using $\left[\M_{v},\SS_{v}\right]=0$),
\[
-\E\cdot\delta\A=\d\left(\d\y_{v}\cdot\De h_{v}-\left(\y_{v}\cdot\delta\M_{v}-\left(\SS_{v}+\left[\M_{v},\y_{v}\right]\right)\cdot\De h_{v}\right)\d\phi_{v}\right),
\]
where we choose as above the polarization $\d\y_{v}\cdot\d\De h_{v}=-\d\left(\d\y_{v}\cdot\De h_{v}\right)$
for the first term. Now we may integrate. Remembering that $\partial v^{*}=\partial_{0}v^{*}\cup\partial_{R}v^{*}$,
we get two contributions, one from the inner boundary $\partial_{0}v^{*}$
and one (with opposite orientation and thus a minus sign) from the
outer boundary $\partial_{R}v^{*}$: 
\[
\Theta_{v^{*}}=\Theta_{\partial_{R}v^{*}}-\Theta_{\partial_{0}v^{*}},
\]
where 
\[
\Theta_{\partial_{R}v^{*}}\equiv\int_{\partial_{R}v^{*}}\left(\d\y_{v}\cdot\De h_{v}-\left(\y_{v}\cdot\delta\M_{v}-\left(\SS_{v}+\left[\M_{v},\y_{v}\right]\right)\cdot\De h_{v}\right)\d\phi_{v}\right).
\]
As above, we will need some simplifications in order to integrate
$\Theta_{\partial_{R}v^{*}}$, which we will do in Sec. \ref{subsec:Rearranging-sums}.
The expression for $\Theta_{\partial_{0}v^{*}}$ is similar except
that the boundary condition at $r=0$ implies that $\left(h_{v},y_{v}\right)|_{r=0}$
are constant and equal to $\left(h_{v}\left(v\right),y_{v}\left(v\right)\right)$.
The integrand may then be trivially integrated, and we get 
\[
-\Theta_{\partial_{0}v^{*}}=\y_{v}\left(v\right)\cdot\delta\M_{v}-\left(\SS_{v}+\left[\M_{v},\y_{v}\left(v\right)\right]\right)\cdot\De\left(h_{v}\left(v\right)\right).
\]

\subsubsection{Summary}

The total symplectic potential now takes the form of a sum of three
contributions: 
\begin{equation}
\Theta=\sum_{c}\Theta_{c}+\sum_{v}\Theta_{\partial_{R}v^{*}}+\sum_{v}\Theta_{\partial_{0}v^{*}},\label{eq:Theta-before-symplification}
\end{equation}
where

\begin{align}
\Theta_{c} & =\int_{\partial\ct}\d\y_{c}\cdot\De h_{c},\label{eq:Theta_c}\\
\Theta_{\partial_{R}v^{*}} & =\int_{\partial_{R}v^{*}}\left(\d\y_{v}\cdot\De h_{v}-\left(\y_{v}\cdot\delta\M_{v}-\left(\SS_{v}+\left[\M_{v},\y_{v}\right]\right)\cdot\De h_{v}\right)\d\phi_{v}\right),\\
\Theta_{\partial_{0}v^{*}} & =\y_{v}\left(v\right)\cdot\delta\M_{v}-\left(\SS_{v}+\left[\M_{v},\y_{v}\left(v\right)\right]\right)\cdot\De\left(h_{v}\left(v\right)\right).
\end{align}

\subsection{\label{subsec:Rearranging-sums}Rearranging the Sums and Integrals}

From the previous section, we see that the total symplectic potential
can be written in terms of the variables $(\HH_{c},\HH_{v},\MMb_{v})$
purely as a sum of line integrals plus vertex contributions. In order
to simplify this expression, we can break the line integrals into
a sum of individual contributions along the boundaries.

\begin{figure}[!h]
\begin{centering}
\input{Figure-5-CellBoundary.tex} 
\par\end{centering}
\caption{\label{fig:CellBoundary}The blue square in the center is the node
$c^{*}$. It is dual to the cell $c$, outlined in black. In this
simple example, we have $N=3$ vertices $v_{1},v_{2},v_{3}$ along
the boundary $\partial c$, dual to 3 disks $v_{1}^{*},v_{2}^{*},v_{3}^{*}$.
Only the wedge $v_{i}^{*}\cap c$ is shown for each disk. After removing
the wedges from $c$ we obtain the truncated cell $\protect\ct$,
in dashed blue. The cell $c$ is adjacent to 3 cells $c_{i}$ (not
shown) dual to the 3 nodes $c_{i}^{*}$, in blue. The boundary $\partial\protect\ct$,
in dashed blue, thus consists of 3 arcs $\left(v_{i}c\right)$ and
3 truncated edges $\left[cc_{i}\right]$.}
\end{figure}
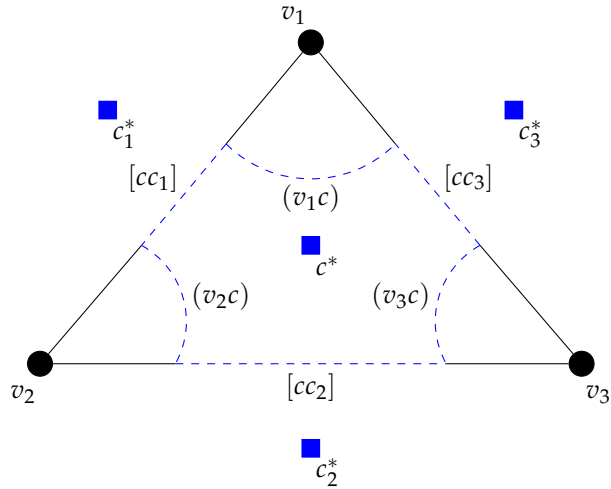

Let us recall the construction of Sec. \ref{subsec:Continuity-Conditions-Disks}.
Consider a cell $c$ with $N$ vertices $v_{1},\ldots,v_{N}$ along
its boundary. Each vertex $v_{i}$ is dual to a punctured disk $v_{i}^{*}$,
and the intersection of the (outer) boundary of that disk with the
cell $c$ is the arc 
\[
\left(v_{i}c\right)\equiv\partial_{R}v_{i}^{*}\cap c.
\]
When we remove the intersections of the (full) disks with the cell
$c$, we obtain the truncated cells 
\[
\ct\equiv c\backslash\bigcup_{i=1}^{N}D_{v_{i}}.
\]
Now, to the cell $c$ there are $N$ adjacent cells $c_{1},\ldots,c_{N}$,
truncated into $\ct_{1},\ldots,\ct_{N}$, and each such truncated
cell intersects $\ct$ at a \emph{truncated edge}, denoted with square
brackets: 
\[
\left[cc_{i}\right]\equiv\ct\cap\ct_{i}.
\]
We thus see that the boundary $\partial\ct$ of the truncated cell
may be decomposed into a union of truncated edges and arcs: 
\[
\partial\ct=\bigcup_{i=1}^{N}\left(\left[cc_{i}\right]\cup\left(v_{i}c\right)\right).
\]
This is illustrated in Fig. \ref{fig:CellBoundary}. Similarly, given
a punctured disk $v^{*}$ surrounded by $N$ cells $c_{1},\ldots,c_{N}$,
its (outer) boundary can be decomposed as a union of arcs: 
\[
\partial_{R}v^{*}=\bigcup_{i=1}^{N}\left(vc_{i}\right).
\]

Accordingly, we can now rearrange the first two sums in \eqref{eq:Theta-before-symplification},
decomposing the sums over the boundaries $\partial\ct$ and $\partial_{R}v^{*}$
into sums over individual truncated edges and arcs: 
\begin{equation}
\Theta=\sum_{\left[cc'\right]}\Theta_{cc'}+\sum_{\left(vc\right)}\Theta_{\left(vc\right)}+\sum_{v}\Theta_{\partial_{0}v^{*}}.\label{eq:sum-edges-arcs}
\end{equation}
The first sum is over all truncated edges $\left[cc'\right]$ for
all pairs of adjacent cells $c$ and $c'$, the second sum is over
all the arcs $\left(vc\right)$ for all pairs of adjacent vertices
$v$ and cells $c$, and the third sum is over all the vertices $v$.

To find the contributions from the edges and arcs, we assume that
$\Sigma$ is an oriented surface and choose the counterclockwise orientation
of the boundary of each cell. Each edge $\left[cc'\right]$ is counted
twice, once from the direction of $c$ as part of the integral over
$\partial\ct$ from $\Theta_{c}$ and once from the direction of $c'$
as part of the integral over $\partial\ct'$ from $\Theta_{c'}$,
with opposite orientation resulting in a relative minus sign: 
\begin{equation}
\Theta_{cc'}=\int_{\left[cc'\right]}\left(\d\y_{c}\cdot\De h_{c}-\d\y_{c'}\cdot\De h_{c'}\right).\label{eq:edget}
\end{equation}
Similarly, each arc $\left(vc\right)$ is counted twice, once from
the direction of $v$ as part of the integral over $\partial v^{*}$
from $\Theta_{\partial_{R}v^{*}}$ and once from the direction of
$c$ as part of the integral over $\partial\ct$ from $\Theta_{c}$,
again with opposite orientation: 
\begin{equation}
\Theta_{\left(vc\right)}=\int_{\left(vc\right)}\left(\d\y_{v}\cdot\De h_{v}-\left(\y_{v}\cdot\delta\M_{v}-\left(\SS_{v}+\left[\M_{v},\y_{v}\right]\right)\cdot\De h_{v}\right)\d\phi_{v}-\d\y_{c}\cdot\De h_{c}\right).\label{eq:vertext}
\end{equation}

\subsection{\label{subsec:Simplifying-the-Edge}Simplifying the Edge and Arc
Contributions}

\subsubsection{The Edge Contributions}

The calculation of the edge symplectic potential $\Theta_{cc'}$ is
similar to \cite{Dupuis:2017otn}, and will serve as a warm-up to
set the stage for the evaluation of the arc symplectic potential below.
One first needs to recall the continuity relations \eqref{eq:continuity-cc-split}:
\[
h_{c'}\left(x\right)=h_{c'c}h_{c}\left(x\right)\sp\y_{c'}\left(x\right)=h_{c'c}\left(\y_{c}\left(x\right)-\y_{c}^{c'}\right)h_{cc'}\sp x\in\left(cc'\right).
\]
Plugging this into \eqref{eq:edget}, we get 
\[
\Theta_{cc'}=\int_{\left[cc'\right]}\left(\d\y_{c}\cdot\De h_{c}-h_{c'c}\d\y_{c}h_{cc'}\cdot\De\left(h_{c'c}h_{c}\right)\right),
\]
where we used the fact that $h_{c'c}$ and $\y_{c}^{c'}$ are constant
(do not depend on $x$) and thus annihilated by $\d$. Next, from
the useful identity \eqref{eq:Delta-useful-identity} we have 
\[
\De\left(h_{c'c}h_{c}\right)=h_{c'c}\left(\De h_{c}-\De h_{c}^{c'}\right)h_{cc'},
\]
where we have used the notation $\De h_{c}^{c'}\equiv\delta h_{cc'}h_{c'c}$,
introduced in \eqref{eq:proper-subscript}, which emphasizes that
it is an algebra element based at $c$. This allows us to cancel $h_{cc'}$
using the cyclicity of the trace and then cancel the first term, simplifying
this expression to 
\[
\Theta_{cc'}=\De h_{c}^{c'}\cdot\int_{\left[cc'\right]}\d\y_{c},
\]
where we took $\De h_{c}^{c'}$ out of the integral since it is constant.
We will perform the final integration in Sec. \ref{subsec:The-Holonomy-Flux-Algebra}.

\subsubsection{\label{subsec:The-Arc-Contributions}The Arc Contributions}

We can now evaluate the arc contribution \eqref{eq:vertext}. One
first recalls the continuity relations \eqref{eq:continuity-h_cv}
and \eqref{eq:continuity-y_cv}:

\begin{align*}
h_{c}\left(x\right) & =h_{cv}\e^{\M_{v}\phi_{v}\left(x\right)}h_{v}\left(x\right)\sp x\in\left(vc\right),\\
\y_{c}\left(x\right) & =h_{cv}\left(\e^{\M_{v}\phi_{v}\left(x\right)}\left(\y_{v}\left(x\right)+\SS_{v}\phi_{v}\left(x\right)\right)\e^{-\M_{v}\phi_{v}\left(x\right)}-\y_{v}^{c}\right)h_{vc}\sp x\in\left(vc\right).
\end{align*}
From these relations, and using the fact that $h_{vc}$ and $\y_{v}^{c}$
are constant, we find that 
\[
\De h_{c}=h_{cv}\left(\e^{\M_{v}\phi_{v}}\left(\De h_{v}+\delta\M_{v}\phi_{v}\right)\e^{-\M_{v}\phi_{v}}-\De h_{v}^{c}\right)h_{cv}^{-1},
\]
\[
\d\y_{c}=h_{cv}\e^{\M_{v}\phi_{v}}\left(\d\y_{v}+\left(\SS_{v}+\left[\M_{v},\y_{v}\right]\right)\d\phi_{v}\right)\e^{-\M_{v}\phi_{v}}h_{vc},
\]
where we have denoted again $\De h_{v}^{c}\equiv\delta h_{vc}h_{cv}$
since the variational differential is based at $v$. Recall that the
arc symplectic potential \eqref{eq:vertext} was 
\[
\Theta_{\left(vc\right)}=\int_{\left(vc\right)}\left(\d\y_{v}\cdot\De h_{v}-\left(\y_{v}\cdot\delta\M_{v}-\left(\SS_{v}+\left[\M_{v},\y_{v}\right]\right)\cdot\De h_{v}\right)\d\phi_{v}-\d\y_{c}\cdot\De h_{c}\right).
\]
Replacing the expressions for $\d\y_{c}$ and $\De h_{c}$ in the
last term, we get after simplification 
\begin{equation}
\Theta_{\left(vc\right)}=\De h_{v}^{c}\cdot\int_{\left(vc\right)}\d\left(\e^{\M_{v}\phi_{v}}\y_{v}\e^{-\M_{v}\phi_{v}}+\SS_{v}\phi_{v}\right)-\delta\M_{v}\cdot\int_{\left(vc\right)}\d\left(\y_{v}\phi_{v}+\hf\SS_{v}\phi_{v}^{2}\right).\label{eq:arc-term}
\end{equation}
We see that the first term depends on $c$, while the second one does
not. Thus we can perform the sum over arcs around each disk in the
second term (since we had $\sum_{\left(vc\right)}\Theta_{\left(vc\right)}$
in \eqref{eq:sum-edges-arcs}) and turn it instead into a sum over
vertices with an integral along the outer boundary $\partial_{R}v^{*}$
for each vertex. In other words, we define: 
\[
\sum_{\left(vc\right)}\Theta_{\left(vc\right)}\equiv\sum_{\left(vc\right)}\Theta_{vc}+\sum_{v}\Theta_{\left(v\right)},
\]
where

\begin{eqnarray}
\Theta_{vc} & \equiv & \De h_{v}^{c}\cdot\int_{\left(vc\right)}\d\left(\e^{\M_{v}\phi_{v}}\y_{v}\e^{-\M_{v}\phi_{v}}+\SS_{v}\phi_{v}\right),\\
\Theta_{\left(v\right)} & \equiv & -\delta\M_{v}\cdot\int_{\partial_{R}v^{*}}\d\left(\y_{v}\phi_{v}+\hf\SS_{v}\phi_{v}^{2}\right).
\end{eqnarray}

\subsubsection{\label{subsec:The-Vertex-Contribution}The Vertex Contribution}

In fact, we may easily integrate $\Theta_{\left(v\right)}$, bearing
in mind the definition of the punctured disk in Sec. \ref{punc}:
\[
\Theta_{\left(v\right)}=-\delta\M_{v}\cdot\left(\y_{v}\left(v_{0}\right)+\alpha_{v}\SS_{v}\right),
\]
where the reader will recall that $v_{0}$ was the intersection of
the cut with the boundary circle at $r_{v}=R$ and $\phi_{v}=\alpha_{v}-\tfrac{1}{2}$
defines the angle assign to the cut. We may now add this contribution
to the term $\Theta_{\partial_{0}v^{*}}$ from \eqref{eq:Theta_c},
obtaining 
\[
\Theta_{v}\equiv\Theta_{\partial_{0}v^{*}}+\Theta_{\left(v\right)}=\left(\y_{v}\left(v\right)-\y_{v}\left(v_{0}\right)-\alpha_{v}\SS_{v}\right)\cdot\delta\M_{v}-\left(\SS_{v}+\left[\M_{v},\y_{v}\left(v\right)\right]\right)\cdot\De\left(h_{v}\left(v\right)\right).
\]
We can finally simplify this expression further. Let us first decompose
$\y_{v}\left(v_{0}\right)$ into a component parallel to $\M_{v}$
(with $\M_{v}^{2}\equiv\M_{v}\cdot\M_{v}$) and another orthogonal
to it: 
\begin{equation}
\y_{v}\left(v_{0}\right)\equiv\y_{v}^{\parallel}\left(v_{0}\right)+\y_{v}^{\perp}\left(v_{0}\right),
\end{equation}
where
\[
\y_{v}^{\parallel}\left(v_{0}\right)\equiv\left(\frac{\y_{v}\left(v_{0}\right)\cdot\M_{v}}{\M_{v}^{2}}\right)\M_{v}.
\]
The term $\y_{v}\left(v_{0}\right)\cdot\delta\M_{v}$ only depends
on the component $\y_{v}^{\parallel}\left(v_{0}\right)$. On the other
hand, the term $\left[\M_{v},\y_{v}\left(v\right)\right]$ is left
invariant if we translate $\y_{v}\left(v\right)$ by any component
parallel to $\M_{v}$. Hence, we can rewrite the vertex potential
as 
\begin{equation}
\Theta_{v}=\X_{v}\cdot\delta\M_{v}-\left(\SS_{v}+\left[\M_{v},\X_{v}\right]\right)\cdot\De\left(h_{v}\left(v\right)\right),\label{thetavr}
\end{equation}
where we have introduced the particle relative position
\begin{equation}
\X_{v}\equiv\y_{v}\left(v\right)-\y_{v}^{\parallel}\left(v_{0}\right)-\alpha_{v}\SS_{v}.\label{eq:defpos}
\end{equation}
The term proportional to $\alpha_{v}$ can be eliminated by a symmetry
transformation of the type \eqref{eq:beta-symmetry-decomp}.

The expression \eqref{thetavr} is the usual expression for the symplectic
potential of a relativistic particle with mass $\M_{v}$ and spin
$\SS_{v}$ \cite{Kirillov,Rempel:2015foa}. In particular, if we introduce
the particle momentum $\p_{v}$, angular momentum\footnote{The definition of $\j_{v}$ given here agrees with the definition
\eqref{eq:momenta} given earlier since $\X_{v}$ differs from $\y_{v}(v)$
only by a translation in the Cartan, which commutes with $\M_{v}$.} $\j_{v}$ and position $\q_{v}$,
\[
\p_{v}\equiv h_{v}^{-1}\left(v\right)\M_{v}h_{v}\left(v\right),\qquad\j_{v}\equiv h_{v}^{-1}\left(v\right)\left(\SS_{v}+\left[\M_{v},\X_{v}\right]\right)h_{v}\left(v\right),\qquad\q_{v}\equiv h_{v}^{-1}\left(v\right)\X_{v}h_{v}\left(v\right),
\]
we can show that the following commutation relations are satisfied\footnote{To clarify the notation, the subscript $v$ denotes the vertex as
usual, while $i$, $j$, $k$ are the Lie algebra indices.}:

\[
\{p_{vi},q_{v}^{j}\}=\delta_{i}^{j},\qquad\{j_{v}^{i},q_{v}^{j}\}=C^{ij}{}_{k}q_{v}^{k},\qquad\{j_{v}^{i},p_{vj}\}=-C^{ik}{}_{j}p_{vk},
\]
where $q_{v}^{i},p_{vi},j_{v}^{i}$ are the components of $\q_{v},\p_{v},\j_{v}$
with respect to the basis $\ta_{i}$ of $\mfg$ or the dual basis
$\ta^{i}$ of $\mfg^{*}$ (according to the index placement), and
$C^{ij}{}_{k}$ are the structure constants such that $\left[\ta^{i},\ta^{j}\right]=C^{ij}{}_{k}\ta^{k}$.
This shows that, as expected, the momentum $\p_{v}$ is the generator
of translations, the angular momentum $\j_{v}$ is the generator of
rotations, and $\q_{v}$ is the particle's position.

It can be useful to express the effective particle potential in terms
of position and momentum: 
\[
\Theta_{v}=\q_{v}\cdot\delta\p_{v}-\SS_{v}\cdot\De\left(h_{v}\left(v\right)\right).
\]
More properties of the vertex potential are presented in Appendix
\ref{sec:The-Relativistic-Particle}.

\subsection{Final Integration}

In the previous section we have established that the total symplectic
potential is given by
\[
\Theta=\sum_{\left[cc'\right]}\Theta_{cc'}-\sum_{\left(vc\right)}\Theta_{vc}+\sum_{v}\Theta_{v},
\]
where $\Theta_{cc'}$ and $\Theta_{vc}$ are simple line integrals
\begin{equation}
\Theta_{cc'}=\De h_{c}^{c'}\cdot\int_{\left[cc'\right]}\d\y_{c}\sp\Theta_{vc}=\De h_{v}^{c}\cdot\int_{\left(vc\right)}\d\left(\e^{\M_{v}\phi_{v}}\y_{v}\e^{-\M_{v}\phi_{v}}+\SS_{v}\phi_{v}\right),\label{eq:Theta_cc-vc}
\end{equation}
while $\Theta_{v}$ is the relativistic particle potential \eqref{thetavr}.
We are left with the integration and study of the edge and arc potentials.

\subsubsection{Source and Target Points}

\begin{figure}[!h]
\begin{centering}
\input{Figure-6-IntersectionPoints.tex} 
\par\end{centering}
\caption{\label{fig:IntersectionPoints}The intersection points (red circles)
of truncated edges and arcs along the oriented boundary $\partial\protect\ct$
(blue arrows).}
\end{figure}
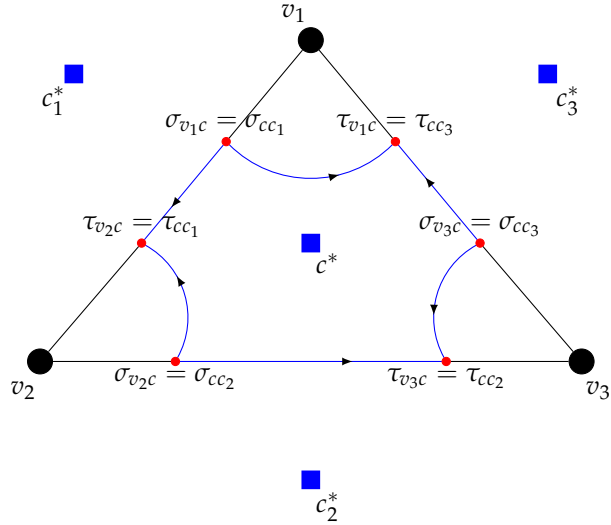

We have obtained, in \eqref{eq:Theta_cc-vc}, two integrals of total
differentials over the truncated edges $\left[cc'\right]$ and the
arcs $\left(vc\right)$. These integrals are trivial; all we need
is to label the source and target points of each edge and arc. One
needs to recall that the truncated edges are oriented form the point
of view of $c$, while the arcs are oriented from the point of view
of $v$ so they have opposite relative orientations as shown in Fig.
\ref{fig:IntersectionPoints}.

These corner points are located along the edges of $\Gamma$ and are
determined by the radius $R$ of the disks. We will label the source
and target points of the edge $\left[cc'\right]$ as $\sigma_{cc'}$
and $\tau_{cc'}$ respectively, and the source and target points of
the arc $\left(vc\right)$ as $\sigma_{vc}$ and $\tau_{vc}$ respectively,
where $\sigma$ stands for ``source'' and $\tau$ for ``target''.
In other words,
\[
\left[cc'\right]\equiv\left[\sigma_{cc'}\tau_{cc'}\right]\sp\left(vc\right)\equiv\left[\sigma_{vc}\tau_{vc}\right].
\]
Note that these labels are not unique, since the source (target) of
an edge is also the source (target) of an arc. More precisely, let
us consider a cell $c=\left[v_{1},\cdots,v_{N}\right]$, where $v_{i}$
denote the boundary vertices. This cell is bounded by $N$ other cells
$c_{1},\cdots,c_{N}$, which are such that $c\cap c_{i}=\left(v_{i}v_{i+1}\right)$.
We then have that 
\[
\sigma_{v_{i}c}=\sigma_{cc_{i}},\qquad\tau_{v_{i+1}c}=\tau_{cc_{i}}.
\]
This labeling is illustrated in Fig. \ref{fig:IntersectionPoints}.

\subsubsection{\label{subsec:The-Holonomy-Flux-Algebra}The Holonomy-Flux Algebra}

Using these labels, we can perform the integrations in \eqref{eq:Theta_cc-vc}.
We define \emph{fluxes} associated to the edges $\left[cc'\right]$
and arcs $\left(vc\right)$ bounding the cell $c$ in terms of the
corner variables:
\begin{align}
\X_{c}^{c'} & \equiv\int_{\left[cc'\right]}\d\y_{c}=\y_{c}\left(\tau_{cc'}\right)-\y_{c}\left(\sigma_{cc'}\right),\label{eq:flux-cc}\\
\X_{v}^{c} & \equiv\int_{\left(vc\right)}\d\left(\e^{\M_{v}\phi_{v}}\y_{v}\e^{-\M_{v}\phi_{v}}+\SS_{v}\phi_{v}\right)\label{eq:flux-vc}\\
 & =\left(\e^{\M_{v}\phi_{v}}\y_{v}\e^{-\M_{v}\phi_{v}}+\SS_{v}\phi_{v}\right)(\tau_{vc})-\left(\e^{\M_{v}\phi_{v}}\y_{v}\e^{-\M_{v}\phi_{v}}+\SS_{v}\phi_{v}\right)(\sigma_{vc}).
\end{align}
This allows us to write the edges potential simply as 
\begin{equation}
{\Theta_{cc'}\equiv\De h_{c}^{c'}\cdot\X_{c}^{c'}\sp\Theta_{vc}\equiv\De h_{v}^{c}\cdot\X_{v}^{c}.}\label{edget}
\end{equation}

We see that these two terms correspond to the familiar \emph{holonomy-flux
phase space} for each edge and arc. $\Theta_{cc'}$ is associated
to each link $\left(cc'\right)^{*}$ in the spin network graph $\Gamma^{*}$,
while $\Theta_{vc}$ is associated to each segment $\left(vc\right)^{*}$
connecting a vertex with a node. The holonomy-flux phase space is
the cotangent bundle $T^{*}G$, and it is the phase space of (classical)
spin networks in loop quantum gravity in the case $G=\SUT$.

One can prove a relation between a flux and its ``inverse''. First,
note that $\sigma_{cc'}=\tau_{c'c}$. Thus, we have from the continuity
condition \eqref{eq:continuity-cc-split}: 
\begin{align*}
\X_{c'}^{c} & =\y_{c'}\left(\tau_{c'c}\right)-\y_{c'}\left(\sigma_{c'c}\right)=-\left(\y_{c'}\left(\tau_{cc'}\right)-\y_{c'}\left(\sigma_{cc'}\right)\right)\\
 & =-h_{c'c}\left(\y_{c}\left(\tau_{cc'}\right)-\y_{c}\left(\sigma_{cc'}\right)\right)h_{cc'}=-h_{c'c}\X_{c}^{c'}h_{cc'}.
\end{align*}
Similarly, from \eqref{eq:continuity-y_cv} we find, that the arc
flux as viewed from the point of view of the cell $c$ is 
\begin{equation}
\X_{c}^{v}\equiv-\int_{\left(vc\right)}\d\y_{c}=-\left(\y_{c}\left(\tau_{vc}\right)-\y_{c}\left(\sigma_{vc}\right)\right)=-h_{cv}\X_{v}^{c}h_{vc},\label{eq:flux-cv}
\end{equation}
where the minus sign comes from the opposite orientation of the arc
when viewed from $c$ instead of $v^{*}$.

In conclusion, the fluxes satisfy the relations 
\begin{equation}
\X_{c'}^{c}=-h_{c'c}\X_{c}^{c'}h_{cc'}\sp\X_{c}^{v}=-h_{cv}\X_{v}^{c}h_{vc}.\label{eq:X-inverse}
\end{equation}
Note that, if we view $\y_{c}\left(\tau_{cc'}\right)$ and $\y_{c}\left(\sigma_{cc'}\right)$
as the relative position of the corner $\tau_{cc'}$ from the node
$c^{*}$ to the points $\tau_{cc'}$ and $\sigma_{cc'}$ respectively,
then the difference $\X_{c}^{c'}=\y_{c}\left(\tau_{cc'}\right)-\y_{c}\left(\sigma_{cc'}\right)$
is a translation vector from $\sigma_{cc'}$ to $\tau_{cc'}$. In
other words, $\X_{c}^{c'}$ is simply a (translational) holonomy along
the truncated edge $\left[cc'\right]$ dual to the link $\left(cc'\right)^{*}$.
Similarly, $\X_{c}^{v}$ is a vector from $\sigma_{vc}$ to $\tau_{vc}$,
along the arc $\left(vc\right)$.

Let us label the holonomies $h_{\ell}\equiv h_{cc'}$ and fluxes $\X_{\ell}\equiv\X_{c}^{c'}$,
where $\ell\equiv\left(cc'\right)^{*}$ is a link in the spin network
graph $\Gamma^{*}$, dual to the edge $\left(cc'\right)\in\Gamma$.
Let us also decompose the flux into components, $\X_{\ell}\equiv X_{\ell}^{i}\ta_{i}$.
Then, from the first term in $\Theta$, one can derive the well-known
\emph{holonomy-flux Poisson algebra}:

\[
\{h_{\ell},h_{\ell'}\}=0\sp\{X_{\ell}^{i},X_{\ell'}^{j}\}=\delta_{\ell\ell'}\epsilon_{k}^{ij}X_{\ell}^{k}\sp\{X_{\ell}^{i},h_{\ell'}\}=\delta_{\ell\ell'}\ta^{i}h_{\ell}.
\]
This concludes the construction of the total symplectic potential.
It can be written solely in terms of discrete data without any integrals,
with contributions from truncated edges, arcs, and vertices, as follows:
\begin{equation}
\boxed{\Theta=\sum_{\left[cc'\right]}\De h_{c}^{c'}\cdot\X_{c}^{c'}-\sum_{\left(vc\right)}\De h_{c}^{v}\cdot\X_{c}^{v}+\sum_{v}\left(\X_{v}\cdot\delta\M_{v}-\left(\SS_{v}+\left[\M_{v},\X_{v}\right]\right)\cdot\De\left(h_{v}\left(v\right)\right)\right).}
\end{equation}

\section{\label{sec:The-Gauss-and}The Gauss and Curvature Constraints}

In the continuum theory of 2+1 gravity, we have the curvature constraint
$\F=0$ and the torsion constraint $\T=0$. These constraints are
modified when we add curvature and torsion defects; as we have seen
in \eqref{eq:F-T-delta}, at least naively, a delta function is added
to the right-hand side of these constraints: $\F\bl_{v^{*}}=\p_{v}\thinspace\delta\left(v\right)$
and $\T\bl_{v^{*}}=\j_{v}\,\delta\left(v\right)$. We will now show
how these constraints are recovered in our formalism to obtain the
discrete Gauss (torsion) constraint and the discrete curvature constraint.

\subsection{\label{subsec:The-Gauss-Constraint}The Gauss Constraint}

\subsubsection{On the Cells}

Recall that on the truncated cell $\ct$ we have 
\[
\A\bl_{\ct}=h_{c}^{-1}\d h_{c}\sp\E\bl_{\ct}=h_{c}^{-1}\d\y_{c}h_{c}\sp\T\bl_{\ct}=\d_{\A}\E\bl_{\ct}=0.
\]
Using the identity 
\[
\d\left(h_{c}\E h_{c}^{-1}\right)=h_{c}\left(\d_{\A}\E\right)h_{c}^{-1},
\]
we can define the quantity $\G_{c}$, which represents the Gauss law
integrated over the truncated cell $\ct$: 
\[
\G_{c}\equiv\int_{\ct}h_{c}\left(\d_{\A}\E\right)h_{c}^{-1}=\int_{\ct}\d\left(h_{c}\E h_{c}^{-1}\right)=\int_{\partial\ct}h_{c}\E h_{c}^{-1}=\int_{\partial\ct}\d\y_{c}.
\]
The Gauss law $\d_{\A}\E=0$ translates into the condition $\G_{c}=0$.

As illustrated in Fig. \ref{fig:CellBoundary}, the boundary $\partial\ct$
consists of truncated edges $\left[cc'\right]$ and arcs $\left(vc\right)$.
Thus 
\[
\G_{c}=\sum_{c'\ni c}\int_{\left[cc'\right]}\d\y_{c}+\sum_{v\ni c}\int_{\left(vc\right)}\d\y_{c},
\]
where $c'\ni c$ means ``all cells $c'$ adjacent to $c$'' and
$v\ni c$ means ``all vertices $v$ adjacent to $c$''. Using the
fluxes defined in \eqref{eq:flux-cc} and \eqref{eq:flux-cv}, we
get 
\[
\G_{c}=\sum_{c'\ni c}\X_{c}^{c'}-\sum_{v\ni c}\X_{c}^{v}=0,
\]
where, as explained before, the minus sign in the second sum comes
from the fact that we are looking at the arcs from the cell $c$,
not from the vertex $v$, and thus the orientation of the integral
is opposite.

In the limit where we shrink the radius $R$ of the punctured disks
to zero the points $\y_{c}\left(\tau_{vc}\right)$ and $\y_{c}\left(\tau_{vc}\right)$
are identified, the arc flux $\X_{c}^{v}$ vanishes, and we recover
the usual loop gravity Gauss constraint. It is well known, however,
that under coarse-graining the naive Gauss constraint is not preserved
\cite{Livine:2013gna,Livine:2017xww}. In our context, this can be
interpreted as opening a small disk around the vertices, which carries
additional fluxes.

\subsubsection{On the Disks}

Similarly, recall that on the \textit{punctured} disks we have
\[
\A\bl_{v^{*}}=h_{v}^{-1}\d h_{v}+(h_{v}^{-1}\M_{v}h_{v})\d\phi_{v}\sp\E\bl_{v^{*}}=h_{v}^{-1}\left(\d\y_{v}+\left(\SS_{v}+\left[\M_{v},\y_{v}\right]\right)\d\phi_{v}\right)h_{v},
\]
and $\d_{\A}\E\bl_{v^{*}}=0$. It will be convenient in this section
to introduce non-periodic variables 
\[
u_{v}\equiv\e^{\M_{v}\phi_{v}}h_{v}\sp\w_{v}\equiv\e^{\M_{v}\phi_{v}}\left(\y_{v}+\SS_{v}\phi_{v}\right)\e^{-\M_{v}\phi_{v}},
\]
in terms of which the connection and frame field can be simply expressed
as $\A\bl_{v^{*}}=u_{v}^{-1}\d u_{v}$ and $\E\bl_{v^{*}}=u_{v}^{-1}\d\w_{v}u_{v}.$
Using the identity 
\[
\d\left(u_{v}\E u_{v}^{-1}\right)=u_{v}\left(\d_{\A}\E\right)u_{v}^{-1},
\]
we can evaluate the Gauss constraint $\G_{v}$ inside the punctured
disk $v^{*}$: 
\[
\G_{v}\equiv\int_{v^{*}}u_{v}\left(\d_{\A}\E\right)u_{v}^{-1}=\int_{v^{*}}\d\left(u_{v}\E u_{v}^{-1}\right)=\int_{\partial v^{*}}u_{v}\E u_{v}^{-1}=\int_{\partial v^{*}}\d\w_{v}.
\]
This splits into contributions from the inner and outer boundaries,
with opposite signs, and (since we are now using non-periodic variables)
from the cut: 
\[
\G_{v}=\int_{\partial_{0}v^{*}}\d\w_{v}+\int_{\partial_{R}v^{*}}\d\w_{v}+\int_{C_{v}}\d\w_{v}.
\]
On the inner boundary $\partial_{0}v^{*}$, we use the fact that $\y_{v}$
takes the constant value $\y_{v}\left(v\right)$ to obtain
\begin{align*}
\int_{\partial_{0}v^{*}}\d\w_{v} & =\e^{\M_{v}\phi_{v}}\left(\y_{v}\left(v\right)+\SS_{v}\phi_{v}\right)\e^{-\M_{v}\phi_{v}}\bgl_{\phi_{v}=\alpha_{v}-\hf}^{\alpha_{v}+\hf}\\
 & =\SS_{v}+\e^{\M_{v}\left(\alpha_{v}-\hf\right)}\left(\e^{\M_{v}}\y_{v}\left(v\right)\e^{-\M_{v}}-\y_{v}\left(v\right)\right)\e^{-\M_{v}\left(\alpha_{v}-\hf\right)}.
\end{align*}
On the outer boundary $\partial_{R}v^{*}$, we split the integral
into separate integrals over each arc $\left(vc\right)=\left(\sigma_{vc}\tau_{vc}\right)$
around $v^{*}$ and use the definition of the flux \eqref{eq:flux-vc}:
\[
\int_{\partial_{R}v^{*}}\d\w_{v}=\sum_{c\in v}\int_{\left(vc\right)}\d\w_{v}=\sum_{c\in v}\X_{v}^{c}.
\]
On the cut $C_{v}$, we have contributions from both sides, one at
$\phi_{v}=\alpha_{v}-\hf$ and another at $\phi_{v}=\alpha_{v}+\hf$
with opposite orientation. Since $\d\phi_{v}=0$ on the cut, we have:
\[
\d\w_{v}\bl_{C_{v}}=\e^{\M_{v}\phi_{v}}\d\y_{v}\e^{-\M_{v}\phi_{v}},
\]
and thus 
\begin{align*}
\int_{C_{v}}\d\w_{v} & =\int_{r=0}^{R}\left(\e^{\M_{v}\phi_{v}}\d\y_{v}\e^{-\M_{v}\phi_{v}}\bgl_{\phi_{v}=\alpha_{v}-\hf}^{\phi_{v}=\alpha_{v}+\hf}\right)\\
 & =\e^{\M_{v}\left(\alpha_{v}-\hf\right)}\left(\e^{\M_{v}}\left(\y_{v}\left(v_{0}\right)-\y_{v}\left(v\right)\right)\e^{-\M_{v}}-\left(\y_{v}\left(v_{0}\right)-\y_{v}\left(v\right)\right)\right)\e^{-\M_{v}\left(\alpha_{v}-\hf\right)},
\end{align*}
since $\y_{v}$ has the value $\y_{v}\left(v_{0}\right)$ at $r=R$
and $\y_{v}\left(v\right)$ at $r=0$ on the cut.

Adding up the integrals, we find that the Gauss constraint on the
disk is 
\[
\G_{v}=\SS_{v}+\e^{\M_{v}\left(\alpha_{v}-\hf\right)}\left(\e^{\M_{v}}\y_{v}\left(v_{0}\right)\e^{-\M_{v}}-\y_{v}\left(v_{0}\right)\right)\e^{-\M_{v}\left(\alpha_{v}-\hf\right)}-\sum_{c\in v}\X_{v}^{c}=0.
\]
The validity of this equation can now be checked from the definition
of the fluxes. By performing the sum explicitly and using the fact
that $\tau_{vc_{i}}=\sigma_{vc_{i+1}}$ where $c_{i}$, $i\in\left\{ 1,\ldots,N\right\} $
are the cells around the disk $v^{*}$ and 
\[
\phi_{v}\left(\sigma_{vc_{N+1}}\right)\equiv\phi_{v}\left(\sigma_{vc_{1}}\right)+1,
\]
we see that this constraint is satisfied identically. Indeed, using
the fact that $\y_{v}$ is periodic, choosing without loss of generality
$\phi\left(\sigma_{vc_{1}}\right)\equiv\alpha_{v}-1/2$ (that is,
the first arc start at the cut), and recalling that $\y_{v}=\y_{v}\left(v_{0}\right)$
at the cut, we obtain 
\begin{equation}
\sum_{c\in v}\X_{v}^{c}=\SS_{v}+\e^{\M_{v}\left(\alpha_{v}-\hf\right)}\left(\e^{\M_{v}}\y_{v}\left(v_{0}\right)\e^{-\M_{v}}-\y_{v}\left(v_{0}\right)\right)\e^{-\M_{v}\left(\alpha_{v}-\hf\right)}.\label{Gaussv2}
\end{equation}
Remember that we have the decomposition $\y_{v}(v_{0})=\y_{v}^{\parallel}\left(v_{0}\right)+\y_{v}^{\perp}\left(v_{0}\right)$.
When $\M_{v}\ne0$, the previous equation defines the value of $\y_{v}^{\perp}\left(v_{0}\right)$
in terms of the sum of fluxes $\sum_{c\in v}\X_{v}^{c}$.

\subsection{The Curvature Constraint}

In the previous section we have expressed the Gauss constraints satisfied
by the fluxes, which follow from the definition of the fluxes in terms
of the translational holonomy variables. Here we do the same for the
curvature constraint. We want to find the relations between the discrete
holonomies $h_{vc},h_{cc'}$ and the mass parameters $\M_{v}$ which
express that the curvature is concentrated on the vertices.

The connection on the truncated cell $\ct$ is 
\[
\A\bl_{\ct^{*}}=h_{c}^{-1}\d h_{c}\soosp\F\bl_{\ct^{*}}=0.
\]
Taking the rotational part of \eqref{eq:H-from-A}, we see that the
associated holonomies $h_{c}$ from a point $x$ to another point
$y$ inside $c$ are given by $\pexp\int_{x}^{y}\A=h_{c}^{-1}\left(x\right)h_{c}\left(y\right).$
One can use the relations \eqref{eq:continuity-cc-split} to evaluate
the holonomy along a path from $x\in c$ to $y\in c'$, where $c$
and $c'$ are adjacent cells: 
\begin{equation}
\pexp\int_{x}^{y}\A=h_{c}^{-1}\left(x\right)h_{cc'}h_{c'}\left(y\right).\label{defhol2}
\end{equation}

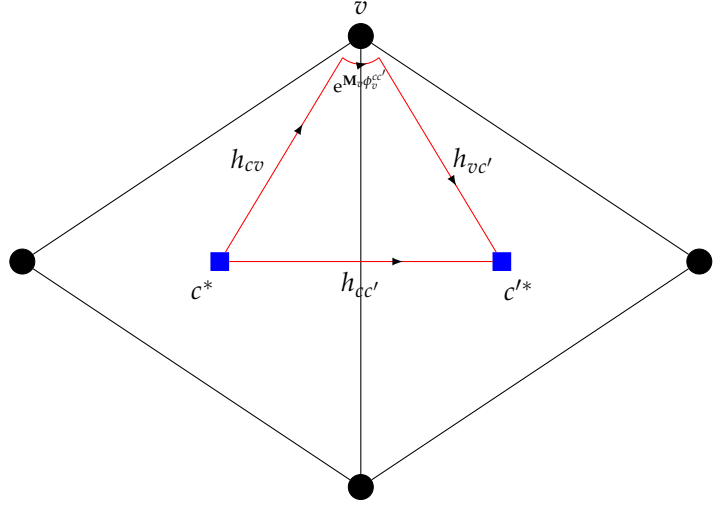
\begin{figure}[!h]
\begin{centering}
\input{Figure-7-Curvature.tex} 
\par\end{centering}
\caption{\label{fig:Curvature}The holonomy from $c^{*}$ to $c^{\prime*}$
going either directly or through the vertex $v$.}
\end{figure}

Similarly, the connection on the \textit{punctured} disk is 
\[
\A\bl_{v^{*}}=\left(\e^{\M_{v}\phi_{v}}h_{v}\right)^{-1}\d\left(\e^{\M_{v}\phi_{v}}h_{v}\right),
\]
and the associated holonomy from a point $x\in v^{*}$ to a point
$y\in v^{*}$ is given by 
\begin{equation}
\pexp\int_{x}^{y}\A=h_{v}^{-1}\left(x\right)\e^{\M_{v}\left(\phi_{v}\left(y\right)-\phi_{v}\left(x\right)\right)}h_{v}\left(y\right),\label{defhol3}
\end{equation}
while the holonomy between a point $x\in v^{*}$ and a point $y\in c$
in an adjacent cell $c$ is 
\begin{equation}
\pexp\int_{x}^{y}\A=h_{v}^{-1}\left(x\right)\e^{-\M_{v}\phi_{v}\left(x\right)}h_{vc}h_{c}\left(y\right).\label{defhol5}
\end{equation}
Using this, we can evaluate in the holonomy of the curve shown in
Fig. \ref{fig:Curvature}. This curve goes from a point $x\in c$
to a point $y\in c'$, where both cells $c,c'$ are around $v$. By
the flatness condition we can evaluate the holonomy from $x$ to $y$
in two ways, and we get 
\[
\pexp\int_{x}^{y}\A=h_{v}^{-1}\left(x\right)\e^{\M_{v}\phi_{v}^{cc'}}h_{v}\left(y\right)=h_{v}^{-1}\left(x\right)h_{vc}h_{cc'}h_{c'v}h_{v}\left(y\right),
\]
where we denoted $\phi_{v}^{cc'}\equiv\phi_{v}\left(y\right)-\phi_{v}\left(x\right)$.
We conclude that the flatness of the connection outside the cells
implies the following relationship among the discrete holonomies:
\begin{equation}
h_{cc'}=h_{cv}\e^{\M_{v}\phi_{v}^{cc'}}h_{vc'}.\label{relh}
\end{equation}
In particular, let the vertex $v$ be surrounded by $N$ cells $c_{1},\ldots,c_{N}$,
and take $c_{1}\equiv c_{N+1}\equiv c$. If we form a loop of discrete
holonomies going from each cell to the next, we find
\[
h_{cc_{2}}\cdots h_{c_{N}c}=h_{cv}\e^{\M_{v}}h_{vc},
\]
since $\phi_{v}^{cc_{2}}+\cdots+\phi_{v}^{c_{n}c}=1$. By moving all
of the terms to the left-hand side, we obtain the curvature constraint:
\[
F_{v}\equiv h_{vc}h_{cc_{2}}\cdots h_{c_{N}c}h_{cv}\e^{-\M_{v}}=1.
\]

\subsection{Generators of Symmetries}

In conclusion, we have obtained two Gauss constraints on the cells
$c$ and disks $v^{*}$,
\begin{equation}
\G_{c}\equiv\sum_{c'\ni c}\X_{c}^{c'}-\sum_{v\ni c}\X_{c}^{v}=0,\label{eq:Gauss-con}
\end{equation}
\[
\G_{v}\equiv\SS_{v}+\e^{\M_{v}\left(\alpha_{v}-\hf\right)}\left(\e^{\M_{v}}\y_{v}\left(v_{0}\right)\e^{-\M_{v}}-\y_{v}\left(v_{0}\right)\right)\e^{-\M_{v}\left(\alpha_{v}-\hf\right)}-\sum_{c\in v}\X_{v}^{c}=0,
\]
and a curvature constraint at the vertex $v$,
\begin{equation}
F_{v}\equiv h_{vc}h_{cc_{2}}\cdots h_{c_{N}c}h_{cv}\e^{-\M_{v}}=1.\label{eq:curv-con}
\end{equation}
Note that all of these constraints are satisfied identically in our
construction, as shown above. We will now see that the Gauss constraint
generates the rotational part of the left translation symmetry transformation
given by \eqref{eq:symmetry-c} and \eqref{eq:symmetry-v}. In order
to do so, we look for transformations $\left(\delta_{\be_{c}},\delta_{\be_{v}},\delta_{\x_{v}}\right)$
such that 
\[
I_{\delta_{\be_{c}}}\Omega=-\be_{c}\cdot\delta\G_{c},\qquad I_{\delta_{\be_{v}}}\Omega=-\be_{v}\cdot\delta\G_{v},\qquad I_{\delta_{\x_{v}}}\Omega=-\x_{v}\cdot\De F_{c}.
\]
We will need the explicit expression for the total symplectic form
$\Omega\equiv\delta\Theta$: 
\begin{equation}
\Omega=\sum_{\left(cc'\right)}\Omega_{cc'}-\sum_{\left(vc\right)}\Omega_{vc}+\sum_{v}\Omega_{v},\label{result}
\end{equation}
where the arc and edge contributions are
\begin{align*}
\Omega_{cc'} & \equiv\hf\left[\De h_{c}^{c'},\De h_{c}^{c'}\right]\cdot\X_{c}^{c'}-\De h_{c}^{c'}\cdot\delta\X_{c}^{c'},\\
\Omega_{vc} & \equiv\hf\left[\De h_{v}^{c},\De h_{v}^{c}\right]\cdot\X_{v}^{c}-\De h_{v}^{c}\cdot\delta\X_{v}^{c},
\end{align*}
while the vertex contribution is 
\begin{align*}
\Omega_{v} & \equiv\delta\X_{v}\cdot\delta\M_{v}-\left(\delta\SS_{v}+\left[\delta\M_{v},\X_{v}\right]+\left[\M_{v},\delta\X_{v}\right]\right)\cdot\De\left(h_{v}\left(v\right)\right)+\\
 & \qquad-\hf\left(\SS_{v}+\left[\M_{v},\X_{v}\right]\right)\cdot\left[\De\left(h_{v}\left(v\right)\right),\De\left(h_{v}\left(v\right)\right)\right].
\end{align*}

\subsubsection{The Gauss Constraint at the Nodes}

Consider the infinitesimal version of the symmetry transformation
\eqref{eq:symmetry-c} acting on $h_{cc'}$, $h_{cv}$, $\y_{c}$
and $\y_{v}$, with $g_{c}\equiv\e^{\be_{c}}$: 
\[
\delta_{\left(\z_{c},\be_{c}\right)}h_{cc'}=\be_{c}h_{cc'}\sp\delta_{\left(\z_{c},\be_{c}\right)}h_{cv}=\be_{c}h_{cv},\qquad\delta_{\left(\z_{c},\be_{c}\right)}\y_{c}=\z_{c}+\left[\be_{c},\y_{c}\right],\qquad\delta_{\left(\z_{c},\be_{c}\right)}\y_{v}=0.
\]
From \eqref{eq:flux-cc} and \eqref{eq:flux-cv}, we see that the
fluxes $\X_{c}^{c'}$ and $\X_{c}^{v}$ transform as follows: 
\[
\delta_{\left(\z_{c},\be_{c}\right)}\X_{c}^{c'}=\left[\be_{c},\X_{c}^{c'}\right]\sp\delta_{\left(\z_{c},\be_{c}\right)}\X_{c}^{v}=\left[\be_{c},\X_{c}^{v}\right].
\]
Note that the translation parameter $\z_{c}$ cancels out, so this
transformation is in fact a pure rotation. Also note that this transformation
only affects holonomies and fluxes which involve the specific cell
$c$ with respect to which we are performing the transformation.

Applying the transformation to $\Omega$, we find 
\[
I_{\delta_{\left(\z_{c},\be_{c}\right)}}\Omega=-\be_{c}\cdot\delta\left(\sum_{c'\ni c}\X_{c}^{c'}-\sum_{v\ni c}\X_{c}^{v}\right)=-\be_{c}\cdot\delta\G_{c},
\]
and thus we have proven that the cell Gauss constraint $\G_{c}$ generates
rotations at the nodes. Since the translation parameter $\z_{c}$
cancels out in this calculation, the translational edge mode symmetry
is pure gauge.

\subsubsection{The Gauss Constraint at the Vertices}

In order to analyze the Gauss constraint at the vertex one first has
to recognize that only the part of $\G_{v}$ along the Cartan subalgebra
is a constraint. The part orthogonal to it can simply be viewed as
a definition of $\y_{v}^{\perp}\left(v_{0}\right)$, as emphasized
earlier. Therefore, at the vertex, the only constraint is $\be_{v}\cdot\G_{v}=0$
for $\beta_{v}\in\mfh$.

We consider the infinitesimal version of the symmetry transformation
\eqref{eq:symmetry-v} acting on $h_{c}$, $h_{v}$, $\y_{c}$ and
$\y_{v}$, with $g_{v}\equiv\e^{\be_{v}}$: 
\[
\delta_{\left(\z_{v},\be_{v}\right)}h_{c}=0\sp\delta_{\left(\z_{v},\be_{v}\right)}h_{v}=\be_{v}h_{v}\sp\delta_{\left(\z_{c},\be_{c}\right)}\y_{c}=0\sp\delta_{\left(\z_{v},\be_{v}\right)}\y_{v}=\z_{v}+\left[\be_{v},\y_{v}\right].
\]
Note that $\z_{v}$ and $\be_{v}$ are both in the Cartan subalgebra.
From \eqref{eq:sym-vc}, \eqref{eq:flux-cc} and \eqref{eq:flux-vc},
we see that the holonomies $h_{cc'}$, $h_{vc}$, and $h_{v}(v)$
and the fluxes $\X_{c}^{c'}$, $\X_{v}^{c}$ and $\X_{v}$ transform
as follows:
\[
\delta_{\left(\z_{v},\be_{v}\right)}h_{cc'}=0\sp\delta_{\left(\z_{v},\be_{v}\right)}h_{vc}=\be_{v}h_{vc}\sp\delta_{\left(\z_{v},\be_{v}\right)}h_{v}\left(v\right)=\be_{v}h_{v}\left(v\right),
\]
\[
\delta_{\left(\z_{v},\be_{v}\right)}\X_{c}^{c'}=0\sp\delta_{\left(\z_{v},\be_{v}\right)}\X_{v}^{c}=\left[\be_{v},\X_{v}^{c}\right]\sp\delta_{\left(\z_{v},\be_{v}\right)}\X_{v}=\left[\be_{v},\X_{v}\right],
\]
where we used the fact that $\z_{v}$ and $\be_{v}$ commute with
$\M_{v}$ and $\SS_{v}$. Again, the translation parameter $\z_{v}$
cancels out, so the transformation is in fact a pure rotation on the
holonomies and fluxes. Also note that this transformation only affects
holonomies and fluxes which involve the specific vertex $v$ with
respect to which we are performing the transformation.

Applying the transformation to $\Omega$, we find 
\[
I_{\delta_{\left(\z_{v},\be_{v}\right)}}\Omega=-\be_{v}\cdot\delta\left(\SS_{v}-\sum_{c\in v}\X_{v}^{c}\right),
\]
where we have again used the fact that $\z_{v}$ and $\be_{v}$ commute
with $\M_{v}$ and $\SS_{v}$, as well as the Jacobi identity. Importantly,
since $\left[\be_{v},\e^{\M_{v}}\right]=0$ and $\delta\be_{v}=0$,
we have that 
\[
\be_{v}\cdot\delta\left(\e^{\M_{v}}\y_{v_{0}}\left(v\right)\e^{-\M_{v}}-\y_{v_{0}}\left(v\right)\right)=0,
\]
where we used the fact that the trace in the definition of the dot
product is cyclic. Therefore we see that, in fact, 
\[
I_{\delta_{\left(\z_{v},\be_{v}\right)}}\Omega=-\be_{v}\cdot\delta\G_{v}.
\]
Thus, we have proven that the disk Gauss constraint $\G_{v}$ generates
rotations at the vertices. Again, since the translation parameter
$\z_{v}$ cancels out in this calculation, the edge mode translations
at the vertices are pure gauge.

\subsubsection{The Curvature Constraint}

We are left with the translation symmetry generated by the curvature
constraint \eqref{eq:curv-con}. As before, let the vertex $v$ be
surrounded by $N$ cells $c_{1},\ldots,c_{N}$, and take $c_{1}\equiv c_{N+1}\equiv c$.
Then we define
\[
H_{vc}\equiv h_{vc}h_{cc_{2}}\cdots h_{c_{N}c}\sp H_{cv}\equiv H_{vc}^{-1},
\]
such that \eqref{eq:curv-con} becomes $F_{v}=H_{vc}h_{cv}\e^{-\M_{v}}=1.$
Now, let $\x_{v}$ be a transformation parameter, and let us consider
the transformation 
\[
\delta_{\x_{v}}\X_{c}^{c'}=-H_{cv}\x_{v}H_{vc},\qquad\delta_{\x_{v}}\X_{v}^{c}=\x_{v}-\e^{-\M_{v}}\x_{v}\e^{\M_{v}},\qquad\delta_{\x_{v}}\X_{v}=-\x_{v}^{\parallel},
\]
where we denoted
\[
\x_{v}^{\parallel}\equiv\left(\frac{\x_{v}\cdot\M_{v}}{\M_{v}^{2}}\right)\M_{v},
\]
and where the only fluxes affected are $\X_{v}^{c}$, $\X_{v}$ and
$\X_{c}^{c'}$ corresponding to links surrounding the vertex $v$.
Then we find that
\[
I_{\delta_{\x_{v}}}\Omega_{cc'}=-\x_{v}\cdot H_{vc}\De h_{c}^{c'}H_{cv}\sp I_{\delta_{\x_{v}}}\Omega_{v}=-\x_{v}\cdot\delta\M_{v},
\]
\[
I_{\delta_{\x_{v}}}\Omega_{vc}=\x_{v}\cdot\left(\De h_{v}^{c}-\e^{\M_{v}}\De h_{v}^{c}\e^{-\M_{v}}\right),
\]
and the total symplectic form transforms as
\begin{align*}
I_{\delta_{\x_{v}}}\Omega & =-\x_{v}\cdot\left(\sum_{i=1}^{N}H_{vc_{i}}\De h_{c_{i}}^{c_{i+1}}H_{c_{i}v}+\De h_{v}^{c}-\e^{\M_{v}}\De h_{v}^{c}\e^{-\M_{v}}-\delta\M_{v}\right)\\
 & =-\x_{v}\cdot\De F_{v},
\end{align*}
which shows that $\delta_{\x_{v}}$ is the translation symmetry associated
with $F_{v}$. Quite remarkably, this symmetry can be used to simplify
the expression for $\X_{v}$ and the vertex Gauss constraint. Let
us consider the transformation parameter 
\[
\x_{v}\equiv-\frac{1}{N_{v}}\left(\e^{\M_{v}\left(\alpha_{v}+\hf\right)}\y_{v}^{\perp}\left(v_{0}\right)\e^{-\M_{v}\left(\alpha_{v}+\hf\right)}\right)+\y_{v}^{\parallel}\left(v_{0}\right),
\]
where $N_{v}$ is the number of cells around $v$. Note that
\[
\x_{v}-\e^{-\M_{v}}\x_{v}\e^{\M_{v}}=\x_{v}^{\perp}-\e^{-\M_{v}}\x_{v}^{\perp}\e^{\M_{v}},
\]
since the parallel part $\x_{v}^{\parallel}$ commutes with $\M_{v}$.
We define the new set of fluxes $\tilde{\X}\equiv\X+\delta_{\x_{v}}\X$.
Then from the expressions \eqref{eq:defpos} and \eqref{Gaussv2},
we can see that the transformed fluxes at the vertices and arcs satisfy
the simpler relations
\[
\tilde{\X}_{v}=\y_{v}\left(v\right)-\alpha_{v}\SS_{v},\qquad\sum_{c\ni v}\tilde{\X}_{v}^{c}=\SS_{v}.
\]
Taking $\alpha_{v}=0$, which we can always do due to the symmetry
transformation \eqref{eq:beta-symmetry-decomp}, we may simplify even
more and obtain simply $\tilde{\X}_{v}=\y_{v}\left(v\right)$.

One has to beware that the transformed fluxes $\tilde{\X}_{c}^{c'}$
on the links no longer satisfy the inverse relations \eqref{eq:X-inverse}:
$\tilde{\X}_{c}^{c'}\ne-h_{cc'}\tilde{\X}_{c'}^{c}h_{c'c}$, since
the expressions on both sides of this equation now involve the differences
between $\x_{v}$ and $\x_{v'}$.

\section{\label{sec:Summary-and-Outlook}Summary and Outlook}

Two types of transformations acting on the group elements arise in
our discretization. Right translations $\HH\mt\HH\GG$, with $\GG$
a group-valued 0-form, correspond to the familiar gauge transformation
of the connection. However, left translations $\HH\mt\GG\HH$, with
$\GG$ constant and possibly restricted to the Cartan subgroup, are
instead a symmetry which leaves the connection invariant.

The appearance of gauge-invariant observables that transform non-trivially
under a new global symmetry is understood in the continuum as the
appearance of new \emph{edge mode} degrees of freedom related to the
presence of boundaries appearing in the subdivision of a large gauge
system into subsystems \cite{Donnelly:2016auv,Rovelli:2013fga}. This
point is usually overlooked and it is often assumed that we can work
in ``a gauge'' where we fixed the elements $\HH_{c}\left(c^{*}\right)$
and $\HH_{v}\left(v\right)$ to the identity.

What this procedure often ignores is the fact that gauge transformations
are transformations in the kernel of the symplectic structure, while
symmetries are not. So before we can postulate that we can ``gauge
fix'' a symmetry, we have to ensure that the variable we are gauge
fixing does not possess a conjugate variable. Otherwise, ``gauge''
fixing would imply reducing the number of degrees of freedom. This
means that we cannot decide beforehand if an edge mode variable can
be dismissed as a gauge variable -- not until we understand its role
in the canonical structure.

We have established, from first principles, a connection between the
continuous 2+1 gravity phase space, constrained to be flat and torsionless
outside defects, and the loop gravity phase space. We have shown that
the vertices carry additional degrees of freedom, behaving as a collection
of relativistic particles coupled to gravity.

This provides a new picture, where pure 2+1 gravity can be equivalently
described using defects carrying ``Poincaré'' charges. This opens
a new perspective on how one can address the continuum limit, similar
to the one proposed in \cite{Delcamp:2016yix}, and opens up the possibility
to directly connect the holonomy-flux variables with the torsion and
curvature holonomies measuring the presence of these defects.

This work opens many new directions to explore. One of the main challenges
is to develop a similar picture in the 3+1-dimensional case. We expect
that it is possible to rewrite the 3+1-dimensional loop gravity phase
space in terms of pointlike defects. Note that, in this case, the
defects are expected to be of codimension 3 instead of codimension
2. The proof of this conjecture is left for future work \cite{barak4d}.

Finally, it is interesting to revisit the choice of polarization that
we made when constructing the symplectic potential. Another natural
choice is the \emph{dual loop polarization} introduced in \cite{Dupuis:2017otn}.
In that case, one expects the piecewise-flat geometry phase space
to reduce to that of (the classical version of) \emph{group networks}
\cite{clement}. In this dual picture, holonomies and fluxes switch
places: the fluxes are on the links and the holonomies are on their
dual edges. This dual formulation will be discussed in a companion
paper, in preparation \cite{barak3d}.

\subsection*{Acknowledgments}

We would like to thank Marc Geiller, Hal Haggard, Etera Livine, Kasia
Rejzner, Aldo Riello, Vasudev Shyam and Wolfgang Wieland for helpful
discussions.

This research was supported in part by Perimeter Institute for Theoretical
Physics. Research at Perimeter Institute is supported by the Government
of Canada through the Department of Innovation, Science and Economic
Development Canada and by the Province of Ontario through the Ministry
of Research, Innovation and Science.

\appendix

\section{\label{sec:The-Maurer-Cartan-Form}The Maurer-Cartan Form on Field
Space}

We define the \emph{Maurer-Cartan form} on field space as follows:
\begin{equation}
\De g\equiv\delta gg^{-1},\label{eq:MC-form}
\end{equation}
where $g$ is a Lie-group-valued 0-form and $\delta$ is the variation,
or differential on field space. Note the bold font on $\De$, denoting
that $\De g$ is valued in the corresponding Lie algebra.

Seen as an operator, $\De$ satisfies the ``Leibniz rule'' 
\begin{equation}
\De\left(hg\right)=\De h+h\left(\De g\right)h^{-1},\label{eq:Delta-Leibniz-rule}
\end{equation}
and the inversion rule 
\begin{equation}
\De(g^{-1})=-g^{-1}\left(\De g\right)g.\label{eq:Delta-inversion-rule}
\end{equation}
Combining them together, we get the useful identity 
\begin{equation}
\De\left(h^{-1}g\right)=h^{-1}\left(\De g-\De h\right)h.\label{eq:Delta-useful-identity}
\end{equation}
Also, it is easy to see that 
\begin{equation}
\delta\left(\De g\right)=\hf\left[\De g,\De g\right].\label{eq:delta-Delta-g}
\end{equation}
Sometimes we will denote a holonomy as, for example, $h_{cc'}$ where
$c$ is the source cell and $c'$ is the target cell for parallel
transport. In this case, our notational convention ensures that the
subscripts are always compatible with adjacent subscripts, as in $h_{c'c}\d\y_{c}^{c'}h_{cc'}$
for example. However, since we have $\De h_{cc'}=\delta h_{cc'}h_{c'c}$,
the proper subscript for this expression is $c$, since it is located
at $c$, while $c'$ is just an internal point which the two holonomies
happen to pass through. We will thus employ, in such cases, the more
appropriate notation 
\begin{equation}
\De h_{c}^{c'}\equiv\De\left(h_{cc'}\right),\label{eq:proper-subscript}
\end{equation}
where the internal point $c'$ is now a superscript, much like in
the notation for $\y_{c}^{c'}$; indeed, the latter was employed for
a similar reason.

\section{\label{sec:Evaluation-of-CS}Evaluation of the Chern-Simons Symplectic
Potential}

The connection $\AAb$ inside the disk $D$ is\textit{\emph{ 
\[
\AAb=\HH^{-1}\LLb\HH+\HH^{-1}\d\HH,
\]
where $\LLb$ is the }}\textit{Lagrangian connection}\textit{\emph{
such that $\LLb\cdot\LLb=0$ and $\HH$ is a $G$-valued 0-form. Its
variation is 
\[
\delta\AAb=\HH^{-1}\left(\delta\LLb+\d_{\LLb}\De\HH\right)\HH,
\]
where $\d_{\LLb}$ denotes the }}\textit{covariant differential}\textit{\emph{
$\d_{\LLb}\equiv\d+[\LLb,\ ]$, and we have used the shorthand notation
$\De\HH\equiv\delta\HH\HH^{-1}$ for the Maurer-Cartan form, introduced
in Appendix }}\ref{sec:The-Maurer-Cartan-Form}.

Let us evaluate the Chern-Simons symplectic form. We have
\[
\omega\left(\AAb\right)=2\delta\LLb\cdot\d_{\LLb}\De\HH+\d_{\LLb}\De\HH\cdot\d_{\LLb}\De\HH,
\]
where we used the fact that $\delta\LLb\cdot\delta\LLb=0$. Now, the
curvature associated to $\LLb$ is $\FFb\left(\LLb\right)\equiv\d\LLb+\hf\left[\LLb,\LLb\right]$,
its variation is $\delta\FFb\left(\LLb\right)=\d_{\LLb}\delta\LLb$,
and it satisfies the Bianchi identity $\d_{\LLb}^{2}=\left[\FFb\left(\LLb\right),\ \right]$.
Using the graded Leibniz rule, we find 
\[
\delta\LLb\cdot\d_{\LLb}\De\HH=\delta\FFb\left(\LLb\right)\cdot\De\HH-\d\left(\delta\LLb\cdot\De\HH\right).
\]
In addition, we have on the one hand 
\[
\d\left(\De\HH\cdot\d_{\LLb}\De\HH\right)=\d\left(\De\HH\cdot\d\De\HH\right)+\d\left(\De\HH\cdot\left[\LLb,\De\HH\right]\right),
\]
and on the other hand 
\[
\d\left(\De\HH\cdot\d_{\LLb}\De\HH\right)=\d_{\LLb}\De\HH\cdot\d_{\LLb}\De\HH+\De\HH\cdot\left[\FFb\left(\LLb\right),\De\HH\right],
\]
and thus 
\[
\d_{\LLb}\De\HH\cdot\d_{\LLb}\De\HH=\d\left(\De\HH\cdot\d\De\HH\right)+\d\left(\De\HH\cdot\left[\LLb,\De\HH\right]\right)-\De\HH\cdot\left[\FFb\left(\LLb\right),\De\HH\right].
\]
Plugging in, we get 
\[
\omega\left(\AAb\right)=2\delta\FFb\left(\LLb\right)\cdot\De\HH-2\d\left(\delta\LLb\cdot\De\HH\right)+\d\left(\De\HH\cdot\d\De\HH\right)+\d\left(\De\HH\cdot\left[\LLb,\De\HH\right]\right)-\De\HH\cdot\left[\FFb\left(\LLb\right),\De\HH\right].
\]
Now, from \eqref{eq:delta-Delta-g} we know that 
\[
\delta\De\HH=\hf\left[\De\HH,\De\HH\right],
\]
and thus, this time using the graded Leibniz rule on field space,
we find 
\[
\delta\LLb\cdot\De\HH=\delta\left(\LLb\cdot\De\HH\right)-\hf\LLb\cdot\left[\De\HH,\De\HH\right],
\]
\[
\delta\FFb\left(\LLb\right)\cdot\De\HH=\delta\left(\FFb\left(\LLb\right)\cdot\De\HH\right)-\hf\FFb\left(\LLb\right)\cdot\left[\De\HH,\De\HH\right].
\]
Plugging into $\omega\left(\AAb\right)$, the triple products all
cancel\footnote{Since $\De\HH\cdot\left[\LLb,\De\HH\right]=-\LLb\cdot\left[\De\HH,\De\HH\right]$
and $\De\HH\cdot\left[\FFb\left(\LLb\right),\De\HH\right]=-\FFb\left(\LLb\right)\cdot\left[\De\HH,\De\HH\right]$
due to graded commutativity on field space.} and we get the general expression 
\begin{equation}
\omega\left(\AAb\right)=2\delta\left(\FFb\left(\LLb\right)\cdot\De\HH\right)+\d\left(\De\HH\cdot\d\De\HH\right)-2\d\delta\left(\LLb\cdot\De\HH\right).\label{eq:omega-A}
\end{equation}
Next, we specialize to the case where the Lagrangian connection $\LLb$
is an Abelian connection of the form considered above: 
\[
\LLb\equiv\MMb\thinspace\d\phi\sp\FFb\left(\LLb\right)=\MMb\thinspace\delta\left(v\right)\sp\MMb\in\mfdh,
\]
where $v$ is the center of the disk $D$. Integrating \eqref{eq:omega-A}
over the disk, we get 
\[
\Omega_{D}\left(\AAb\right)\equiv\int_{D}\omega\left(\AAb\right)=2\delta\left(\MMb\cdot\De\HH\left(v\right)\right)+\oint_{\partial D}\left(\De\HH\cdot\d\De\HH-2\delta\left(\MMb\cdot\De\HH\right)\d\phi\right).
\]
Using the useful identity \eqref{eq:Delta-useful-identity} we find
\[
\De\HH=\HH\left(v\right)\De\left(\HH\left(v\right)^{-1}\HH\right)\HH\left(v\right)^{-1}+\De\HH\left(v\right),
\]
which allows us to write 
\begin{align*}
\Omega_{D}\left(\AAb\right) & =2\delta\left(\MMb\cdot\De\HH\left(v\right)\right)+\oint_{\partial D}\De\HH\cdot\d\De\HH+\\
 & \qquad-2\oint_{\partial D}\delta\left(\HH\left(v\right)^{-1}\MMb\HH\left(v\right)\cdot\De\left(\HH\left(v\right)^{-1}\HH\right)+\MMb\cdot\De\HH\left(v\right)\right)\d\phi.
\end{align*}
Since $\delta\left(\MMb\cdot\De\HH\left(v\right)\right)$ is a constant
with respect to $\phi$, the second term in the second integral becomes
trivial and we see that it exactly cancels the first term (recall
that $\int\d\phi=1$). We are thus left with 
\[
\Omega_{D}\left(\AAb\right)=\oint_{\partial D}\De\HH\cdot\d\De\HH-2\oint_{\partial D}\delta\left(\HH\left(v\right)^{-1}\MMb\HH\left(v\right)\cdot\De\left(\HH\left(v\right)^{-1}\HH\right)\right)\d\phi.
\]
Next, we introduce the $\DG$-valued 0-form $\tHH\left(x\right)\equiv\HH\left(v\right)^{-1}\HH\left(x\right)$,
which satisfies $\tHH\left(v\right)=1$. Then we can write the symplectic
form as 
\[
\Omega_{D}\left(\AAb\right)=\oint_{\partial D}\De\left(\HH\left(v\right)\tHH\right)\cdot\d\De\left(\HH\left(v\right)\tHH\right)-2\oint_{\partial D}\delta\left(\HH\left(v\right)^{-1}\MMb\HH\left(v\right)\cdot\De\tHH\right)\d\phi.
\]
Finally, from \eqref{eq:Delta-Leibniz-rule} and \eqref{eq:Delta-useful-identity}
we see that 
\[
\De\left(\HH\left(v\right)\tHH\right)=\HH\left(v\right)\left(\De\tHH-\De\left(\HH\left(v\right)^{-1}\right)\right)\HH\left(v\right)^{-1}\sp\d\De\left(\HH\left(v\right)\tHH\right)=\HH\left(v\right)\d\De\tHH\HH\left(v\right)^{-1},
\]
since $\HH\left(v\right)$ is constant. Thus we obtain 
\[
\Omega_{D}\left(\AAb\right)=\oint_{\partial D}\De\tHH\cdot\d\De\tHH-\oint_{\partial D}\d\left(\De\left(\HH\left(v\right)^{-1}\right)\cdot\De\tHH\right)-2\oint_{\partial D}\delta\left(\HH\left(v\right)^{-1}\MMb\HH\left(v\right)\cdot\De\tHH\right)\d\phi.
\]
However, the second term vanishes since $\tHH$ is a periodic function
on the circle $\partial D$, and thus we obtain the final expression:
\[
\Omega_{D}\left(\AAb\right)=\oint_{\partial D}\De\tHH\cdot\d\De\tHH-2\oint_{\partial D}\delta\left(\HH\left(v\right)^{-1}\MMb\HH\left(v\right)\cdot\De\tHH\right)\d\phi.
\]

\section{\label{sec:The-Relativistic-Particle}The Relativistic Particle}

In this Appendix we study the relativistic particle symplectic potential
\begin{equation}
\Theta\equiv\X\cdot\delta\M-\left(\SS+\left[\M,\X\right]\right)\cdot\De h,\label{eq:relativistic-particle-potential}
\end{equation}
with the symplectic form 
\[
\Omega\equiv\delta\Theta=\delta\X\cdot\delta\M-\left(\delta\SS+\left[\delta\M,\X\right]+\left[\M,\delta\X\right]\right)\cdot\De h-\hf\left(\SS+\left[\M,\X\right]\right)\cdot\left[\De h,\De h\right].
\]
We define the momentum $\p$ and angular momentum $\j$ of the particle:
\[
\p\equiv h^{-1}\M h\in\mfg^{*},\qquad\j\equiv h^{-1}\left(\SS+\left[\M,\X\right]\right)h\in\mfg,
\]
which have the variational differentials 
\[
\delta\p=h^{-1}\left(\delta\M+\left[\M,\De h\right]\right)h,
\]
\[
\delta\j=h^{-1}\left(\delta\SS+\left[\delta\M,\X\right]+\left[\M,\delta\X\right]+\left[\SS+\left[\M,\X\right],\De h\right]\right)h.
\]
We also define the ``position'' 
\[
\q\equiv h^{-1}\X h\in\mfg,
\]
in terms of which the symplectic potential may be written as 
\[
\Theta=\q\cdot\delta\p-\SS\cdot\De h.
\]

\subsection{Right Translations (Gauge Transformations)}

Let 
\[
\HH\equiv\left(\X,h\right)=e^{\X}h\in\DG\sp\GG\equiv\left(g,\x\right)\in\DG\soosp\HH\GG=\e^{\X+h\x h^{-1}}hg.
\]
This is a right translation, with parameter $\GG$, of the group element
$\HH$, which corresponds to a gauge transformation: 
\[
h\mt hg,\qquad\X\mt\X+h\x h^{-1},\qquad\M\mt\M\sp\SS\mt\SS.
\]
It is interesting to translate this action onto the physical variables
$(\p,\q,\j)$ which transform as 
\[
\p\to h^{-1}\p h,\qquad\q\to\x+h^{-1}\q h,\qquad\j\to h^{-1}\j h.
\]
This shows that the parameter $h$ labels a rotation of the physical
variables, while $\x$ labels a translation of the physical position
$\q$. Taking $g\equiv\e^{\al}$, we may consider transformations
labeled by $\x+\al\in\mfdg$ with $\x\in\mfg^{*}$ a translation parameter
and $\al\in\mfg$ a rotation parameter, given by the infinitesimal
version of the gauge transformation: 
\[
\delta_{\left(\x,\al\right)}h=h\al,\qquad\delta_{\left(\x,\al\right)}\X=h\x h^{-1},\qquad\delta_{\left(\x,\al\right)}\M=0\sp\delta_{\left(\x,\al\right)}\SS=0.
\]
Let $I$ denote the interior product on field space, associated with
the variational exterior derivative $\delta$. Then one finds that
this transformation is Hamiltonian: 
\[
I_{\delta_{\left(\x,\al\right)}}\Omega=-\delta H_{\left(\x,\al\right)}\sp H_{\left(\x,\al\right)}\equiv-\left(\p\cdot\x+\j\cdot\al\right).
\]
This shows that the variable conjugated to $\p$ is the ``position''
$\q\equiv h^{-1}\y h$, while the angular momentum $\j$ generates
right translations on $G$. The Poisson bracket between two such Hamiltonians
is given by 
\[
\left\{ H_{\left(\x,\al\right)},H_{\left(\x'.\al'\right)}\right\} =\delta_{\left(\x,\al\right)}H_{\left(\x',\al'\right)}=H_{\left(\left[\al,\x'\right]+\left[\x,\al'\right],\left[\al,\al'\right]\right)},
\]
which reproduces, as expected, the symmetry algebra $\mfdg$.

\subsection{Left Translations (Symmetry Transformations)}

Similarly, let 
\[
\HH\equiv\left(h,\X\right)\in\DG\sp\GG\equiv\left(g,\z\right)\in\DG\soosp\GG\HH=\e^{\z+g\X g^{-1}}gh.
\]
This is a left translation, with parameter $\GG$, of the group element
$\HH$, which corresponds to a symmetry that leaves the connection
invariant: 
\[
h\mt gh,\qquad\X\mt\z+g\X g^{-1},\qquad\M\mt g\M g^{-1},\qquad\SS\mt g\left(\SS+\left[\z,\M\right]\right)g^{-1}.
\]
Note that it commutes with the right translation. The infinitesimal
transformation, with $g\equiv\e^{\be}$, is 
\begin{equation}
\delta_{\left(\z,\be\right)}h=\be h,\qquad\delta_{\left(\z,\be\right)}\X=\z+\left[\be,\X\right],\qquad\delta_{\left(\z,\be\right)}\M=\left[\be,\M\right],\qquad\delta_{\left(\z,\be\right)}\SS=\left[\be,\SS\right]+\left[\z,\M\right].\label{eq:transf-z-beta}
\end{equation}
Once again, we can prove that this transformation is Hamiltonian:
\[
I_{\delta_{\left(\z,\be\right)}}\Omega=-\delta H_{\left(\z,\be\right)}\sp H_{\left(\z,\be\right)}\equiv-\left(\M\cdot\z+\SS\cdot\be\right).
\]
This follows from the fact that 
\[
\delta_{\left(\z,\be\right)}\left(\SS+\left[\M,\X\right]\right)=\left[\be,\SS+\left[\M,\X\right]\right],
\]
which implies that these transformations leave the momentum and angular
momentum invariant: $\delta_{\left(\z,\be\right)}\p=0=\delta_{\left(\z,\be\right)}\j.$

\subsection{Restriction to the Cartan Subalgebra}

In the case discussed in this paper, where $\M\in\mfh^{*}$ and $\SS\in\mfh$
are in the Cartan subalgebra, we need to restrict the parameter of
the left translation transformation to be in $\mfdh$. A particular
class of transformations of this type is when the parameter is itself
a function of $\M$ and $\SS$, which we shall denote $F\left(\M,\SS\right)$.
One finds that the infinitesimal transformation 
\begin{equation}
\delta_{F}h=\frac{\partial F}{\partial\SS}h\sp\delta_{F}\y=\frac{\partial F}{\partial\M}+\left[\frac{\partial F}{\partial\SS},\X\right]\sp\delta_{F}\M=0\sp\delta_{F}\SS=0,\label{eq:transf-F}
\end{equation}
is Hamiltonian: 
\[
I_{\delta_{F}}\Omega=-\delta H_{F}\sp H_{F}\equiv-F\left(\M,\SS\right).
\]
In particular, taking 
\[
F\left(\M,\SS\right)\equiv\frac{\xi}{2}\M^{2}+\chi\M\cdot\SS\sp\xi,\chi\in\BBR,
\]
we obtain the Hamiltonian transformation 
\begin{equation}
\delta_{F}h=\M\chi h\sp\delta_{F}\X=\M\xi+\left(\SS+\left[\M,\X\right]\right)\chi\sp\delta_{F}\M=0\sp\delta_{F}\SS=0,\label{eq:transf-F-Casimir}
\end{equation}
corresponding to \eqref{eq:transf-z-beta} with 
\[
\z=\frac{\partial F}{\partial\M}=\M\xi+\SS\chi\sp\be=\frac{\partial F}{\partial\SS}=\M\chi.
\]
This may be integrated to 
\[
h\mt\e^{\M\chi}h,\qquad\X\mt\e^{\M\chi}\left(\M\xi+\SS\chi+\X\right)\e^{-\M\chi},\qquad\M\mt\M,\qquad\SS\mt\SS.
\]
The Hamiltonians $\M^{2}$ and $\M\cdot\SS$ represent the Casimir
invariants of the algebra $\mfdg$.

\section{\label{sec:A-Quicker-Derivation}A Quicker Derivation of the Symplectic
Potential}

Using the non-periodic variables
\[
u_{v}\equiv\e^{\M_{v}\phi_{v}}h_{v}\sp\w_{v}\equiv\e^{\M_{v}\phi_{v}}\left(\y_{v}+\SS_{v}\phi_{v}\right)\e^{-\M_{v}\phi_{v}},
\]
which were defined in Sec. \ref{subsec:The-Gauss-Constraint}, we
may perform the calculation of Sec. \ref{sec:Discretizing-the-Symplectic}
in a quicker and clearer way. The symplectic potential is given as
before by
\[
\Theta=\sum_{c}\Theta_{c}+\sum_{v}\Theta_{v^{*}}\sp\Theta_{c}\equiv-\int_{\ct}\E\cdot\delta\A,\qquad\Theta_{v^{*}}\equiv-\int_{v^{*}}\E\cdot\delta\A.
\]
On the cells, the calculation is the same, and we obtain as before
\[
\Theta_{c}=\int_{\partial\ct}\d\y_{c}\cdot\De h_{c}.
\]
On the disks, we have
\[
\A\bl_{v^{*}}=u_{v}^{-1}\d u_{v}\sp\E\bl_{v^{*}}=u_{v}^{-1}\d\w_{v}u_{v}\sp\delta\A\bl_{v^{*}}=u_{v}^{-1}\left(\d\De u_{v}\right)u_{v},
\]
and thus
\[
\Theta_{v^{*}}=-\int_{v^{*}}\E\cdot\delta\A=-\int_{v^{*}}\d\w_{v}\cdot\d\De u_{v}=\int_{v^{*}}\d\left(\d\w_{v}\cdot\De u_{v}\right)=\int_{\partial v^{*}}\d\w_{v}\cdot\De u_{v}.
\]
The boundary of the punctured disk decomposes into $\partial v^{*}=\partial_{0}v^{*}\cup\partial_{R}v^{*}\cup C_{v}$.
Since our variables are now non-periodic, we must also integrate on
the cut $C_{v}$, which we did not need to do before. Thus
\[
\Theta_{v^{*}}\equiv\Theta_{\partial_{R}v^{*}}-\Theta_{\partial_{0}v^{*}}-\Theta_{C_{v}},
\]
where
\[
\Theta_{\partial_{R}v^{*}}\equiv\int_{\partial_{R}v^{*}}\d\w_{v}\cdot\De u_{v}\sp\Theta_{\partial_{0}v^{*}}\equiv\int_{\partial_{0}v^{*}}\d\w_{v}\cdot\De u_{v}\sp\Theta_{C_{v}}\equiv\int_{\partial_{0}v^{*}}\d\w_{v}\cdot\De u_{v}.
\]
Now, we have
\[
\d\w_{v}=\e^{\M_{v}\phi_{v}}\left(\d\y_{v}+\left(\SS_{v}+\left[\M_{v},\y_{v}\right]\right)\d\phi_{v}\right)\e^{-\M_{v}\phi_{v}},
\]
and from \eqref{eq:Delta-Leibniz-rule} we find
\[
\De u_{v}=\De\left(\e^{\M_{v}\phi_{v}}h_{v}\right)=\e^{\M_{v}\phi_{v}}\left(\delta\M_{v}\phi_{v}+\De h_{v}\right)\e^{-\M_{v}\phi_{v}}.
\]
Thus we may write the integrand as
\[
\d\w_{v}\cdot\De u_{v}=\d\y_{v}\cdot\left(\delta\M_{v}\phi_{v}+\De h_{v}\right)+\left(\SS_{v}\cdot\delta\M_{v}\phi_{v}+\left(\SS_{v}+\left[\M_{v},\y_{v}\right]\right)\cdot\De h_{v}\right)\d\phi_{v}.
\]
Integrating this over the inner boundary $\partial_{0}v^{*}$ is easy,
since the integrand is evaluated at the vertex $v$, and $\y_{v}\left(v\right)$
and $h_{v}\left(v\right)$ are constant with respect to $\phi_{v}$,
by assumption. The integral is from $\phi_{v}=\alpha_{v}-1/2$ to
$\phi_{v}=\alpha_{v}+1/2$, and we immediately get:
\[
\Theta_{\partial_{0}v^{*}}=\alpha_{v}\SS_{v}\cdot\delta\M_{v}+\left(\SS_{v}+\left[\M_{v},\y_{v}\left(v\right)\right]\right)\cdot\De\left(h_{v}\left(v\right)\right).
\]
On the cut $C_{v}$, we have contributions from both sides, one at
$\phi_{v}=\alpha_{v}-1/2$ and another at $\phi_{v}=\alpha_{v}+1/2$,
with opposite orientation. Since $\d\phi_{v}=0$ on the cut, only
the term $\d\y_{v}\cdot\left(\delta\M_{v}\phi_{v}+\De h_{v}\right)$
contributes, and we get:
\begin{align*}
\Theta_{C_{v}} & =\int_{r=0}^{R}\left(\d\y_{v}\cdot\left(\delta\M_{v}\phi_{v}+\De h_{v}\right)\bgl_{\phi_{v}=\alpha_{v}+1/2}-\d\y_{v}\cdot\left(\delta\M_{v}\phi_{v}+\De h_{v}\right)\bgl_{\phi_{v}=\alpha_{v}-1/2}\right)\\
 & =\int_{r=0}^{R}\d\y_{v}\cdot\delta\M_{v}=\left(\y_{v}\left(v_{0}\right)-\y_{v}\left(v\right)\right)\cdot\delta\M_{v},
\end{align*}
since $\y_{v}$ has the value $\y_{v}\left(v_{0}\right)$ at $r=R$
on the cut and $\y_{v}\left(v\right)$ at $r=0$. The vertex symplectic
potential is then obtained by defining $\Theta_{v}\equiv-\left(\Theta_{\partial_{0}v^{*}}+\Theta_{C_{v}}\right)$:
\[
\Theta_{v}=\left(\y_{v}\left(v\right)-\y_{v}\left(v_{0}\right)-\alpha_{v}\SS_{v}\right)\cdot\delta\M_{v}-\left(\SS_{v}+\left[\M_{v},\y_{v}\left(v\right)\right]\right)\cdot\De\left(h_{v}\left(v\right)\right).
\]
In this way, we have immediately obtained $\Theta_{v}$ right from
the beginning via the integration on the inner boundary and the cut,
without ever having to invoke the continuity conditions or go through
the trouble of collecting terms from different arcs later on, as we
did in the main text (see Secs. \ref{subsec:The-Arc-Contributions}
and \ref{subsec:The-Vertex-Contribution}).

To find the rest of the symplectic potential, we split it into contributions
from the edges and arcs: 
\[
\Theta=\sum_{\left[cc'\right]}\Theta_{cc'}+\sum_{\left(vc\right)}\Theta_{vc}+\sum_{v}\Theta_{v},
\]
where
\[
\Theta_{cc'}\equiv\int_{\left[cc'\right]}\left(\d\y_{c}\cdot\De h_{c}-\d\y_{c'}\cdot\De h_{c'}\right)\sp\Theta_{vc}=\int_{\left(vc\right)}\left(\d\w_{v}\cdot\De u_{v}-\d\y_{c}\cdot\De h_{c}\right).
\]
For the edges, we use the continuity conditions as we did in Sec.
\ref{subsec:Simplifying-the-Edge} and get
\[
\Theta_{cc'}=\De h_{c}^{c'}\cdot\int_{\left[cc'\right]}\d\y_{c}.
\]
For the arcs we may now use much simpler continuity conditions formulated
in terms of $u_{v}$ and $\w_{v}$,
\[
h_{c}\left(x\right)=h_{cv}u_{v}\left(x\right)\sp\y_{c}\left(x\right)=h_{cv}\left(\w_{v}\left(x\right)-\y_{v}^{c}\right)h_{vc}\sp x\in\left(vc\right),
\]
and thus we may simplify $\Theta_{vc}$ in exactly the same way as
we did for $\Theta_{cc'}$, and we immediately obtain
\[
\Theta_{\left(vc\right)}=\De h_{v}^{c}\cdot\int_{\left(vc\right)}\d\w_{v}.
\]
We have thus reproduced the result \eqref{eq:Theta_cc-vc} without
having to go through the many steps we took in the main text. However,
the calculation in the main text is more explicit, and thus leaves
less room for error; the fact that we have obtained the same result
in both calculations is a good consistency check.

\bibliographystyle{Utphys}
\phantomsection\addcontentsline{toc}{section}{\refname}\bibliography{2P1-LQG-On-The-Edge}

\end{document}

%% file: Figure-1-Triangle.tex
\begin{tikzpicture}[scale=0.8]
	\begin{pgfonlayer}{nodelayer}
		\node [style=none] (0) at (0.25, -1.9) {$c^{*}$};
		\node [style=none] (1) at (0, -0.45) {$v$};
		\node [style=none] (2) at (1.5, 0.6) {$c^{\prime*}$};
		\node [style=none] (3) at (5.25, -3.4) {$v'$};
		\node [style=Vertex] (4) at (0, -0) {};
		\node [style=Vertex] (5) at (0, 4) {};
		\node [style=Vertex] (6) at (-5, -3) {};
		\node [style=Vertex] (7) at (5, -3) {};
		\node [style=Node] (8) at (0, -1.5) {};
		\node [style=Node] (9) at (1.25, 1) {};
		\node [style=Node] (10) at (-1.25, 1) {};
		\node [style=none] (11) at (-3.5, 3.25) {};
		\node [style=none] (12) at (3.5, 3.25) {};
		\node [style=none] (13) at (0, -4.5) {};
	\end{pgfonlayer}
	\begin{pgfonlayer}{edgelayer}
		\draw [style=Edge] (5) to (6);
		\draw [style=Edge] (6) to (7);
		\draw [style=Edge] (7) to (5);
		\draw [style=Edge] (5) to (4);
		\draw [style=Edge] (4) to (7);
		\draw [style=Edge] (4) to (6);
		\draw [style=Link] (8) to (9);
		\draw [style=Link] (9) to (10);
		\draw [style=Link] (10) to (8);
		\draw [style=Link] (10) to (11.center);
		\draw [style=Link] (9) to (12.center);
		\draw [style=Link] (8) to (13.center);
	\end{pgfonlayer}
\end{tikzpicture}

%% file: Figure-2-Disk.tex
\begin{tikzpicture}
	\node [style=Vertex] at (0, 0) {};
	\node [style=none] at (0, -0.33) {$v$};
	\node [style=none] at (0, 0.75) {$\partial_{0}v^{*}$};
	\node [style=none] at (0, 3.25) {$\partial_{R}v^{*}$};
	\node [style=none] at (1.5, 0) {$C_{v}$};
	\centerarc [style=SegmentArrow] (0,0) (330:30:0.5);
	\centerarc [style=SegmentArrow] (0,0) (5:355:3);
	\draw [red,thick] ($({0.5*cos(30)},{0.5*sin(30)})$) -- node {\midarrow} ($({3*cos(5)},{0.5*sin(30)})$);
	\draw [red,thick]  ($({0.5*cos(330)},{0.5*sin(330)})$) -- node {\midarrowop} ($({3*cos(355)},{0.5*sin(330)})$);
	\node [style=Vertex] () at (3, 0) {};
	\node [style=none] () at (3.33, -0.33) {$v_{0}$};
\end{tikzpicture}

%% file: Figure-3-ContinuityNN.tex
\begin{tikzpicture}
	\begin{pgfonlayer}{nodelayer}
		\node [style=Vertex] (0) at (0, -2) {};
		\node [style=Vertex] (1) at (0, 2) {};
		\node [style=Vertex] (2) at (-3, -0) {};
		\node [style=Vertex] (3) at (3, -0) {};
		\node [style=Node] (4) at (-1.25, -0) {};
		\node [style=Node] (5) at (1.25, -0) {};
		\node [style=none] (6) at (-1.5, -0.3) {$c^{*}$};
		\node [style=none] (7) at (1.5, -0.3) {$c^{\prime*}$};
		\node [style=none] (8) at (0, -0.75) {};
		\node [style=none] (9) at (0, 0.25) {$(cc')$};
		\node [style=none] (10) at (-0.75, -0.87) {$\mathcal{H}_{c}(x)$};
		\node [style=none] (11) at (0.75, -0.87) {$\mathcal{H}_{c'}^{-1}(x)$};
	\end{pgfonlayer}
	\begin{pgfonlayer}{edgelayer}
		\draw [style=Edge] (1) to (2);
		\draw [style=Edge] (2) to (0);
		\draw [style=Edge] (0) to (3);
		\draw [style=Edge] (3) to (1);
		\draw [style=Edge] (1) to (0);
		\draw [style=SegmentArrow] (4) to (8.center);
		\draw [style=SegmentArrow] (8.center) to (5);
	\end{pgfonlayer}
\end{tikzpicture}

%% file: Figure-4-ContinuityNV.tex
\begin{tikzpicture}
	\begin{pgfonlayer}{nodelayer}
		\node [style=Node] (0) at (0, -1) {};
		\node [style=Vertex] (1) at (0, 3) {};
		\node [style=Vertex] (2) at (-4, -1.75) {};
		\node [style=Vertex] (3) at (4, -1.75) {};
		\node [style=none] (4) at (0.25, -1.25) {$c^{*}$};
		\node [style=none] (5) at (-0.25, 3.25) {$v$};
		\node [style=none] (6) at (-1.25, 1.5) {};
		\node [style=none] (7) at (1.25, 1.5) {};
		\node [style=none] (8) at (-1, 1.29) {};
		\node [style=none] (9) at (0.33, 0.725) {$(vc)$};
		\node [style=none] (12) at (-1.25, 0.25) {$\mathcal{H}_{c}^{-1}(x)$};
		\node [style=none] (13) at (-1.2, 2.1) {$\mathcal{H}_{v}(x)$};
	\end{pgfonlayer}
	\begin{pgfonlayer}{edgelayer}
		\draw [style=Edge] (1) to (2);
		\draw [style=Edge] (2) to (3);
		\draw [style=Edge] (3) to (1);
		\draw [style=Edge, bend right=45, looseness=1.00] (6.center) to (7.center);
		\draw [style=SegmentArrow] (8.center) to (0);
		\draw [style=SegmentArrow] (1) to (8.center);
	\end{pgfonlayer}
\end{tikzpicture}

%% file: Figure-5-CellBoundary.tex
\begin{tikzpicture}[scale=0.9]
	\begin{pgfonlayer}{nodelayer}
		\node [style=Node] (0) at (0, -0) {};
		\node [style=Vertex] (1) at (0, 3) {};
		\node [style=Vertex] (2) at (-4, -1.75) {};
		\node [style=Vertex] (3) at (4, -1.75) {};
		\node [style=none] (4) at (0.25, -0.3) {$c^{*}$};
		\node [style=none] (5) at (-0.25, 3.4) {$v_1$};
		\node [style=none] (6) at (-1.25, 1.5) {};
		\node [style=none] (7) at (1.25, 1.5) {};
		\node [style=none] (8) at (2.5, -0) {};
		\node [style=none] (9) at (2, -1.75) {};
		\node [style=none] (10) at (-2.5, -0) {};
		\node [style=none] (11) at (-2, -1.75) {};
		\node [style=none] (12) at (-4.25, -2.2) {$v_2$};
		\node [style=none] (13) at (4.25, -2.2) {$v_3$};
		\node [style=none] (14) at (0, 0.7) {$(v_1c)$};
		\node [style=none] (15) at (-1.33, -0.75) {$(v_2c)$};
		\node [style=none] (16) at (1.33, -0.75) {$(v_3c)$};
		\node [style=none] (17) at (-2.33, 1) {$[cc_1]$};
		\node [style=none] (18) at (0, -2.1) {$[cc_2]$};
		\node [style=none] (19) at (2.33, 1) {$[cc_3]$};
		\node [style=Node] (20) at (3, 2) {};
		\node [style=Node] (21) at (-3, 2) {};
		\node [style=Node] (22) at (0, -3) {};
		\node [style=none] (23) at (3.25, 1.65) {$c_3^{*}$};
		\node [style=none] (24) at (-2.75, 1.65) {$c_1^{*}$};
		\node [style=none] (25) at (0.25, -3.35) {$c_2^{*}$};
	\end{pgfonlayer}
	\begin{pgfonlayer}{edgelayer}
		\draw [style=Link, bend right=45, looseness=1.00] (6.center) to (7.center);
		\draw [style=Link, bend left=45, looseness=1.00] (10.center) to (11.center);
		\draw [style=Link, bend right=45, looseness=1.00] (8.center) to (9.center);
		\draw [style=Link] (8.center) to (7.center);
		\draw [style=Link] (6.center) to (10.center);
		\draw [style=Link] (11.center) to (9.center);
		\draw [style=Edge] (1) to (6.center);
		\draw [style=Edge] (10.center) to (2);
		\draw [style=Edge] (2) to (11.center);
		\draw [style=Edge] (9.center) to (3);
		\draw [style=Edge] (3) to (8.center);
		\draw [style=Edge] (7.center) to (1);
	\end{pgfonlayer}
\end{tikzpicture}

%% file: Figure-6-IntersectionPoints.tex
\begin{tikzpicture}[scale=0.9]
	\begin{pgfonlayer}{nodelayer}
		\node [style=Node] (0) at (0, -0) {};
		\node [style=Vertex] (1) at (0, 3) {};
		\node [style=Vertex] (2) at (-4, -1.75) {};
		\node [style=Vertex] (3) at (4, -1.75) {};
		\node [style=none] (4) at (0.25, -0.3) {$c^{*}$};
		\node [style=none] (5) at (-0.25, 3.4) {$v_1$};
		\node [style=Point] (6) at (-1.25, 1.5) {};
		\node [style=Point] (7) at (1.25, 1.5) {};
		\node [style=Point] (8) at (2.5, -0) {};
		\node [style=Point] (9) at (2, -1.75) {};
		\node [style=Point] (10) at (-2.5, -0) {};
		\node [style=Point] (11) at (-2, -1.75) {};
		\node [style=none] (12) at (-4.25, -2.17) {$v_2$};
		\node [style=none] (13) at (4.25, -2.17) {$v_3$};
		\node [style=Node] (14) at (3.5, 2.5) {};
		\node [style=Node] (15) at (-3.5, 2.5) {};
		\node [style=Node] (16) at (0, -3.5) {};
		\node [style=none] (17) at (3.8, 2.1) {$c_3^{*}$};
		\node [style=none] (18) at (-3.8, 2.1) {$c_1^{*}$};
		\node [style=none] (19) at (0.25, -3.9) {$c_2^{*}$};
		\node [style=none] (20) at (1.25, 1.75) {$\tau_{v_1c}=\tau_{cc_3}$};
		\node [style=none] (21) at (-1.25, 1.75) {$\sigma_{v_1c}=\sigma_{cc_1}$};
		\node [style=none] (22) at (-2.5, 0.25) {$\tau_{v_2c}=\tau_{cc_1}$};
		\node [style=none] (23) at (-2, -2) {$\sigma_{v_2c}=\sigma_{cc_2}$};
		\node [style=none] (24) at (2, -2) {$\tau_{v_3c}=\tau_{cc_2}$};
		\node [style=none] (25) at (2.5, 0.25) {$\sigma_{v_3c}=\sigma_{cc_3}$};
	\end{pgfonlayer}
	\begin{pgfonlayer}{edgelayer}
		\draw [style=LinkArrow, bend right=45, looseness=1.00] (6) to (7);
		\draw [style=LinkArrow, bend right=45, looseness=1.00] (11) to (10);
		\draw [style=LinkArrow, bend right=45, looseness=1.00] (8) to (9);
		\draw [style=LinkArrow] (8) to (7);
		\draw [style=LinkArrow] (6) to (10);
		\draw [style=LinkArrow] (11) to (9);
		\draw [style=Edge] (1) to (6);
		\draw [style=Edge] (10) to (2);
		\draw [style=Edge] (2) to (11);
		\draw [style=Edge] (9) to (3);
		\draw [style=Edge] (3) to (8);
		\draw [style=Edge] (7) to (1);
	\end{pgfonlayer}
\end{tikzpicture}

%% file: Figure-7-Curvature.tex
\begin{tikzpicture}[scale=1.5]
	\node [style=Vertex] (vbottom) at (0, -2) {};
	\node [style=Vertex] (vtop) at (0, 2) {};
	\node [style=Vertex] (vleft) at (-3, -0) {};
	\node [style=Vertex] (vright) at (3, -0) {};
	\node [style=Node] (cleft) at (-1.25, -0) {};
	\node [style=Node] (cright) at (1.25, -0) {};
	\node [style=none] at (0, 2.25) {$v$};
	\node [style=none] at (-1.4, -0.25) {$c^{*}$};
	\node [style=none] at (1.4, -0.25) {$c^{\prime*}$};
	\node [style=none] at (0, -0.2) {$h_{cc'}$};
	\node [style=none] at (-1, 0.9) {$h_{cv}$};
	\node [style=none] at (1, 0.9) {$h_{vc'}$};
	\node [style=none, scale=0.7] at (0, 1.6) {$\mathrm{e}^{\mathbf{M}_v\phi_{v}^{cc'}}$};
	\draw [style=Edge] (vtop) to (vleft);
	\draw [style=Edge] (vleft) to (vbottom);
	\draw [style=Edge] (vbottom) to (vright);
	\draw [style=Edge] (vright) to (vtop);
	\draw [style=Edge] (vtop) to (vbottom);
	\draw [style=SegmentArrow] (cleft) to (cright);
	\centerarc [style=SegmentArrow] (0,2) (230:310:0.25);
	\draw [style=SegmentArrow] (cleft) to ($(0,2)+({0.25*cos(230)},{0.25*sin(230)})$);
	\draw [style=SegmentArrow] ($(0,2)+({0.25*cos(310)},{0.25*sin(310)})$) to (cright);
\end{tikzpicture}